\documentclass{article}
\usepackage{amssymb}

\begin{document}
\author{Anatoly Konechny${\,}^{1}$\footnote{Research supported by the Director, Office of Energy 
Research, Office of High Energy and Nuclear Physics, Division of High 
Energy Physics of the U.S. Department of Energy under Contract 
DE-AC03-76SF00098 and in part by the National Science Foundation grant PHY-95-14797.} \enspace  and  Albert Schwarz${\,}^{2}$
\footnote{Research supported in part by NSF grant DMS-9801009}
 \\
${}^{1}\,$Department of Physics, University of California Berkeley \\
and \\
Theoretical Physics Group, Mail Stop 50A-5101\\
LBNL, Berkeley, CA 94720 USA \\ 
konechny@thsrv.lbl.gov\\
\\
${}^{2}\,$Department of Mathematics, University of California Davis\\
Davis, CA 95616 USA\\
  schwarz@math.ucdavis.edu}
\title{\bf Introduction to M(atrix) theory and noncommutative geometry }
\maketitle
\begin{abstract}
Noncommutative geometry is based on an idea that an associative algebra 
can be regarded as  "an algebra of functions on a noncommutative space".
The major contribution to noncommutative geometry was made by A. Connes,
who, in particular, analyzed Yang-Mills theories on noncommutative spaces,
using important notions that were  introduced in his papers (connection, Chern character, etc).
It was found recently that Yang-Mills theories on noncommutative spaces
appear naturally in string/M-theory; the  notions and results
of noncommutative geometry were applied very successfully to the problems
of physics.

In this paper we give a mostly self-contained review of some 
aspects of M(atrix) theory, of Connes' noncommutative geometry
and of applications of noncommutative geometry to M(atrix) theory. 
The topics include introduction to BFSS and IKKT matrix models, 
compactifications on noncommutative tori, a review of basic notions of 
noncommutative geometry with a detailed discussion of   noncommutative tori, 
Morita equivalence and $SO(d,d|{\mathbb Z})$-duality, an elementary discussion 
of instantons and noncommutative orbifolds. The review is primarily intended 
for physicists who would like to learn some basic techniques of noncommutative 
geometry and how they can be applied in string theory  and to mathematicians 
who would like to learn about some new problems arising in 
theoretical physics.    
\end{abstract}
\large
\tableofcontents

\newpage

\section{Introduction}
First of all we would like to give an exposition of some basic facts about M(atrix) theory 
that is completely independent of any string theory textbooks. We will consider M(atrix) model 
as a starting point and we will show that string theory can be obtained from it. More precisely 
M(atrix) theory should be considered as a nonperturbative formulation of string theory.

Our second goal is to give an exposition of Connes'
differential noncommutative
geometry and to show that it arises 
very naturally in the framework of M(atrix) theory. We will show that noncommutative geometry can be 
used in the consideration of dualities, in analysis of BPS states, etc.

We address the present  review to  mathematicians who would like to learn about some mathematical problems 
arising at the forefront of modern theoretical physics and to physicists who would like to study some basic
notions of noncommutative geometry and see how they can be applied to  physics. 
We do not assume that a  mathematician reading this review  has any preliminary knowledge of string theory or 
M-theory. However some acquaintance with the basic notions of supersymmetry is desired (for example see 
IAS school lectures \cite{IAS}). 
Also  we do not suppose that a reader - physicist has  any prior knowledge of noncommutative geometry.

Let us describe very briefly the place of M(atrix) theory in a recent development. (The exposition below is 
addressed primarily to mathematicians.) 
In the mid nineties a new era in String theory began with explorations of nonperturbative effects in the theory. 
New remarkable objects called D-branes were discovered. It was found  that these objects play fundamental role in 
establishing various duality relations between the known five consistent superstring theories. Those dualities 
led to a conjecture that all of those theories can be obtained as limiting cases of some hypothetical unifying 
theory that was christened M-theory. (There are different versions of interpretation of the letter M in the 
name of the theory, such as Mystery, Mother, Membrane.) Not much is known precisely about this theory. 
 It is supposed to live in 11 space-time dimensions
and  it has to be invariant under 11-dimensional Poincare group. The low energy limit of M-theory is known to be 
11-dimensional supergravity theory. After compactification on a circle M-theory describes a 10-dimensional 
superstring. Of course a mathematician would be unable to work with a theory having such a vague definition. 
Nevertheless, physicists manage to obtain many consistent and beautiful results. More than that, to the delight 
of mathematicians, one of physicists conjectures  can  be formulated in a precise mathematical way. 
It was conjectured that M-theory can be formulated in the framework of matrix quantum mechanics. This conjecture 
is known by the name ``M(atrix) theory''. Originally it was formulated in terms of  reduction of 10-dimensional super Yang-Mills  
theory (SYM) to 1+0 dimensions (this means that one considers a 10D theory in which all fields are independent of 
spatial coordinates). Another form of M(atrix) theory was later proposed in \cite{IKKT}. In that version the 10D 
SYM is reduced to a point.

The relation of M(atrix) theory to String/M-theory is not completely clear but there exist some impressive consistency checks. 
It was found in \cite{CDS} that noncommutative geometry arises very naturally in the framework of M(atrix) theory. 
It was shown later that the notions and theorems of noncommutative geometry can be used very efficiently to analyze 
dualities, BPS spectrum, etc. After this paper the appearance of noncommutative geometry was understood from 
various other viewpoints. The subject is currently very popular and  the number of papers trying to 
relate noncommutative geometry and String/M-theory grows exponentially. We will not try (and are not able) to review 
any essential part of those developments. Also we will not attempt to explain the ``string part of the story'' 
(we refer the reader to papers \cite{Banks_rev}, \cite{Banks_TASI}, \cite{Susskind_rev}, \cite{Taylor_rev1}, 
\cite{Taylor_rev2}, \cite{Bigatti}  for a review of M(atrix) theory and its relation with string theory and to paper 
\cite{SeibWitt} for a very clear explanation of how noncommutative space-time geometry emerges in the framework of string theory). 

Almost all of physics papers dealing with the notion of noncommutative space do not use the well developed apparatus 
of noncommutative geometry \cite{Connesbook}. In particular the Yang-Mills theory on noncommutative spaces is usually considered  
only on free modules (corresponding to trivial vector bundles) and  more general projective modules (nontrivial 
bundles) are  disregarded. One of our main  goals is to give an accessible exposition of some of the important 
notions and theorems of noncommutative geometry and to show how they can be applied to concrete problems in physics.
We hope to convince a reader-physicist that noncommutative geometry 
is not more complicated than the commutative one. 
As a matter of fact in many instances it turns out to be simpler than the last one. (The commutative geometry 
can be considered as a degeneration of a noncommutative geometry and a generic situation is often simpler 
than a degenerate one.)  
However, the reader
should have in mind that we give an exposition only of a very small part
of noncommutative geometry; many  notions and results we need can be
found in the very first  paper by A. Connes devoted to this subject
\cite{Connes1}. If the reader would like to study noncommutative
geometry more thoroughly, he should turn to the beautiful exposition of it in the book
\cite{Connesbook} and to papers \cite{Connesrev1}, \cite{Connesrev2} for recent reviews.

The present text constitutes only a part of what we are planning to write. 
We restrict ourselves mostly to the theory and applications of unital algebras (algebras having a unit element) 
with the main example being an algebra of functions on a noncommutative torus. In the commutative case unital 
algebras correspond to algebras of functions on compact spaces. Thus, we will be saying very little about 
noncommutative ${\mathbb R}^{d}$ spaces. The theory of non-unital algebras is also well developed and equally 
important but it is more complicated. We are planning to consider non-unital algebras in the second part of this review.
In particular, we we are going to discuss in it solitons and instantons on noncommutative  ${\mathbb R}^{d}$.

 Still we hope the present text will be useful for the growing audience of mathematicians and physicists interested 
in applications of noncommutative geometry to physics. We hope to write the second part of the review in the nearest future.  

%%%%%%%%%%%%%%%%%%%%%%%%%%%%%%%%%%%%%%%%%%%%%%%%%%%%%%%%%%%%%%%%%%%%%%%%%%%%%%%%%%%%%%%%%%%%%%%%%%%%%%%%%%%%%%%%%

\section{Yang-Mills theory reduced to a point}
Let us consider gauge fields $A_{\mu}(x)$ on the space ${\mathbb R}^{d}$. A gauge field can be considered as a one-form 
$A=A_{\mu}dx^{\mu}$ on ${\mathbb R}^{d}$ taking values in the Lie algebra $\bf g$ of the gauge group $G$. 
The field strength of the gauge field 
$A$ can be defined as a $\bf g$-valued two-form $F=\frac{1}{2}F_{\mu \nu} dx^{\mu}dx^{\nu}$ where 
$F_{\mu \nu} = \partial_{\mu} A_{\nu} - \partial_{\nu} A_{\mu} + [A_{\mu}, A_{\nu}]$. 
Let us further assume that  the Lie algebra $\bf g$ is equipped with an invariant inner product $\langle , \rangle$, i.e. 
$\langle [a, b], c \rangle = \langle a , [b, c] \rangle$ for any $a, b, c \in {\bf g}$. For example if ${\bf g} = u(N)$  
then one defines  $\langle a , b \rangle = {\rm Tr} a^{*} b$.  
 If the space ${\mathbb R}^{d}$ is provided 
with Minkowski or Euclidean metric one can define a Yang-Mills  action functional using the formula
\begin{equation}\label{YM}
S =  \frac{1}{4}\int \langle F_{\mu \nu} , F^{\mu \nu} \rangle dV \equiv \frac{1}{4} (F , F) \, .
\end{equation}
This functional is invariant under gauge transformations
\begin{equation} \label{gtr}
A_{\mu} \mapsto g^{-1}A_{\mu}g +  g^{-1}(\partial_{\mu} g) 
\end{equation}
where $g(x)$ is a $G$-valued function on ${\mathbb R}^{d}$.
We would like to consider  the reduction of    Yang-Mills functional     to a point. 
This means that we consider the functional  (\ref{YM}) on constant fields (and disregard the infinite volume element that 
appears from integration). This gives us a functional defined on the $d$-tuples $(X_{1}, X_{2}, \dots , X_{d})$, 
$X_{\mu} \in {\bf g}$
\begin{equation} \label{YMonapoint}
S[X] = \frac{1}{4} \langle [X_{\mu}, X_{\nu}], [X^{\mu}, X^{\nu}]  \rangle \, .
\end{equation}
This functional is invariant under  transformations
$$
X_{\mu} \mapsto g^{-1}X_{\mu}g 
$$
where $g\in G$, i.e. 
 $X_{i}$'s transform  by means of  the adjoint action of the Lie group $G$. 
This invariance is a remnant of the gauge invariance (\ref{gtr}).

In particular  let us consider a $2n$-dimensional symplectic manifold  $(\cal M , \omega )$  and take $\bf g$ 
to be a space of smooth real valued functions on $\cal M$ with a Lie algebra structure defined by  Poisson brackets. 
An invariant inner product of two elements $X, Y\in {\bf g}$ is defined by the formula
$$
\langle X ,  Y \rangle = \int XY\omega^{n} \, . 
$$
Substituting this data in the general formula (\ref{YMonapoint}) we obtain 
\begin{equation} \label{1}
S[X,\omega ] =  \frac{1}{4}\int_{\cal M} \{ X^{\mu}, X^{\nu}\}^{2}   \omega^{n} 
\end{equation}
where $\{ . , .\}$ stands for  Poisson brackets. One can allow the symplectic structure to vary, thus 
we stressed in the notation $S[X,\omega ]$ that this is a functional of 
both $X$ and $\omega$. 
As we want to establish a connection of the reduced theory (\ref{YMonapoint}) with a string theory 
we will specialize to the case when $\cal M$ is a two-dimensional surface
 and  will change the notation for this case from $\cal M$ to $\Sigma$.  

 Without losing gauge invariance the   functional (\ref{1}) can be generalized as 
\begin{equation} \label{2}
S[X,\omega ] =  \frac{\alpha}{4} \int_{\Sigma} \{ X^{\mu}, X^{\nu}\}^{2}   \omega + \frac{\beta}{2} \int_{\Sigma}   \omega 
\end{equation}
where $\alpha$ and $\beta$ are some fixed numerical factors.
The $d$ elements  $X_{1}, X_{2}, \dots , X_{d}$ define a mapping 
  $X: \Sigma \to {\mathbb R}^{d}$ that gives an embedding of the manifold $\cal M$ into ${\mathbb R}^{d}$. 
Thus, $S[X,\omega ]$ can be considered as  a functional that depends on a two-dimensional surface (the world-sheet of a string) 
embedded into (space-time) ${\mathbb R}^{d}$ and on a symplectic form $\omega$.  
In local coordinates 
 $(\sigma_{0}, \sigma_{1})$ on $\Sigma$ the symplectic form can be written as  
$\omega = \omega(\sigma_{0}, \sigma_{1})d\sigma_{0}d\sigma_{1}$. 
  Varying $\omega(\sigma_{0}, \sigma_{1})$ we get the equation of motion 
$$
-\frac{\alpha}{4\omega^{2}} (\partial_{a}X^{\mu}\partial_{b}X^{\nu}\epsilon^{ab})^{2} + \frac{\beta}{2} = 0
$$
that gives
$$
\omega(\sigma_{0}, \sigma_{1}) = \sqrt{\alpha/\beta det(\partial_{a}X^{\mu}\partial_{b}X_{\mu})} \, .
$$ 
Substituting this solution into (\ref{2}) we obtain 
\begin{equation} \label{bstr}
S[\Sigma ] = \sqrt{\alpha \beta } \int_{\Sigma} \sqrt{ det(\partial_{a}X^{\mu}\partial_{b}X_{\mu})}  \, . 
\end{equation}
This is a Nambu-Goto  action of  bosonic string with tension $T = \pm \sqrt{\alpha \beta}$ where the sign depends 
on the type of the metric 
in ${\mathbb R}^{d}$. The functionals (\ref{2}) and (\ref{bstr}) are classically equivalent in 
the sense that  they have equivalent (isomorphic) spaces of classical solutions.

We have shown that the action functional of bosonic string theory is related to a particular case of 
functional (\ref{YMonapoint}) for some choice of the Lie algebra $\bf g$.  It is important to realize that 
the action functional (\ref{2}) can be obtained, in some sense, as a limit as $N\to \infty$ of functions 
\begin{equation} \label{S_N}
S[X] = \frac{\alpha}{4} {\rm tr}\, [X^{\mu}, X^{\nu}]^{2} + \frac{\beta}{2}{\rm tr}\, 1
\end{equation}
where $X_{\mu}$ are $N\times N$ hermitian matrices. To justify this statement we note that in a quantization 
procedure we assign to every function $f(\sigma )$ on a symplectic manifold $\cal M$ an operator $\hat f $ acting in 
some  Hilbert space.  This  assignment is such that for $\hbar \to 0$ the Poisson bracket of two functions $f(\sigma)$ 
and $g(\sigma)$ corresponds to $\hbar^{-1}[\hat f, \hat g]$ and 
$$
(2\pi \hbar )^{-dim {\cal M}/2} \int f(\sigma) \omega^{dim {\cal M}/2}  
$$
corresponds to ${\rm tr} \hat f$. In the case at hand the manifold $\cal M$ is compact and two-dimensional. Therefore, 
after the quantization we obtain a Hilbert space of   finite dimension 
$$
N = {\rm tr} \, 1 \approx \frac{Vol({\cal M})}{2\pi \hbar} \, .
$$
For a rigorous treatment of these questions in the case when $\cal M$ is a K\"ahler manifold we refer the reader 
to paper \cite{Bord_etal}.

Considering operators in this finite dimensional Hilbert space as matrices and noticing that the limit $N\to \infty$ 
corresponds to the limit $\hbar \to 0$ we obtain the action functional (\ref{2}) from the function (\ref{S_N}) that 
can be considered as a version of  (\ref{YMonapoint}) in the limit 
$N\to \infty$.

One can construct N=1 supersymmetric Yang-Mills theories (SYM) in (space-time) dimensions $d=10, 6, 4, 3$ \cite{SYM}.
(Note that these dimensions correspond to the dimensions of division algebras shifted by 2). 
In particular in 10=9+1-dimensional Minkowski space we can write down an action functional for N=1 SYM as 
\begin{equation} \label{10dSYM}
S =  \int dV\left[ -\frac{1}{4} \langle F_{\mu \nu} , F^{\mu \nu} \rangle  +  
\frac{1}{2}\langle \psi^{\alpha} ,   \sigma^{\mu}_{\alpha \beta} [\nabla_{\mu} , \psi^{\beta}] \rangle   \right] \, .
\end{equation}
Let us explain the conventions we use. 
 The vector indices indices run from 0 to 9 and are raised and lowered by means of the standard  flat Minkowski metric tensor $\eta_{\mu \nu}$,  
$\eta_{00} = -1$, $\eta_{ij} = \delta_{ij}$, $i,j =1, \dots, 9$. 
The fields $\psi^{\alpha}$ are anticommuting, taking values in  $\bf g$. 
 The index $\alpha=1, \dots , 16$ corresponds  to a Majorana-Weyl spinor representation of  
$SO(9,1)$. The upper/lower  spinor indices correspond to spinors of positive/negative chirality. 
So we will distinguish systematically between the upper and lower spinor indices. 
The ten-dimensional $32\times 32$ Gamma-matrices $\Gamma_{\mu}$ are constructed in terms of symmetric $16\times 16$ matrices 
$\sigma^{\alpha \beta}_{\mu}$, $(\sigma_{\mu})_{\alpha \beta}$ as 
$$
\Gamma_{\mu} = \left( 
\begin{array}{cc}
0 & \sigma^{\alpha \beta}_{\mu} \\
(\sigma_{\mu})_{\alpha \beta} & 0 
\end{array} \right) 
$$ 
The  matrices $\sigma^{\alpha \beta}_{\mu}$, $(\sigma_{\mu})_{\alpha \beta}$ satisfy 
$$
\sigma_{\mu}^{\alpha \gamma}(\sigma_{\nu})_{\gamma \beta} + \sigma_{\nu}^{\alpha \gamma}(\sigma_{\mu})_{\gamma \beta} = 
2\eta_{\mu \nu} \delta^{\alpha}_{\beta}
$$
as well as the important Fierz identity 
\begin{equation} \label{Fierz} 
 (\sigma_{\mu})_{\alpha \beta}\sigma^{\mu}_{\gamma \delta} +
  (\sigma_{\mu})_{\gamma \alpha}\sigma^{\mu}_{\beta \delta} + 
 (\sigma_{\mu})_{\beta \gamma}\sigma^{\mu}_{\alpha \delta}
= 0 \, .
\end{equation}

This action is invariant under gauge transformations 
$$
A_{\mu} \mapsto g^{-1}A_{\mu}g +  g^{-1}(\partial_{\mu} g) \, , \quad \psi^{\alpha} \mapsto g^{-1} \psi^{\alpha} g
$$
and supersymmetry transformations 
$$
\delta_{\epsilon} A_{\mu}  = \epsilon^{\alpha} (\sigma_{\mu})_{\alpha \beta} \psi^{\beta} \, , \quad
\delta_{\epsilon}  \psi^{\alpha}  = \frac{1}{2}{(\sigma^{\mu \nu})^{\alpha}}_{\beta} \epsilon^{\beta} F_{\mu \nu}
$$
where $\sigma^{\mu \nu} = \sigma^{\mu}\sigma^{\nu} - \eta^{\mu \nu}$ and  $\epsilon^{\alpha}$ 
is a constant Majorana-Weyl spinor parameterizing the transformation.

One can reduce  functional (\ref{10dSYM}) to a point and obtain a functional 
\begin{equation} \label{supred}
S[X, \psi] =  -\frac{1}{4} \langle [X_{\mu}, X_{\nu}], [X^{\mu}, X^{\nu}]  \rangle + 
\frac{1}{2}\langle \psi^{\alpha} , \sigma^{\mu}_{\alpha \beta} [X_{\mu} , \psi^{\beta}] \rangle
\end{equation} 
which is invariant under ``gauge'' transformations 
$$
X_{\mu} \mapsto g^{-1}X_{\mu}g\, , \qquad \psi^{\alpha} \mapsto g^{-1} \psi^{\alpha} g
$$
and supersymmetry transformations 
\begin{equation} \label{susy1}
\delta_{\epsilon} X_{\mu} =  \epsilon \sigma_{\mu} \psi \, , \qquad 
\delta_{\epsilon} \psi =  \frac{1}{2}[X_{\mu}, X_{\nu}] \sigma^{\mu \nu} \epsilon \, .
\end{equation}
In addition to this there are trivial supersymmetry transformations
\begin{equation} \label{susy2}
\tilde \delta_{\epsilon} X_{\mu} = 0 \, , \qquad \tilde \delta_{\epsilon} \psi = \epsilon \, .
\end{equation}

As we have seen above, for a particular choice of the Lie algebra $\bf g$ the bosonic analogue of (\ref{supred}) is 
related to the bosonic string action functional. It is clear then that the whole action (\ref{supred}) is related to 
some sort of superstring. 
Indeed as it was  shown in \cite{IKKT} the type IIB superstring action functional in the
Green - Schwarz formulation in a particular gauge (so called Schild gauge \cite{Schild}) takes the following form
\begin{equation} \label{supschild}
S_{Schild} = \int  [ \alpha(\frac{1}{4}\{ X^{\mu}, X^{\nu}\}^{2} 
+\frac{1}{2} \psi \sigma^{\mu}\{ X_{\mu}, \psi \}) + \beta ] \omega 
\end{equation} 
where $\omega$ is  a  worldsheet volume element   
$ \omega = \sqrt{g}d\sigma_{0} d\sigma_{1}$, $g$ is the determinant of the worldsheet metric. 
This action can be obtained from (\ref{supred}) by choosing $\bf g$ as in the bosonic case 
 to be a Lie algebra of functions on a two dimensional surface $\Sigma$ equipped with a symplectic 
structure $\omega$, and adding a term proportional to $\int \omega$.

%%%%%%%%%%%%%%%%%%%%%%%%%%%%%%%%%%%%%%%%%%%%%%%%%%%%%%%%%%%%%%%%%%%%%%%%%%%%%%%%%%%%%%%%%%%%%%%%%%%%%%%%%%%%%%%%%%%%%%%%%%%%%%%%%%%%%%%%%%%%%%
%%%%%%%%%%%%%%%%%%%%%%%%%%%                 M A T R I X             M O D E L S                              %%%%%%%%%%%%%%%%%%%%%%%%%%%%%%%%%
%%%%%%%%%%%%%%%%%%%%%%%%%%%%%%%%%%%%%%%%%%%%%%%%%%%%%%%%%%%%%%%%%%%%%%%%%%%%%%%%%%%%%%%%%%%%%%%%%%%%%%%%%%%%%%%%%%%%%%%%%%%%%%%%%%%%%%%%%%%%%%
\section{ Matrix models}
\subsection{IKKT matrix model}
We start with a U(N) 10-dimensional super Yang-Mills  theory on ${\mathbb R}^{9+ 1}$ defined by (\ref{10dSYM}). 
 We would like to look at reductions of this theory to (0 + 1)-dimensions and to a point. 
The reduction to a point was discussed above. It leads to a functional   
\begin{equation} \label{IKKT}
S= -\alpha \left( \frac{1}{4}{\rm Tr}\,  [X^{\mu}, X^{\nu} ]^{2} + 
\frac{1}{2}{\rm Tr} \, \psi^{\alpha}  \sigma^{\mu}_{\alpha \beta} [X_{\mu} , \psi^{\beta}] \right) + \beta {\rm Tr} 1
\end{equation}
which is called IKKT matrix model and was introduced in \cite{IKKT}. 
Here $X^{\mu}$, $\mu = 0, \dots , 9$ and  $\psi^{\alpha}$, $\alpha=1, \dots, 16$ are $N\times N$ hermitian matrices 
of  even and odd Grassmann parity respectively. 
(Strictly speaking the last term in (\ref{IKKT}) 
has no analog in Yang-Mills theory on a noncompact space and its addition should be considered as a possible generalization.) 
Here $\alpha$ and $\beta$ are constants. 
One can use the considerations above to prove that  IIB superstring action can be obtained from (\ref{IKKT})
by means of some limiting procedure. Moreover it can be argued that  (\ref{IKKT}) gives a nonperturbative 
definition of IIB superstring theory. We refer the reader to  review \cite{IIBrev} and references therein for a 
 discussion of this proposal. The above mentioned limiting procedure and its relevance to string theory in Schild gauge  
were also discussed in papers \cite{FFZachos1}, \cite{FZachos}, \cite{FFZachos2}.

\subsection{BFSS matrix quantum mechanics}
Reduction to  (0 + 1)-dimensions leads to the so called BFSS matrix model \cite{BFSS}. The Lagrangian of this theory is 
\begin{equation} \label{BFSS}
L = {\rm Tr} \left( \frac{1}{2}\nabla_{t}X^{i} \nabla_{t}X_{i} - \frac{1}{4} [X_{i}, X_{j}]^{2} + \frac{1}{2} \psi \sigma^{0}\nabla_{t}\psi 
+ \frac{1}{2}\psi^{\dagger}\sigma^{i}[\psi , X_{i}] \right)
\end{equation}
where $\nabla_{t} = \partial_{t} + X_{0}$, indices $i$, $j$ run from 1 to 9. 
This Lagrangian is invariant with respect to gauge transformations 
$$
\delta X_{i} = [X_{i}, g] \, , \quad \delta X_{0} = \partial_{t} g + [X_{0}, g] \, , \quad \delta \psi^{\alpha} = [\psi^{\alpha}, g]  
$$
where $g=g(t)\in U(N)$ and with respect to two kinds of supersymmetry transformations.
Transformations of the first kind  have the form   
\begin{equation} \label{dynsusy}
\delta_{\epsilon} X_{i} = \epsilon \sigma_{i} \psi \, , \qquad 
\delta_{\epsilon} \psi =  \frac{1}{2}[X_{i}, X_{j}] \sigma^{i j} \epsilon + [\nabla_{t},X_{j}]\sigma^{0j}\epsilon  \, . 
\end{equation}
and transformations of the second kind   read
\begin{equation}\label{kinsusy}
\tilde \delta_{\epsilon} X_{i} = \delta X_{0} = 0 \, , \quad \tilde \delta_{\epsilon} \psi = \epsilon \, .
\end{equation}
The Lagrangian (\ref{BFSS}) is also invariant under $SO(9)$ rotations and under  
translations 
\begin{equation} \label{transl}
X_{i} \mapsto X_{i} + c_{i}
\end{equation}
 where $c_{i}$ belong to the center of $u(N)$. 

This model was first considered in \cite{ClaudsonHalp} as a maximally supersymmetric gauge quantum mechanics. 
In  \cite{BFSS} a large N limit of this model  was considered as a main ingredient of the so called M(atrix) Theory  
conjecture put forward as a nonperturbative definition of M-theory. Parts of this conjecture can be tested by studying  quantum mechanical 
properties of the model for finite N.

 Instead of the matrix notations employed in (\ref{BFSS}) it is convenient to fix an orthonormal basis $T_{a}\in u(N)$, $a=1, \dots , N^{2}$
in the Lie algebra $u(N)$ with respect to the invariant inner product $\langle a, b \rangle =-2{\rm Tr}ab$ which is chosen to be positive definite. 
Denote $f_{abc} = \langle T_{a}, [T_{b}, T_{c}]\rangle$  the structure constants in the chosen basis.  
One can write now the classical Hamiltonian corresponding to (\ref{BFSS}) in the following form
\begin{equation} \label{H}
H =   \frac{1}{2}\Pi^{i}_{a}\Pi^{i}_{a}  + \frac{1}{4}f_{abc}X^{b}_{i}X^{c}_{j}f_{ade}X^{d}_{i}X^{e}_{j} -\frac{i}{2}
 f_{abc}X_{a}^{j}\psi_{b}^{\alpha}(\sigma_{j})_{\alpha \beta}\psi_{c}^{\beta} 
-G_{a} X^{a}_{0} 
\end{equation}  
where $G_{a} = f_{abc}(X^{b}_{j}\Pi^{j}_{c} - \frac{i}{2}\psi_{b}^{\alpha}\psi_{c}^{\alpha})$.
 Here  we assume that there is a summation over all of the repeating indices running over the following sets of values: 
$i, j= 1, \dots ,9$, $a,b, ... = 1,\dots , N^{2}$, and
 the variables $\psi^{\alpha}$ are assumed to be in a real representation with index $\alpha = 1, \dots, 16$.
Also notice that for simplicity we took the metric coming  from ${\mathbb R}^{9}$ to be given by identity matrix. 
The expression (\ref{H})  means that we have a  constrained classical system with Hamiltonian 
\begin{equation} \label{h}
h(X, \Pi, \psi) = \frac{1}{2}\Pi^{i}_{a}\Pi^{i}_{a}  + \frac{1}{4}f_{abc}X^{b}_{i}X^{c}_{j}f_{ade}X^{d}_{i}X^{e}_{j} -\frac{i}{2}
 f_{abc}X_{a}^{j}\psi^{b}_{\alpha}\sigma_{j}^{\alpha \beta}\psi^{c}_{\beta} 
\end{equation} 
and constraints 
\begin{equation} \label{const}
G_{a}(X, \Pi, \psi) = f_{abc}(X^{b}_{j}\Pi^{j}_{c} - \frac{i}{2}\psi_{b}^{\alpha}\psi_{c}^{\alpha}) = 0
\end{equation} 
($X_{0}$ plays a role of Lagrange multiplier).

After quantization we obtain a quantum system with a Hamiltonian $\hat H$ given by (\ref{h}) with $\Pi^{i}_{a}$, $X_{i}^{a}$, $\psi_{a}^{\alpha}$ 
replaced by operators  $\pi^{i}_{a}$, $x_{i}^{a}$, $\hat \psi_{a}^{\alpha}$    satisfying (anti)commutation relations 
$$
 [x_{j}^{a}, \pi^{k}_{b}] =  i\delta_{j}^{k}\delta_{b}^{a}   \, , \qquad 
\{\hat \psi_{a}^{\alpha} , \hat \psi_{b}^{\beta} \} = \delta^{\alpha \beta} \delta_{ab} \, .
$$ 
One can realize these operators in terms of differential operators in the space $E=L_{2}({\mathbb R}^{9N^{2}})\otimes F$ 
where $F$ is the fermionic Fock space. The space  $L_{2}({\mathbb R}^{9N^{2}})$ can be considered as a space of functions 
depending on variables $X_{i}^{a}$, $i=1,\dots , 9$ and $F$ is a finite-dimensional space of irreducible representation 
of the anticommutation relations for $\hat \psi^{\alpha}_{a}$ (i.e. of the Clifford algebra). 
To take into account constraints (\ref{const}) we should 
restrict ourselves to the subspace $E^{phys}$ of $E$ consisting of vectors $v$ satisfying $\hat G_{a} v =0$ where $\hat G_{a}$ stands 
for the  quantum mechanical operator corresponding to the function (\ref{const}). 
(This condition can be interpreted as a gauge invariance of vectors in $E^{phys}$.)
The Hamiltonian $\hat H$ on   $E^{phys}$ is supersymmetric. More precisely it commutes with supersymmetry 
generators 
$$
Q^{\alpha}= (\Gamma^{j}\hat \psi_{a})^{\alpha}\pi_{a}^{j} + \frac{1}{2}f_{abc}(\sigma^{jk}\hat \psi_{a})^{\alpha}x_{b}^{j}x_{c}^{k}
$$ 
that are quantum counterparts of the classical supersymmetries (\ref{dynsusy}).
On the space $E^{phys}$ these generators satisfy 
\begin{equation} \label{salg}
\{ Q^{\alpha} , Q^{\beta} \} = 2\delta^{\alpha \beta} \hat H \, .
\end{equation}
It follows from these relations that the Hamiltonian $\hat H$ is nonnegative on $E^{phys}$. 

%%%%%%%%%%%%%%%%%%%%%%%%%%%%%%%%%%%%%%%%%%%%%%%%%%%%%%%%%%%%%%%%%%%%%%%%%%%%%%%%%%%%%%%%%%%%%%%%%%%%%%%%%%%%%%%%%%%%

\subsection{Bound states and scattering in BFSS model}
Let us fix one of the supersymmetry generators $Q$. Then on the space $E^{phys}$ we have $\hat H=Q^{2}$. The operator $Q$ 
is self-adjoint. It follows then that  the operator $Q^{2}$ is nonnegative. The expression 
$$
V = -\frac{1}{4}[X_{i}, X_{j}][X^{i}, X^{j}]= \frac{1}{4}f_{abc}X^{b}_{i}X^{c}_{j}f_{ade}X^{d}_{i}X^{e}_{j}
$$
entering (\ref{H}) plays the role of the potential energy. It is clear that $min (V) = 0$ and that the minimum is 
achieved when the matrices $X_{i}$ all commute between themselves. Denote the set of  9-tuples of commuting hermitian matrices by $R$. 
Commuting hermitian matrices can be simultaneously diagonalized by means of a unitary transformation. This means that 
every point of $R$ is gauge equivalent to  a point consisting of diagonal matrices. The set $R$ is unbounded, that is 
we can have commuting matrices with arbitrarily large entries. We may say that the graph of $V$ has valleys surrounded 
by potential walls that become steeper and steeper as one goes along the valley to  infinity.  
This description can be given a more precise meaning as follows. Consider the function $V$ in a neighborhood of a 
point $(X_{1}, \dots , X_{9})\in R$ where $X_{i} = diag (d_{i}^{(1)}, \dots , d_{i}^{(N)})$. Then the second 
derivatives of $V$ in the directions orthogonal to $R$ will be large as 
\begin{equation} \label{limit}
\sqrt{\sum_{k=1}^{N} (d_{i}^{(k)}-d_{j}^{(k)})^{2}  } \to \infty \, .
\end{equation}
To analyze the low lying excitations of our Hamiltonian and the time evolution of wave functions we can apply the 
Born-Oppenheimer method considering the coordinates on $R$ as slow variables and coordinates in the directions 
transverse to $R$ as fast ones. One can check \cite{Kon} that the effective Hamiltonian that governs the dependence 
of a wave function of slow variables in the limit (\ref{limit}) is a free Hamiltonian 
\begin{equation} \label{Hfree}
H_{eff} = -\frac{1}{2} \frac{ \partial^{2}}{\partial D_{i}^{\mu}\partial D_{i}^{\mu}}
\end{equation}
where $D_{i}^{\mu}$, $\mu = 1, \dots , 9$, $i= 1, \dots N$ are orthonormal set of coordinates along $R$. 
 This statement would be  wrong if one would consider only the bosonic part of the Hamiltonian at hand.
An intuitive explanation of the effect is as follows. If we consider the potential for the bosonic degrees of freedom 
as it was pointed above it gets infinitely steep as we go to infinity. This means that the transverse excitations have 
a larger and larger ground energy in the limit (\ref{limit}). In a purely bosonic theory this will not let 
 a finite energy wave packet traveling along the valley  escape to infinity. 
In the supersymmetric theory however the transverse excitations are 
described by a system of supersymmetric oscillators that has a vanishing ground state energy. In fact one can show 
how a wave packet can escape to infinity \cite{deWHN}.  Quantitatively  this fact is reveals itself  in (\ref{Hfree}).

Thus we see that in our approximation the model  describes a system of $N$ free noninteracting identical (bosonic) particles. 
If we take into account subleading terms of the approximation we obtain an interaction between these particles. 
One can talk then about their scattering.

It follows from the above consideration that the Hamiltonian (\ref{H}) has a continuous spectrum starting at zero. 
One can prove however that there are also bound states at threshold. Note that singling out the $u(1)$ part of the Lie 
algebra  $u(N)$ one obtains a decomposition of the Hamiltonian $H$ into two noninteracting parts. The summand corresponding 
to the $u(1)$ part is just a free Hamiltonian. One can prove that the $su(N)$ part of the Hamiltonian has at least 
one normalizable state having zero energy (a normalizable ground state). To prove this assertion it is 
sufficient to show that the operator $Q$ has at least one normalizable zero mode. This follows in its turn from the fact 
that the index of $Q$ is equal to one. The computation of this index is a nontrivial task because the operator $Q$ is 
not Fredholm. For $N=2$ a convincing computation of the index was done in \cite{SethiSt}. For $N>2$ such a justification 
is still missing. A computation of the index for $N>2$ containing some gaps follows from a combination of  results 
obtained in  \cite{MNS}, \cite{GrG1}, \cite{GrG2}. See also \cite{PorrRoz}, \cite{KacSm} for other approaches. 
The fact   that the index is equal to one is a strong indication that the ground state is unique. For $N=2$ 
there are additional arguments in favor of this conjecture \cite{HalpSchwartz}.

One can also consider a problem of scattering of the bound states that  according to the BFSS conjecture 
 describes scattering of 11-dimensional supergravitons. Note that when we single out the $U(1)$ part of the gauge group
the bosonic part of the theory is that of a free nonrelativistic particle and describes a center of mass degree of freedom. 
The $U(1)$ fermions in their turn generate a Fock space of dimension $2^{8} = 256$ that describes supergraviton 
polarization states \cite{sugra}.
 
Consider a set $R_{n_{1}, \dots , n_{k}}$ 
consisting of 9-tuples of hermitian matrices $X_{1}, \dots , X_{9}$ that can be transformed simultaneously into a block-diagonal 
form with blocks of sizes $n_{1}\times n_{1}, \dots , n_{k}\times n_{k}$. Here $n_{1}, \dots , n_{k}$ are natural numbers 
such that their sum is equal to $N$ (the size of matrices $X_{i}$). The set $R$ we considered above coincides with 
$R_{1,1, \dots, 1}$. One can apply the Born-Oppenheimer approximation to the scattering problem regarding coordinates 
along $R_{n_{1}, \dots , n_{k}}$ as slow variables and coordinates in the transverse directions as the fast ones. 
Each block then describes a supergraviton. The asymptotic state in the scattering problem 
should be  a superposition of $k$ ground state wave functions for each block respectively. 
For further discussion of scattering in BFSS Matrix Theory and its comparison with supergravity see papers \cite{BFSSscat}, 
\cite{sugra}, \cite{3grav} and references therein. 

%%%%%%%%%%%%%%%%%%%%%%%%%%%%%%%%%%%%%%%%%%%%%%%%%%%%%%%%%%%%%%%%%%%%%%%%%%%%%%%%%%%%%%%%%%%%%%%%%%%%%%%%%%%%%%%%%%%%%%%%%
%%%%%%%%%%%%%%%%%%                     C O M P A C T I F I C A T I O N S                                     %%%%%%%%%%%%
%%%%%%%%%%%%%%%%%%%%%%%%%%%%%%%%%%%%%%%%%%%%%%%%%%%%%%%%%%%%%%%%%%%%%%%%%%%%%%%%%%%%%%%%%%%%%%%%%%%%%%%%%%%%%%%%%%%%%%%%%

\section{Compactifications}
\subsection{Compactification on a circle. Relation between IKKT and BFSS models} \label{Comp1sec}
Compactifying   the IKKT or BFSS model on a circle in the direction $X_{1}$ means that we would like to define  a restriction of the 
IKKT/BFSS  action to  the space  where an equivalence relation $X_{1} \sim X_{1} + 2\pi R_{1}{\bf 1}$ is satisfied. 
Here $\bf 1$ is the identity matrix and $R_{1}$ is a radius of compactification. By equivalence relation in this context one    
should understand gauge equivalence. Thus, one considers the following equations
\begin{eqnarray} \label{S1}
 UX_{1}U^{-1} &=&  X_{1} + 2\pi R_{1}{\bf 1} \, , \nonumber \\
 UX_{i}U^{-1}&=& X_{i} \, , \enspace i \ne 1 \, , \nonumber \\
U\psi^{\alpha}U^{-1}&=& \psi^{\alpha} 
\end{eqnarray}
where $U$ is a unitary matrix. This equation cannot be satisfied (unless $R_{1}=0$) if $X_{i}$, $\psi^{\alpha}$ are  finite matrices  as 
can be easily seen by taking the trace of both sides of the first equation in (\ref{S1}). However there are solutions in terms 
of operators in infinite-dimensional Hilbert space. Let ${\cal H} = L_{2}(S^{1})\otimes {\cal H}'$ where ${\cal H}'$ is some Hilbert 
space (finite or infinite dimensional). Then we have the following solutions to (\ref{S1}) in terms of operators in $\cal H$
\begin{eqnarray} \label{S1sol}
X_{1} &=& 2\pi i R_{1} \frac{\partial}{\partial \sigma} + A_{1}(\sigma ) \, , \nonumber \\
X_{i}&=& A_{i}(\sigma ) \, , \enspace i  \ne 1\, , \nonumber \\
\psi^{\alpha}&=& \Psi^{\alpha}(\sigma ) \, , \nonumber \\
(Uf)(\sigma ) &=& e^{i\sigma}f(\sigma ) 
\end{eqnarray}  
where $0\le \sigma\le 2\pi $ is a coordinate on $S^{1} = {\mathbb R}/2\pi {\mathbb Z}$, 
$A_{i}(\sigma )$ are operators acting on $\cal H'$ depending on $\sigma$ as on 
a parameter. Here our conventions are such that  $X_{\mu}$, $\Psi^{\alpha}$ are Hermitian operators on $\cal H$. 
One can prove that all other solutions to (\ref{S1}) are gauge equivalent to  solution (\ref{S1sol}). 
 The choice of the direction $X_{1}$ in the above discussion is absolutely inessential. One can replace $X_{1}$ 
by any of $X_{i}$, $i=1, \dots , 9$ in the BFSS or IKKT model. 
Note that for the IKKT model these directions are space-like.

Here we would like to make some general comments about  the metric signature in both models. 
 The BFSS model is a reduction of the 10-dimensional SYM model to 0 + 1 dimensions that breaks 
the original $SO(9,1)$ Lorentz invariance. One ends up with a theory that only possesses the invariance under 
$SO(9)$ spatial  rotations. All of the directions we can compactify  are spatial. 
The IKKT model was defined above as a reduction of 10D SYM to a point. 
The metric inherited from  ${\mathbb R}^{10}$  naturally has  Minkowski signature and the model is $SO(9,1)$ invariant.
To put all of the directions in the discussion of  IKKT compactifications  on equal footing 
we can consider a Euclidean version of the model. This will also  permit  us to  establish a connection between both models. 
 We are going to 
show below that the Euclidean IKKT model compactified on a circle gives the BFSS model at finite temperature.

If one wants to use the Euclidean signature metric in the IKKT action (\ref{IKKT}) one encounters 
 complications related to  the non-existence of Majorana-Weyl fermions in the  Euclidean space. 
However as far as physical quantities such as correlation functions are concerned it is just a technicality. 
The configuration space of the IKKT model is ${\cal M} = (u(N))^{10}\otimes (\Pi u(N))^{16}$ where $\Pi$ stands for the
parity reversion operator (in the sense  of a Grassmann algebra). One can extend the action functional to the 
space ${\cal M}_{\mathbb C} = ({\rm Mat}_{N}({\mathbb C})^{10}\otimes (\Pi {\rm Mat}_{N}({\mathbb C})^{16}$ in such a way that it is 
a holomorphic functional on the complex manifold ${\cal M}_{\mathbb C}$.    
 Any physical quantity can be given in terms of integrals over configuration 
space containing $e^{-S}$ factor. To perform  integration over a complex manifold one needs to specify a real 
cycle (real slice), i.e. impose some reality conditions on the variables. The result then depends only on the homology class 
of a chosen cycle. For a complex supermanifold an integral over a holomorphic function depends only on the even part 
of the chosen cycle. (This is essentially due to the formal algebraic nature of Berezin integral over odd variables.) 
We can work with complex 10-dimensional Weyl spinors. Then the absence of Majorana-Weyl spinors means that there is no 
$SO(10, {\mathbb C})$-invariant real slice in the space of Weyl spinors. 
But this is irrelevant since the result of integration does not depend on a choice of slice for odd variables anyway.    
From now on when discussing the IKKT model we will assume that the Euclidean ${\mathbb R}^{10}$ metric is being used.

One can also construct an action of the compactified theory starting with the original  action (\ref{IKKT}) or (\ref{BFSS}), 
substituting a sequence of  finite-dimensional approximations to   (\ref{S1sol}) and performing a proper limiting procedure. 
Namely  the operator $\frac{\partial}{\partial \sigma}$ in (\ref{S1sol}) can be approximated by means of a finite difference operator 
on a lattice with spacing $a$. Let us consider this approximation applied for the compactification of the IKKT model on a circle 
in the direction $X_{0}$.  Taking $a\to 0$ we obtain after a rescaling of numerical constants $\alpha$, $\beta$ in 
(\ref{IKKT}) the following action functional
\begin{eqnarray}
S&=& R\cdot  \int d\sigma{\rm Tr}\Bigl[ 2 \sum_{i=1}^{9}(\nabla_{0} A_{i}(\sigma))^{2} + \sum_{i,j =1}^{9}[A_{i}(\sigma), A_{j}(\sigma)]^{2} 
+ \nonumber \\ 
&& +  2 \Psi^{\alpha}(\sigma)\sigma^{0}_{\alpha \beta}\nabla_{0}\Psi^{\beta}(\sigma) + 2\sum_{i=1}^{9} \Psi^{\alpha}(\sigma)
\sigma^{i}_{\alpha \beta}[A_{i}(\sigma), \Psi^{\beta}(\sigma)]
 \Bigr]
\end{eqnarray}
where $(\nabla_{0}) f(\sigma) = iR_{0}\frac{\partial f}{\partial \sigma} + [A_{0}, f](\sigma)$ and $R$ is an overall numerical factor. 
We obtained an action functional 
of a matrix quantum mechanics with a compact Euclidean time direction. This is equivalent to considering the BFSS matrix 
quantum mechanics (\ref{BFSS}) at finite temperature. 

%%%%%%%%%%%%%%%%%%%%%%%%%%%%%%%%%%%%%%%%%%%%%%%%%%%%%%%%%%%%%%%%%%%%%%%%%%%%%%%%%%%%%%%%%%%%%%%%%%%%%%%%%%%%%%%%%%%%%%%%%%%
%%%%%%%%%%%%%%%%%%%%%%%%%%%%%%%                 T W O     T O R U S                         %%%%%%%%%%%%%%%%%%%%%%%%%%%%%%%
%%%%%%%%%%%%%%%%%%%%%%%%%%%%%%%%%%%%%%%%%%%%%%%%%%%%%%%%%%%%%%%%%%%%%%%%%%%%%%%%%%%%%%%%%%%%%%%%%%%%%%%%%%%%%%%%%%%%%%%%%%%

\subsection{Compactification on a regular $T^{2}$}
 Similarly to the case of a circle considered 
in the previous section we can write down equations specifying compactification in two 
directions, $X_{1}$ and $X_{2}$ 
\begin{eqnarray} \label{T2}
&&U_{1}X_{1}U_{1}^{-1} = X_{1} + 2\pi R_{1}{\bf 1} \, , \qquad U_{2}X_{2}U_{2}^{-1} = X_{2} + 2\pi R_{2}{\bf 1} \, , \nonumber \\
&&U_{1}X_{i}U_{1}^{-1} = X_{i} \, , \enspace i\ne 1 \, , \enspace  U_{2}X_{i}U_{2}^{-1} = X_{i} \, , \enspace i  \ne  2 \, , \nonumber \\
&& U_{1}\psi^{\alpha}U_{1}^{-1} = U_{2}\psi^{\alpha}U_{2}^{-1}=\psi^{\alpha}  
\end{eqnarray}
where $R_{1}$ and $R_{2}$ are radii of compactification. As above we are going to search for solutions to these equations in 
terms of operators in an infinite-dimensional Hilbert space $\cal H$.

It is straightforward  to derive from (\ref{T2}) that the quantity $U_{1}U_{2}U_{1}^{-1}U_{2}^{-1}$ commutes with all 
$X_{i}$ and $\psi^{\alpha}$. It is natural then to set it to be a scalar operator, i.e. 
\begin{equation} \label{talg}
U_{1}U_{2} = \lambda U_{2}U_{1} 
\end{equation} 
where $\lambda = e^{2\pi i \theta}$ is a complex constant.
Let us start with the case  $\lambda = 1$.  One can take ${\cal H} = L_{2}(T^{2})\oplus {\cal H}'$ where 
$L_{2}(T^{2})\cong L_{2}(S^{1}\times S^{1})$ is the $L_{2}$ space of functions on a two-torus and $\cal H'$ is some other Hilbert space.  
Equation  (\ref{T2}) can be checked to have the following solution  
\begin{eqnarray} \label{T2sol}
&&X_{1} = 2\pi iR_{1}\frac{\partial}{\partial \sigma_{1}} + A_{1}(\sigma_{1}, \sigma_{2})\, , \quad 
X_{2} = 2\pi iR_{2}\frac{\partial}{\partial \sigma_{2}} + A_{2}(\sigma_{1}, \sigma_{2}) \, , \nonumber \\
&&X_{i} = A_{i}(\sigma_{1}, \sigma_{2}) \, , \enspace i\ne 1,2  \, , \qquad \psi^{\alpha} = \Psi^{\alpha}(\sigma_{1}, \sigma_{2})   \, , \nonumber \\
&& (U_{j}f)(\sigma_{1}, \sigma_{2}) = e^{i\sigma_{j}}f(\sigma_{1}, \sigma_{2}) \, , \enspace j=1,2  
\end{eqnarray}
where $0\le \sigma_{1}, \sigma_{2} < 2\pi$ are coordinates on $T^{2}$ and $A_{i}$, $\Psi^{\alpha}$ are functions on $T^{2}$ taking 
values in operators on $\cal H'$. If the space $\cal H'$ is finite-dimensional then the total Hilbert space $\cal H$ can be viewed 
as a space of sections of a topologically trivial vector bundle $E$ over $T^{2}$ with a typical fiber $\cal H'$. The operators $X_{1}$, $X_{2}$ then 
specify a connection on $E$, the operators $X_{i}$, $i\ne 1,2$ and $\Psi^{\alpha}$ are sections of the adjoint bundle $E\otimes E^{*}$ having 
as a fiber the space of linear operators acting in the corresponding fiber of the bundle $E$. The last equation in  (\ref{T2sol}) just
says that $U_{j}$'s are generators of the algebra of functions on $T^{2}$ acting on sections of $E$ by point-wise multiplication. 
One can also construct solutions to (\ref{T2}) in terms of operators acting on a space of sections of topologically nontrivial bundle $E$. 
In this case again the operators $U_{j}$ can be represented by pointwise multiplications by functions $exp(i\sigma_{j})$. 
 We can rewrite the first two equations in (\ref{T2}) as 
$$
X_{j}U_{k} = U_{k}X_{j} -2\pi R_{j} \delta_{jk} U_{k} = U_{k}X_{j} + 2\pi i \frac{\partial U_{k}}{\partial \sigma_{j}} 
$$ 
that is a Leibniz rule meaning that the operators $X_{j}$ are connections on $E$. 
The rest of the equations mean as above that the operators $X_{i}$, $i\ne 1,2$ and $\Psi^{\alpha}$ are sections of the adjoint bundle. 
Hence any connection $\nabla_{j}$ ($\equiv X_{j}=i\nabla_{j}$, $j=1,2$ where the factor of $i$ inserted because of our hermiticity conventions, 
see the general definition of connection in section ...) 
   along with adjoint sections $X_{i}$, $i\ne 1,2$, $\Psi^{\alpha}$ gives a 
solution to (\ref{T2}).
It can be shown that this solution is in some sense generic.

%%%%%%%%%%%%%%%%%%%%%%%%%%%%%%%%%%%%%%%%%%%%%%%%%%%%%%%%%%%%%%%%%%%%%%%%%%%%%%%%%%%%%%%%%%%%%%%%%%%%%%%%%%%%%%%%%%%%%%%%%%%%%%%%%%%%%%

\subsection{Compactification on a noncommutative $T^{2}$ }
We now consider the case $\lambda \ne 1$. The relation 
\begin{equation} \label{nctorus}
U_{1}U_{2} = e^{2\pi i \theta} U_{2}U_{1} 
\end{equation}
is known in the mathematical literature as 
a relation defining the algebra of functions on a  noncommutative two-torus $T_{\theta}$ with noncommutativity parameter $\theta$. 
By definition  two unitary operators in Hilbert space obeying (\ref{nctorus}) specify a representation 
of a noncommutative two-torus $T_{\theta}$, or, in other  words, a module over $T_{\theta}$. See section 4.... for 
more details.
 
 The equations on $X_{1}$, $X_{2}$ can be compactly written as 
\begin{equation} \label{con}
[X_{j}, U_{k}] = -2\pi R_{j}\delta_{jk} U_{k} \, , \enspace j,k = 1,2 \, .    
\end{equation}
 If these equations  are satisfied for  operators $X_{1}$, $X_{2}$ acting on the same Hilbert space $E$ then by definition 
one says that $X_{1}$, $X_{2}$ define  a connection on a module $E$ 
(equations (\ref{con}) can be considered as an analogue of  a Leibniz rule). The rest of the equations (\ref{T2})
say that  $X_{i}$, $i\ne 1,2$ and $\Psi^{\alpha}$ commute with $U_{j}$, $j=1,2$. This means that the corresponding operators 
 are endomorphisms of $T_{\theta}$ module $E$. Thus, we see that solutions to (\ref{T2}) in the case $\lambda =e^{2\pi i \theta}\ne 1$ 
can be obtained in terms of connections and endomorphisms of modules over a noncommutative torus $T_{\theta}$.

Let us give here a concrete example  of a module over $T_{\theta}$ and a connection on it. 
Consider operators $U_{1}$ and $U_{2}$ acting on functions from the Schwartz space ${\cal S}({\mathbb R})$  as 
\begin{equation} \label{mod}
U_{1} : f(x) \mapsto f(x + \theta ) \, , \qquad U_{2}: f(x) \mapsto f(x)e^{2\pi i x} \, .
\end{equation}
It is easy to see that these operators satisfy (\ref{nctorus}). 
Operators obeying the commutation relations  (\ref{con}) can be constructed as  
$$
X_{1}: f(x) \mapsto 2\pi R_{1}x \cdot f(x) \, , \qquad 
 X_{2} : f(x) \mapsto i\frac{d f(x)}{dx} \, . 
$$
However these $X_{1}$ and $X_{2}$ are not the most general solutions. One can add to the particular solutions written above, that 
we will denote as $X_{i}^{*}$, any two operators that commute with $U_{1}$ and $U_{2}$ (endomorphisms). 
We can construct such operators in terms of linear combinations of operators 
$$
Z_{ n_{1}, n_{2} } : f(x) \mapsto f(x + n_{1}) e^{2\pi i n_{2}/\theta} 
$$ 
labeled by a pair of integers $n_{1}$, $n_{2}$.  It can be easily checked that $Z_{ n_{1}, n_{2} }$ 
commute with operators (\ref{mod}).
Thus a general solution to (\ref{con}) can be written as 
$$
X_{i} = X_{i}^{*} + \sum_{n_{1}, n_{2} \in {\mathbb Z}} C_{i}(n_{1}, n_{2}) Z_{ n_{1}, n_{2} }
$$
where $C_{i}(n_{1}, n_{2})$ are number coefficients. The rest of the fields $X_{i}$, $i\ne 1,2$, $\Psi^{\alpha}$ 
can be represented as linear combinations of operators $ Z_{ n_{1}, n_{2} }$ with coefficients of appropriate 
Grassmann parity.

%%%%%%%%%%%%%%%%%%%%%%%%%%%%%%%%%%%%%%%%%%%%%%%%%%%%%%%%%%%%%%%%%%%%%%%%%%%%%%%%%%%%%%%%%%%%%%%%%%%%%%%%%%%%%%%%%%%%%%%%%%%%

\subsection{Compactifications on $T^{d}$ and $T_{\theta}^{d}$} 
The above discussion of the two-dimensional case can be directly generalized to the case of $d$ dimensions. 
In this case we have the following equations 
\begin{eqnarray}\label{Td}
&& U_{j} X_{k} U_{j}^{-1} = X_{k} + \delta_{kj}2\pi R_{k} {\bf 1} \, , \enspace j,k = 1, \dots , d \, ,\nonumber \\
&& U_{j} X_{I} U_{j}^{-1} = X_{I} \, , \enspace I>d \, , \nonumber \\
&& U_{j}\psi^{\alpha}U_{j}^{-1}=\psi^{\alpha} \, .   
\end{eqnarray}
It is a matter of a simple calculation  to check that equations (\ref{Td}) imply that the products $U_{j}U_{k}U_{j}^{-1}U_{k}^{-1}$ 
commute with all $X_{i}$'s. It is natural to  set these  combinations to be constants. One obtains then the following commutation relations     
\begin{equation}\label{dnctori}
U_{j}U_{k} = e^{2\pi i \theta^{jk}} U_{k}U_{j}
\end{equation}
where $\theta^{jk}$ is a constant $d\times d$ matrix. By  taking  the inverse of both sides in (\ref{dnctori}) one immediately finds   
that $\theta^{jk}-\theta^{kj}$ is an integer. Hence, without loss of generality the matrix $\theta^{ij}$ can be 
chosen to be antisymmetric. An algebra generated by $U_{i}$'s satisfying (\ref{dnctori}) is known as an algebra 
of functions on a d-dimensional noncommutative torus $T_{\theta}^{d}$ (where $\theta$ stands for the matrix $\theta^{ij}$). 
 A representation of these commutation relations in terms of operators 
in a Hilbert space specifies  a module $E$ over the algebra $T_{\theta}^{d}$. The equations in the first line in (\ref{Td}) 
mean by definition that $X_{j}$'s define  a connection on $E$. Finally the last two equations in (\ref{Td}) 
are equivalent to saying that $X_{I}$ for $I>d$ and $ \psi^{\alpha}$ are endomorphisms of $E$. 
Below we will construct explicit examples of modules over $T_{\theta}^{d}$ and study connections and endomorphisms thereof.

In the particular case when $\theta_{jk}$ is a matrix with zero (or integer) entries the equations 
(\ref{dnctori}) give a commutative algebra. The whole system of equations (\ref{Td}) can be solved in terms of  
connections and endomorphisms (sections of the adjoint bundle) of a vector bundle over  a $d$-dimensional torus $T^{d}$.

One can restrict the action functional of BFSS or IKKT matrix model to the set of solutions to (\ref{Td}). 
This leads to a supersymmetric Yang-Mills theory on a noncommutative torus (see sections \ref{YMsec} and \ref{SYMsec} for details). 

%%%%%%%%%%%%%%%%%%%%%%%%%%%%%%%%%%%%%%%%%%%%%%%%%%%%%%%%%%%%%%%%%%%%%%%%%%%%%%%%%%%%%%%%%%%%%%%%%%%%%%%%%%%%%%%%%%%%%

\subsection{Noncommutative geometry from a constant curvature background} \label{Backgrsec}

We have shown that noncommutative tori arise very naturally in the consideration of M(atrix) Theory compactifications. 
Another way to obtain a noncommutative geometry from M(atrix) theory is based on the expansion of this 
theory around a certain classical background. The construction is as follows. 
Consider matrices $\hat p_{\mu}$ satisfying the relation
\begin{eqnarray} \label{backgr}
&& [\hat p_{\mu}, \hat p_{\nu}] = iB_{\mu \nu}\cdot \hat 1 \, , \enspace \mbox{if} \enspace 
0\le \mu, \nu \le d-1 \, ,  \\
&&[\hat p_{\mu}, \hat p_{\nu}] = 0 \, , \enspace \mbox{otherwise} \, . 
\end{eqnarray}
Here $B_{\mu \nu}$ is a constant $d\times d$ antisymmetric matrix. We will assume that $B_{\mu \nu}$ is invertible.

The equation (\ref{backgr}) cannot be satisfied by finite-dimensional matrices. 
One can work either with an exact solution to (\ref{backgr}) in terms of infinite-dimensional matrices or 
with  approximate solutions in terms of $N\times N$ matrices with $N\to \infty$. 
We will  work with infinite-dimensional matrices having in mind that all calculations should be justified 
by some limiting procedure.

In the case when $B_{\mu \nu}$ is invertible the operators (infinite dimensional matrices) $p_{\mu}$, $\mu = 0, \dots , d-1$ 
generate a Heisenberg algebra of $d/2$ degrees of freedom. By Stone-Von Neumann theorem it has a unique irreducible representation 
and each representation   breaks into a direct sum (integral) of those. Let us consider the situation when we have an
irreducible  representation $\cal F$ of algebra (\ref{backgr}). We can realize $\cal F$ in terms of operators 
acting on functions defined on  ${\mathbb R}^{d/2}$ in the following way 
$$
\hat p_{1} = b_{1}\partial_{1}, \dots , \hat p_{d/2} = b_{d/2}\partial_{d/2} \, \qquad 
\hat p_{d/2 + 1} = ib_{1}x_{1}, \dots , \hat p_{d} = ib_{d/2}x_{d/2} 
$$
where we assumed that the matrix $B_{\mu \nu}$ is brought to the canonical block-diagonal form  
$$
\left( \begin{array}{cc}
0&\mbox{diag}(b_{1}, \dots , b_{d/2})\\
-\mbox{diag}(b_{1}, \dots , b_{d/2})&0
\end{array} \right)
$$ 
where $b_{\mu}$ are positive numbers.

To a function $\phi(x)\in {\cal S}({\mathbb R}^{d/2})$ we can assign an operator $\hat \phi$ on the representation space $\cal F$ 
by the formula 
\begin{equation}\label{ph}
\hat \phi = \int dk\, \phi(k) e^{ ik_{i}C^{ij}\hat p_{j}} 
\end{equation}
where $\phi(k)$ are Fourier modes of $\phi(x)$, 
 $C^{ij}= (B^{-1})^{ij}$,  and the indices $i,j$ run from $0$ to $d-1$.

We can construct a map from an operator to a function as
$$
\hat \phi \mapsto \phi(x) = \int dk \phi(k) exp (ikx)
$$
where the function $\phi(k)$ corresponds to the representation of operator $\hat \phi$ in the form (\ref{ph}).
Under this mapping the operator product goes into the so called star product
\begin{eqnarray} \label{st_pr}
&&\hat a \hat b \mapsto a(x)\ast b(x) \, , \nonumber \\
&&\phi(x)\ast \chi(x) = \Bigl(e^{\frac{1}{2i}\frac{\partial}{\partial x^{i} }C^{ij}\frac{\partial}{\partial y^{j} }} \phi(x)\chi(y)
\Bigr)_{x=y} \, . 
\end{eqnarray}
The operator trace of $\hat a$ is equal up to a constant factor to the integral of 
the corresponding function. Namely, we have 
\begin{equation} \label{traceeq}
{\rm Tr} \, \hat \phi = \sqrt{det(B)} (2\pi)^{-d/2}\int d^{d}x \phi(x) \, .
\end{equation}

Now let us expand the fields $X_{\mu}$  around the background given by $p_{\mu}$ 
$$
X_{\mu} = \hat p_{\mu} + \hat \phi_{\mu} 
$$
where $\hat \phi_{\mu}$  are  operators that we will assume to be specified by functions $\phi_{\mu}(x)$ 
as in (\ref{ph}).

Applying the above rules (\ref{st_pr}), (\ref{traceeq})  we obtain (after omitting an irrelevant infinite constant) 
that the bosonic part of the IKKT action becomes 
\begin{eqnarray} \label{Sb}
S_{bosonic}&=& -\frac{1}{4}\sqrt{det(B)}(2\pi)^{d/2}\int d^{d}x\, [\nabla_{i}, \nabla_{j}]_{\ast}\ast [\nabla^{i}, \nabla^{j}]_{\ast} +
 \frac{1}{2}[\nabla_{i}, \phi_{I}]_{\ast}\ast [\nabla^{i}, \phi^{I}]_{\ast} + \nonumber \\
&& \frac{1}{4}[\phi_{I}, \phi_{J}]_{\ast}\ast[\phi^{I}, \phi^{J}]_{\ast} 
\end{eqnarray}
where $\nabla_{j}$, $a=0, \dots, d-1$ stand for a noncommutative analog of covariant derivatives defined by the formula 
$$
[\nabla_{j}, \phi ]_{\ast} = i\partial_{j}\phi(x) +  \phi_{j}\ast \phi(x) - \phi\ast \phi_{j}(x) \, .
$$ 
The indices $I, J$ in (\ref{Sb}) run from $d$ to $9$. 
It is easy to check that the correspondence between operators and functions transforms the commutator 
$[\hat p_{j} + \hat \phi_{j}, \hat \phi]$ into $[\nabla_{j}, \phi]_{\ast}$. This remark together with 
formula (\ref{traceeq}) leads to (\ref{Sb}). An analogous procedure can be applied to the fermionic part 
of (\ref{IKKT}); it yields 
\begin{equation} \label{Sf}
S_{fermionic} = \frac{1}{2} \sqrt{det(B)}(2\pi)^{d/2}\int d^{d}x\, \psi\ast \sigma^{j}[\nabla_{j}, \psi]_{\ast} + 
\psi \ast \sigma^{I}[\phi_{I}, \psi]_{\ast} \, .
\end{equation}
 The first term in the action functional (\ref{Sb}) can be called an action functional of 
noncommutative $U(1)$ Yang-Mills  theory on a noncommutative ${\mathbb R}^{d}$ space 
(more formally   this space will be discussed in the next section). 
The sum of the bosonic part (\ref{Sb}) and 
the fermionic part (\ref{Sf}) gives a noncommutative supersymmetric Yang-Mills theory on the same space. 
The above derivation can be easily generalized to the case when the representation space breaks into a direct sum of $N>1$ 
irreducible representations of the Heisenberg algebra (\ref{backgr}). In that case $X_{\mu} = p_{\mu}\cdot \delta_{a}^{b} + 
(\hat \phi_{\mu}){a}^{b}$ where $(\hat \phi_{\mu}){a}^{b}$ is an $N\times N$ matrix whose entries are operators of the form 
(\ref{ph}).
This leads to a noncommutative $U(N)$ Yang-Mills theory.  

%%%%%%%%%%%%%%%%%%%%%%%%%%%%%%%%%%%%%%%%%%%%%%%%%%%%%%%%%%%%%%%%%%%%%%%%%%%%%%%%%%%%%%%%%%%%%%%%%%%%%%%%%%%%%%%%%%%%%%%%%
%%%%%%%%                                N  C      G E O M E T R Y                                                %%%%%%                         
%%%%%%%%%%%%%%%%%%%%%%%%%%%%%%%%%%%%%%%%%%%%%%%%%%%%%%%%%%%%%%%%%%%%%%%%%%%%%%%%%%%%%%%%%%%%%%%%%%%%%%%%%%%%%%%%%%%%%%%%%

\section{Noncommutative geometry}
\subsection{Algebras of functions and vector bundles}
In the next few sections we would like to give a general outline of how 
the analogs of basic objects of ordinary (commutative) 
differential geometry, such as algebras of functions, vector bundles and connections can be defined 
in noncommutative geometry. We will start with algebras of  functions and vector bundles, then 
give an example of a noncommutative space - a quantum ${\mathbb R}^{d}$ space and will proceed  
with  general definitions of endomorphisms and connections. All of these general notions will 
be illustrated by a variety of concrete examples based on noncommutative tori in the subsequent sections.

The key idea of noncommutative geometry is in replacement of commutative algebra of functions 
on a smooth manifold $\cal M$ by a  noncommutative deformation of it. One can consider an algebra 
$C({\cal M})$   of continuous functions and construct its noncommutative deformation. 
The deformed algebra (and in fact any associative noncommutative algebra) can be considered 
as an algebra of functions on a noncommutative space. 
In order to build up noncommutative differential geometry one has to consider  a  deformation of  an algebra 
$C^{\infty}({\cal M})$ of  smooth functions on $\cal M$. 
Conventionally  noncommutative geometry is 
developed in the framework of $C^{*}$-algebras (\cite{Connesbook}). This means that the algebra $A$ 
of functions on a noncommutative space is assumed to be equipped with an involution (see section \ref{Invsec} 
for the precise definition and a discussion of involutive algebras) and a norm 
satisfying certain axioms. 
For most of the aspects of noncommutative geometry we are going to 
discuss  the norm structure will not be essential. We will work therefore with noncommutative 
associative algebras over complex numbers equipped with an involution and a unit element.  
Moreover we will concentrate on the deformations of $C^{\infty}({\cal M})$.

Most of  noncommutative geometry constructions have the following pattern. We give a definition 
of some geometric notion in standard (commutative) geometry in purely algebraic 
terms, using algebras of functions $C({\cal M})$ or $C^{\infty}({\cal M})$. 
To define the corresponding notion for noncommutative spaces we replace in this definition 
the algebra of functions by a noncommutative algebra. Let us show how to use this idea 
to obtain a definition of a vector bundle over a noncommutative space.
Let $\zeta$  be a vector bundle over $\cal M$ specified by a projection  $p: T \to {\cal M}$. 
Its space of sections $E$ is an (infinite dimensional) vector space. 
Any section can be multiplied by a function over $\cal M$ in a pointwise manner. The result is another  section.
Obviously if one does a successive multiplication of a section by two different functions the result is the 
same as a multiplication by the product of these functions.
This means that the space of sections carries a representation of the  algebra of functions. If we consider 
continuous  sections then naturally it carries a representation of the  algebra of continuous functions, 
if we are interested in smooth structures then all  sections and functions should be smooth.   
Phrasing it differently $E$ is a module over the algebra of functions. 
If the bundle $\zeta$ is a trivial bundle of rank $N$ then the space of (smooth for definiteness) sections $E$ is isomorphic 
 as a  vector space to $(C^{\infty}({\cal M}))^{N}$ with a natural action of the algebra $C^{\infty}({\cal M})$. 
This module is called a free module of rank $N$ over $C^{\infty}({\cal M})$.   
There is  a theorem due to Serre and Swan that states that any vector bundle $\zeta$ can be embedded into a trivial one. Moreover this 
trivial bundle can be represented as a direct sum of $\zeta$  and some other vector bundle $\zeta'$ ($\zeta'$ can be 
constructed as an orthogonal complement to $\zeta$ with respect to some inner product).
 This implies  that  the module of sections $E$ can be singled out as a direct summand in the module $(C^{\infty}({\cal M}))^{N}$ 
for some $N$.  Modules that are isomorphic to  direct summands of a free module are called projective.
We consider only free modules of finite rank, correspondingly our projective modules are always finitely generated. 
We see that any vector bundle gives rise to  a  {\it projective module}. Conversely every 
finitely generated projective module 
over   $C^{\infty}({\cal M})$ can be realized in terms of sections of some vector bundle over $\cal M$.
  This leads us to a natural generalization 
of the notion of a vector bundle to  noncommutative geometry. The corresponding object is a (finitely 
generated) projective 
module over an algebra of functions on a noncommutative space.     

%%%%%%%%%%%%%%%%%%%%%%%%%%%%%%%%%%%%%%%%%%%%%%%%%%%%%%%%%%

\subsection{Noncommutative ${\mathbb R}^{d}$ spaces}
As an example of an algebra  of functions on a noncommutative space 
let us consider an associative algebra of operators acting on a Hilbert space $L_{2}({\mathbb R}^{n})$.
One can consider various classes of operators. They correspond to various classes of functions on a 
noncommutative space. One can choose, for example, the class of all bounded operators acting on 
$L_{2}({\mathbb R}^{n})$. This choice is not very convenient, in particular because it does not contain the 
operators $x^{i}$ of multiplication by the coordinate $x^{i}$ and operators $\partial_{i}$ of differentiation 
with respect to $x_{i}$. Here $i$ runs from 1 to $n$. To include those operators we can take the algebra 
of all linear operators acting on the Schwartz space ${\cal S}({\mathbb R}^{n})$. Given a linear operator acting on 
functions of $n$ variables 
one can  assign to it a function of $2n$ variables called a symbol of a linear operator. For example the so called 
Weyl symbol of an operator can be defined by the formula 
$$
A(x, p) = \int d\xi \langle x- \xi/2|\hat A|x + \xi/2 \rangle e^{ip \xi}
$$ 
where $\langle x|\hat A|y \rangle$ stands for the kernel of operator $A$. 
The transition back from the symbol to the operator is called  quantization. 
The Weyl symbol corresponds to symmetric quantization. Notice that one can consider Weyl 
symbols not only for differential operators but also for other classes of operators. 
In particular if $\hat A$ is an integral operator with a kernel belonging to the Schwartz space then 
the Weyl symbol always exists and belongs to the Schwartz space. 
It is easy to check that the Weyl symbol of a product of two operators having Weyl symbols $a(z)$, $b(z)$ 
is given by the formula 
\begin{equation} \label{qplane}
c(z) \equiv  (a\ast b ) (z) = \left( e^{\pi i \theta^{jk}\frac{\partial}{\partial x^{j}}\frac{\partial}{\partial y^{k}}} 
a(x)b(y) \right)_{x=y=z} \, . 
\end{equation}  
Here $z$ stands for the $d=2n$-dimensional vector $(x, p)$, and $\theta^{jk}$ is a $d\times d$ antisymmetric matrix 
\begin{equation} \label{tt}
(\theta^{jk}) = \left( 
\begin{array}{cc}
0 & 1_{n\times n} \\
-1_{n\times n} & 0
\end{array} \right) \, . 
\end{equation}
This star product can be generalized as in  (\ref{st_pr}) for the case when (\ref{tt}) is replaced by an 
 arbitrary nondegenerate antisymmetric matrix $\theta^{jk}$.

It follows from the associativity of operator multiplication that the star product (\ref{qplane}) is 
also associative. Moreover, one can apply formula (\ref{qplane}) in the case when $\theta^{jk}$ is an 
arbitrary antisymmetric matrix and $d$ is not necessarily even. The product of functions defined in this way 
remains associative. It is called  Moyal or star product. 
For nonzero $\theta$ the multiplication (\ref{qplane}) is noncommutative. For example it is easy to calculate 
the following commutation relation
\begin{equation} \label{commutator}
[x^{j}, x^{k}]_{\ast} \equiv x^{j}\ast x^{k} - x^{k}\ast x^{j} = 2\pi i \theta^{jk} \, .
\end{equation}

One can consider various associative algebras of functions where the product is defined by (\ref{qplane}).
For instance one can consider the space of polynomials in $x^{i}$ equipped with the star product, that is the algebra 
generated by hermitian elements $x^{j}$ satisfying   (\ref{commutator}). 
Another option is to consider the space $S({\mathbb R}^{d})$  along with the  star product multiplication as a noncommutative 
deformation of the commutative algebra of Schwartz class functions on ${\mathbb R}^{d}$. 
We denote this algebra ${\mathbb R}^{d}_{\theta}$ and call it an algebra of smooth functions on a noncommutative $d$-dimensional
Euclidean space. 
 One easily sees that when $\theta \to 0$ the $\ast$-multiplication reduces to the usual commutative 
pointwise multiplication of 
functions. This means that the matrix $\theta^{jk}$ can be considered as a parameter of noncommutativity.

Notice that the algebra ${\mathbb R}^{d}_{\theta}$ contains neither the coordinate functions $x^{j}$ nor  the  function identically 
equal to 1 that would play the role of  unit element.

\subsection{Endomorphisms and connections} \label{Endsec}

Along with sections of a vector bundle $\zeta$  that pick a vector in a fiber over every point in the base $\cal M$ 
we can consider sections taking values in linear operators on fibers. Such sections are called endomorphisms 
of $\zeta$. More precisely endomorphisms of $\zeta$ are sections of the vector bundle $End (\zeta ) = \zeta^{*}\otimes \zeta$ 
where $\zeta^{*}$ is the dual bundle (a bundle whose fibers are dual  vector spaces to the corresponding fibers of $\zeta$). 
The space of sections of $End(\zeta )$ is naturally a vector space but in addition to that endomorphisms 
can be multiplied by means of a pointwise composition of the corresponding linear operators acting on  fibers. 
This gives a structure of associative algebra to the space of endomorphisms. A pointwise matrix trace applied to an 
endomorphism of $\zeta$ gives us a function over $\cal M$. Let us assume that  the manifold $\cal M$ is equipped with a volume form.
 If $\cal M$ is compact we can normalize the volume form  so that the volume of $\cal M$ is 1. 
Then, the  composition of matrix trace and an integral applied to an endomorphism yields a number. One can easily see 
that this operation defines a trace on the algebra of endomorphisms.
The trace of the identity endomorphism equals the dimension of the vector bundle $\zeta$. 
In general for any associative algebra $A$ a mapping ${\rm Tr}: A\to {\mathbb C}$ is called a trace 
if it satisfies ${\rm Tr}(ab) = {\rm Tr}(ba)$ for any $a,b \in A$.
If the algebra $A$ is unital (has a unit element) we say that the trace is normalized if 
${\rm Tr}{\bf 1} = 1$.

For example 
one can define a trace on the algebra ${\mathbb R}^{d}_{\theta}$ by means of integration of functions over ${\mathbb R}^{d}$.
The fact that we take functions from the Schwartz space ensures the convergence of integrals. 
It is easy to check that the defining property of trace: ${\rm Tr} f\ast g = {\rm Tr} g\ast f$ is 
satisfied. Note that due to the noncompactness of the underlying commutative   ${\mathbb R}^{d}$ the 
algebra ${\mathbb R}^{d}_{\theta}$ does not contain a unit element. There is no natural way, therefore, 
to normalize the trace. We will see more examples of noncommutative algebras  and traces  on them in the subsequent 
sections. Usually we will assume that our algebras of functions on a noncommutative space contain a unit element.

Consider now a projective module $E$ over an associative noncommutative algebra $A$. A linear operator $Z: E\to E$ is called 
{\it an endomorphism of module $E$}  if it commutes with the action of $A$ on $E$. In other words endomorphisms are 
$A$-linear maps.  Endomorphisms can be added together, 
composed and multiplied by a complex number. This means that they form an associative algebra denoted $End_{A}E$. 
For example consider a free module $E=A^{N}$. Its elements are $N$-tuples of elements from $A$ and the action of $A$ 
is a componentwise multiplication from the left. Endomorphisms of $E$ correspond to  $N\times N$ matrices whose 
entries are elements of $A$. These matrices act on $E$ by multiplication from the right. As an algebra 
$End_{A} A^{N}$ is isomorphic to $Mat_{N}( A^{op})$ where $A^{op}$ is an associative algebra whose elements 
are elements of $A$ and the multiplication $\circ $ is defined as $a\circ b = ba$ where $ba$ is the ordinary product 
in $A$. This description comes from the fact that we used a  right action above.

 A normalized trace on $A$ (an analog of the integral over $\cal M$ in the commutative case)  
 gives rise to a  (non-normalized) trace $\rm Tr$ on the algebra $End_{A}E$ that can be described 
as follows. Since $E$ is projective it  can be realized as a direct 
summand in a free module: $A^{N}= E\oplus E'$. This decomposition determines an 
endomorphism $P: A^{N} \to A^{N}$ projecting $A^{N}$ onto $E$. 
This means that $P^{2}= P$, $Px = x$ for $x\in E$ and $Px' = 0 $ for $x'\in E'$. The endomorphisms of 
module $E$ can be identified with a subalgebra in $Mat_{N}( A^{op})$ of endomorphisms of  $A^{N}$. 
Namely this subalgebra consists of elements of the form $PaP$. The algebra  $Mat_{N}(A^{op})$ has a 
canonical trace $\tau_{N}$ that is a composition of the matrix trace with a given normailized trace on  $A$. 
 Restricting this trace to the corresponding subalgebra we obtain a canonical  trace $\rm Tr$ on $End_{A}E$. 
By definition the number ${\rm Tr} {\bf 1}$ is called {\it the dimension of module $E$ } and is denoted ${\rm dim}(E)$. 
Here ${\bf 1}\in End_{A}E $ is the identity endomorphism. It follows from the definition of ${\rm Tr}$ that 
 the dimension is equal to the trace of the corresponding projector in $Mat_{N}(A^{op})$
\begin{equation} \label{dim}
 {\rm dim}(E) = {\rm Tr} {\bf 1} = \tau_{N} P \, .  
\end{equation}

Now let us turn to connections on a vector bundle. Let $p: T \to {\cal M}$ be an $n$-dimensional vector bundle. 
In local coordinates a connection can be written as a collection of 
 differential  operators $\nabla_{i} = \partial_{i} + A_{i}(x)$, $i = 1, \dots, {\rm dim}{\cal M}$ acting on smooth sections of $T$. 
  Here $A_{i}(x)$ are $N\times N$-matrix valued functions.
Moreover for any vector field $X= X^{i}(x)\partial_{i}$ on $\cal M$ we can consider an operator 
$\nabla_{X} = X^{i}\nabla_{i}$ satisfying   
\begin{equation} \label{comLeib}
\nabla_{X} (f\cdot s) = f\cdot \nabla_{X}(s) + \delta_{X}(f)\cdot s
\end{equation}
for any section $s$ and any function $f\in C^{\infty}({\cal M})$. Here $\delta_{X}(f) = X^{i}\partial_{i} f$. 
In addition to this it is required that $\nabla_{fX} = f\nabla_{X}$ for any $f\in C^{\infty}({\cal M})$. 
This requirement along with (\ref{comLeib}) can be taken as a definition of connection on a vector bundle.
In noncommutative geometry one can give different definitions of connection (see \cite{Connesbook}). 
In the simplest definition we assume that there is a Lie algebra $L$ that acts on associative algebra $A$ 
by means of infinitesimal automorphisms (derivations). This means that we have linear operators 
$\delta_{X} : A \to A$ that linearly depend on $X\in L$ and satisfy   
$$
\delta_{X}(a\cdot b) = (\delta_{X}a)\cdot b + a\cdot (\delta_{X}b)
$$
for any  $a,b \in A$. The mapping $X\mapsto \delta_{X}$ 
is  a Lie algebra homomorphism, i.e. 
$\delta_{[X,Y]} = [\delta_{X}, \delta_{Y} ] $.
{\it A connection $\nabla_{X}$ on an $A$-module (=module over $A$) 
 defined with respect to the Lie algebra $L$ and action $\delta_{X}$}  
is by definition a set of linear operators $\nabla_{X}$, $X\in L$ on $E$ depending linearly on $X$ and 
satisfying the Leibniz rule
\begin{equation} \label{Leib}
\nabla_{X} (a\cdot e) = a\cdot \nabla_{X}(e) + \delta_{X}(a)\cdot e
\end{equation}
for any  $e\in E$ and any  $a\in A$. 
This is a generalization of (\ref{comLeib}). 
It follows from the definition (\ref{Leib}) that for any  two connections $\nabla_{X}$ and $\nabla_{X}'$ 
the difference $\nabla_{X}' - \nabla_{X}$ commutes with the action of $A$ on $E$, i.e. is an endomorphism of $E$. 
 
Hence, if we fix some fiducial connection $\nabla_{i}^{0}$ on $E$ an arbitrary connection has the form 
\begin{equation} \label{fid_conn}
\nabla_{X} = \nabla_{X}^{0} + A_{X}
\end{equation}
where $A_{X} \in End_{A}E$ depend linearly on $X\in L$.

A curvature of  connection $\nabla_{X}$ is a two-form $F_{XY}$ on $L$ with values in linear operators on $E$ 
that measures a deviation of mapping $X\mapsto \nabla_{X}$ from being a Lie algebra homomorphism: 
$$
F_{XY} = [\nabla_{X}, \nabla_{Y}] - \nabla_{[X,Y]} \, .
$$
It is easy to check that $F_{XY}$ commute with the action of $A$ on $E$, i.e. $F_{XY}$ take values 
in the endomorphisms of module $E$. Suppose now that the Lie algebra $L$ is abelian and choose 
a basis $X_{i}$ in $L$. Then we have the curvature tensor 
$$
F_{ij} = [\nabla_{i}, \nabla_{j}] \, , \quad \nabla_{i} \equiv \nabla_{X_{i}} \, .
$$

On any projective module specified by a projector $P: A^{N} \to A^{N}$ one can construct a connection 
in the following way. The set of derivations $\delta_{X}$ acts naturally on the free module $A^{N}$. 
Consider a set of  operators  $\nabla^{LC}_{X} = P\cdot \delta_{X}\cdot P$. Evidently this operators commute 
with $P$ and hence they descend to operators on $E = P\, A^{N}$.  Moreover, because $P$ is an endomorphism 
of $A^{N}$ the operators $\nabla^{LC}_{X}$ satisfy $[\nabla^{LC}_{X}, a] = P\cdot [\delta_{X}, a]\cdot P = 
[\delta_{X}, a]\cdot P$ that  implies (\ref{Leib}) when restricted on $E$. Hence $\nabla^{LC}_{X}$ defines a connection 
on an arbitrary projective module which may be called a Levi-Civita connection. It is easy to check that 
${\rm Tr} [\nabla_{X}^{LC}, \phi] = 0$ for any endomorphism $\phi \in End_{A} E$. It follows from 
this identity and the fact that any two connections differ by an endomorphism that 
\begin{equation} \label{ttt}
{\rm Tr} [\nabla_{X}, \phi ] = 0 
\end{equation}
 fro an arbitrary connection $\nabla_{X}$ and an arbitrary endomorphism $\phi$.

%%%%%%%%%%%%%%%%%%%%%%%%%%%%%%%%%%%%%%%%%%%%%%%%%%%%%%%%%%%%%%
\subsection{Involutive algebras} \label{Invsec}

An operator ${}^{*}$ acting on an associative algebra $A$  is called an (antilinear) involution 
if $(ab)^{*}=b^{*}a^{*}$, $(a + b)^{*}= a^{*} + b^{*}$, $(\lambda a)^{*} = \bar \lambda a^{*}$.
The standard examples of involution are the complex conjugation in the algebra of functions and hermitian 
conjugation in the algebra of operators. An element $a\in A$ is called selfadjoint if $a=a^{*}$ and 
unitary if $aa^{*} = a^{*}a = 1$. 

An involution ${}^{*}$ on $A$ induces an involution on the matrix algebra $Mat_{N}(A)$ by the formula 
$(a^{*})_{ij} = a_{ji}^{*}$. Therefore, we can talk about selfadjoint and unitary elements of $Mat_{N}(A)$. 
Identifying the algebra $End_{A}E$ of endomorphisms of projective module $E$ with matrices of the form 
$eae$ where $a\in Mat_{N}(A)$ and $e$ is a selfadjoint projector we obtain an involution on $End_{A}E$.

If the algebra $A$ is endowed with an involution we can define an $A$-valued inner product in a free 
module $A^{N}$ by the formula 
$$
<(a_{1}, \dots , a_{N}), (b_{1}, \dots , b_{N})>_{A} = \sum_{i=1}^{N}a_{i}^{*}b_{i} \, .
$$
Embedding a projective module into a free module we obtain an $A$-valued inner product $<.,.>_{A}$ 
on a projective module. 
We can also impose a hermiticity condition on a connection on a projective  $A$-module $E$. 
A connection $\nabla_{X}:E\to E$ is said to be compatible with the inner product (unitary) if it satisfies
$$
<\nabla_{X}\xi , \eta>_{A} +  <\xi, \nabla_{X}\eta>_{A} = \delta_{X}(<\xi,\eta >_{A})
$$ 
for any $\xi , \eta \in E$. 
If Tr is a trace on $A$ then ${\rm Tr}<.,.>_{A}$ defines an ordinary $\mathbb C$-valued hermitian inner product.
The above compatibility condition implies that the operators $\nabla_{X}$ is an antihermitian operator. 

 Note that in the ordinary Yang-Mills 
theory choosing $\nabla_{j} = \partial_{j} + iA_{j}$ corresponds to a hermitian Yang-Mills field $A_{j}$. 
 The Yang-Mills field strength
 $F_{jk}^{a} = \partial_{j}A_{k}^{a} - \partial_{k}A_{j}^{a} + f^{a}_{bc}A^{b}_{j}A^{c}_{k}$ is hermitian with respect 
to the invariant inner product on the space of adjoint sections. 
In our conventions the endomorphisms $A_{X}$ 
are antihermitian and so is the curvature $F_{XY}$. 
In all of the subsequent sections we will assume that we work with unitary connections.

 %%%%%%%%%%%%%%%%%%%%%%%%%%%%           N. C.    T O R I            %%%%%%%%%%%%%%%%%%%%%%%%%%%%%%%%%%%%%%%%%%%%%%%%%%%

\subsection{Noncommutative tori} \label{Nctorisec}
Let us consider an algebra generated as a linear space by elements $U_{\bf n}$ where 
$\bf n = (n_{1}, \dots , n_{d})$ is a $d$-dimensional 
vector with integer entries. We assume that multiplication in this algebra is given by the formula 
\begin{equation} \label{un}
U_{\bf n}U_{\bf m} = e^{\pi i n_{j}\theta^{jk}m_{k}} U_{\bf n + m}
\end{equation}
where $\theta^{jk}$ is an antisymmetric $d\times d$ matrix, and summation over repeated indices is assumed. 
Denote this algebra by $T_{\theta}^{d}$.
We introduce an involution $*$ in $T_{\theta}^{d}$ by the rule: $U_{\bf n}^{*} = U_{-\bf n}$. 
The elements $U_{\bf n}$ are assumed to be unitary with respect to this involution, i.e. 
$U^{*}_{\bf n}U_{\bf n} = U_{-\bf n}U_{\bf n} = 1$. Denote $U_{i}=U_{\bf e^{i}}$ where 
${\bf e^{i}} = (e^{i}_{1}, \dots , e_{d}^{i})$, $e^{i}_{j} = \delta_{i}^{j}$. 
These elements satisfy the condition 
\begin{equation} \label{ui}
 U_{j}U_{k} = e^{2\pi i\theta^{jk}}U_{k}U_{j} \, .
\end{equation}
It is simple to check the following identity 
$$
U_{\bf n} = U_{1}^{n_{1}}U_{2}^{n_{2}}\dots U_{d}^{n_{d}}e^{-\pi i \sum_{j<k} n_{j}n_{k}\theta^{jk}} 
$$
where $U_{i}$ taken to a negative power is understood as the corresponding power of the inverse operator 
$U_{i}^{-1} = U_{i}^{*}$. 
Hence, the elements 
$U_{1}, \dots , U_{d}, U_{1}^{*}, \dots , U_{d}^{*}$ are multiplicative generators of $T_{\theta}^{d}$.

A unitary representation of $T_{\theta}^{d}$ (a $T_{\theta}^{d}$-module, or more precisely a left $T_{\theta}^{d}$-module) 
is by definition a set of unitary operators in a Hilbert space $E$ satisfying  (\ref{un}). Clearly in order to specify 
such a representation it suffices to find $d$ unitary operators obeying (\ref{ui}).

There is a canonical normalized trace  on $T_{\theta}^{d}$  specified by the rule 
\begin{equation} \label{Tr}
{\rm Tr}U_{\bf n} = 0 \enspace {\mbox if} \enspace {{\bf n} \ne {\bf 0}} \, , 
\quad {\rm Tr} U_{\bf 0} \equiv {\rm Tr} 1 = 1 \, .
\end{equation}
For $\theta = 0$ we can realize the algebra $T_{\theta}^{d}$ as an algebra of trigonometric polynomials on a 
$d$-dimensional torus $T^{d}$. Namely if $\sigma^{1}, \dots , \sigma^{d} \in [0, 2\pi)$ are angular coordinates 
on $T^{d}$ we can identify $U_{\bf n}$ with $e^{i \sum \sigma^{j}n_{j}}$. Then the trace (\ref{Tr}) corresponds 
to an integral over $T^{d}$ provided the volume of  $T^{d}$ is  1. 
It follows from the general discussion in the previous section that this canonical trace gives raise to a trace 
on the algebra of endomorphisms $End_{T_{\theta}^{d}}E$ of any projective $T_{\theta}^{d}$-module $E$.

For a general $\theta$ we can assign to every linear combination of $U_{\bf n}$ a function on the torus $T^{d}$ 
assuming that $U_{\bf n}$ goes to  $e^{i \sum \sigma^{j}n_{j}}$. However the multiplication in noncommutative 
algebra $T_{\theta}^{d}$ does not correspond to a pointwise multiplication of functions under this map. 
It is easy to check that the appropriate multiplication denoted by $\ast$ for functions on $T_{\theta}^{d}$ is 
given by the formula 
\begin{equation} \label{torus_star}
(f\ast g ) (\sigma) = \left( e^{\pi i \theta^{jk}\frac{\partial}{\partial \sigma^{j}}\frac{\partial}{\partial \nu^{k}}} 
f(\sigma)g(\nu) \right)_{\nu =\sigma} \, . 
\end{equation}
This product is called Moyal product. When $\theta$ goes to zero  it reduces to the 
usual pointwise product of functions. 
In such a way the algebra $T_{\theta}^{d}$ can be considered as a noncommutative deformation of 
the algebra of functions on the commutative torus $T^{d}$. It is called therefore an algebra of functions 
on noncommutative torus. Note that finite linear combinations of $U_{\bf n}$ that we considered until now 
correspond to a very narrow class of functions on $T^{d}$. However we can also consider infinite linear combinations. 
In particular  we can consider linear combinations $\sum_{\bf n} C({\bf n}) U_{\bf n}$ where the coefficients $C({\bf n})$ tend 
to 0 faster than any power of $\| {\bf n} \|$. The algebraic operations can be extended to the set of such linear combinations. 
One can say that this set constitutes 
an algebra of smooth functions on a noncommutative torus (the corresponding functions on 
commutative torus are smooth). From now on we will use the notation $T_{\theta}^{d}$ for this algebra. 
This algebra was first introduced and studied in a seminal paper \cite{Connes1}. 

The canonical normalized trace (\ref{Tr}) can be represented in terms of the ordinary integration of functions on $T^{d}$ 
\begin{equation} \label{trace=int}
{\rm Tr} f = \frac{1}{(2\pi )^{d}} \int  d\sigma^{d} \, f(\sigma ) \, .
\end{equation}
Notice that this trace is a positive functional , i.e. ${\rm Tr}A^{*}A \ge 0$ for any 
$A \in T_{\theta}^{d}$.

%%%%%%%%%%%%%%%%%%%%%%%%%%%%%%%%%%%%%%%%%%%%%%%%%%%%%%%%%%%%%%%%%%%%%%%%%%%%%%%%%%%%%%%%%%%%%%%%%%%%%%%%%%%%%%%

\subsection{Projective modules over noncommutative tori} \label{Projsec}

In this section we are going to describe (finitely generated) projective modules over noncommutative tori.
They were investigated in depth in \cite{RieffelProj}.  
We start with free modules that are analogs of trivial bundles.

Let us consider a free module over $T_{\theta}^{d}$ of rank $N$, i.e.   $E=(T_{\theta}^{d})^{N}$ consisting of $N$-tuples of elements from 
$T_{\theta}^{d}$. ($T_{\theta}^{d}$ acts by means of multiplication from the left.) 
 Endomorphisms of this module can be identified with $N\times N$ 
matrices with entries from $T_{\theta}^{d}$. These matrices act on $(T_{\theta}^{d})^{N}$ by 
means of multiplication from the right. One easily sees that as an algebra $End_{T_{\theta}^{d}}E$ 
is isomorphic to the matrix algebra $Mat_{N}(T_{-\theta}^{d})$ where the minus sign comes from 
the fact that we have a  right action (in the notations of section \ref{Endsec} we have $(T_{\theta}^{d})^{op} \cong T_{-\theta}^{d}$).
Notice that the algebra of endomorphisms is also 
equipped with a trace (it is given by a combination of matrix trace with the trace defined in (\ref{Tr})). 
Of course, this is just a particular example of the general fact that there is a canonical trace on the algebra of endomorphisms 
of any projective module over $T_{\theta}^{d}$.

Now we are going to describe a large class of projective modules over $T_{\theta}$, the 
so called Heisenberg modules. We start with some examples illustrating the main idea and give the most general construction of a Heisenberg 
module at the end of the section.

\noindent {\bf Example 1.} Consider the Schwartz space $E = {\cal S}({\mathbb R})$. On this space we can define the 
following two operators: $U_{1}: E \to E$, $U_{2} : E \to E$ by the formulas 
\begin{equation} \label{R1}
(U_{1}f)(x) = f(x + \gamma) \, , \qquad (U_{1}f)(x) = f(x)e^{2\pi i  \tilde \gamma x}
\end{equation}   
where $\gamma$ and $\tilde \gamma$ are numbers.  One can check that these operators  satisfy 
$$
U_{1}U_{2} = U_{2}U_{1}e^{2\pi i  \tilde \gamma \gamma} \, .
$$ 
This means that these operators represent a $2$-dimensional noncommutative torus $T_{\theta}^{2}$ with 
 $\theta^{12} = -\theta^{21} = \tilde \gamma \gamma$.
One can easily describe endomorphisms of this module. They are generated by the operators $Z_{1}$, $Z_{2}$:
\begin{equation} \label{Z1}
(Z_{1}f)(x) = f(x + \tilde \gamma^{-1}) \, , \qquad (Z_{2} f)(x) = f(x)e^{2\pi i x\gamma^{-1}} \, .
\end{equation}
We see that the algebra $End_{T_{\theta}^{2}}E$ is isomorphic to $T_{\tilde \theta}^{2}$ where 
$\tilde \theta^{12} = \tilde \gamma^{-1}\gamma^{-1}$ that is  $\tilde \theta^{ij} = (\theta^{-1})^{ij}$.

\noindent {\bf Example 2.} Let us show now how this construction can be generalized. 
In the two-dimensional case we can assume that the torus is labeled by a single number $\theta^{12}=\theta \in {\mathbb R}$. 
 Elements of our  module will be functions $\phi_{j}(x) \in {\cal S}({\mathbb R}\times {\mathbb Z}_{m})$, 
$x\in {\mathbb R}$, $j\in  {\mathbb Z}_{m}$ where ${\mathbb Z}_{m}$ is a cyclic group of order $m$. 
Define  the generators 
 
\begin{eqnarray}\label{U's}
  (U_{1}\phi)_{j}(x) &=& \phi_{j-1}\left( x-\frac{n}{m} + \theta \right) \, , \nonumber \\
 (U_{2}\phi)_{j}(x) &=& \phi_{j}(x) e^{2\pi i(x-jn/m)} 
\end{eqnarray}
where $n$ is some integer. 
It is easy to check that  these generators satisfy $U_{1}U_{2} = U_{2}U_{1}e^{2\pi i \theta}$. 
Hence, we constructed  modules  $E_{n,m}$  over $T_{\theta}^{2}$ labeled by two integers $n,m$, with $m>0$.
These modules can be proven to be projective if (and only if) $n -m\theta \ne 0$ that can fail only 
for rational $\theta$. 
 The integers $n$, $m$ are related to topological numbers of the module (see section \ref{Ksec} for a detailed discussion). 
It turns out that, assuming    $\theta$ is irrational, an arbitrary   $T_{\theta}^{2}$-module is 
either a free module or is isomorphic to one of the modules $E_{n,m}$.

In the case when the integers $m$ and $n$ are relatively prime (and $\theta$ is irrational) the endomorphisms of 
module $E_{n,m}$  are generated by operators 
\begin{eqnarray} \label{Z's}
(Z_{1}\phi)_{j}(x) &=& \phi_{j - a}\left( x  - \frac{1}{m} \right) \nonumber \, ,  \\
(Z_{2}\phi)_{j}(x) &=& \phi_{j}(x) exp\left[  2\pi i \left( \frac{x}{n-m\theta} - \frac{j}{m} \right) \right]
\end{eqnarray}
where $a$ is an integer satisfying $an-bm=1$ for some other integer $b$. 
These generators satisfy 
\begin{equation}
Z_{1}Z_{2} = e^{2\pi i \hat \theta} Z_{2}Z_{1} \, , \quad \hat \theta = \frac{b-a\theta}{n-m\theta} \, .  
\end{equation}
That is the endomorphisms themselves constitute a noncommutative torus $T_{\hat \theta}$ with a different noncommutativity 
parameter $\hat \theta$. In general if $g.c.d.(m, n) = D >1$ one can show that the algebra of endomorphisms of $E_{n,m}$  
is  isomorphic to a matrix algebra  $Mat_{D}(T_{\hat \theta})$. (This follows from the fact that $E_{n, m}$ is 
isomorphic to a direct sum of $D$ copies of $E_{n', m'}$, $n' = n/D$, $m' = m/D$.)

\noindent {\bf General construction.} The examples above lead us to the following  general construction.  
 Let $G$ be a direct sum of ${\mathbb R}^{p}$ and an abelian finitely generated group, and let 
$G^{*}$ be  its dual group. The last one is defined as a group of homomorphisms 
$\mu: G \to S^{1}$. We will identify $S^{1}$ with a group of complex numbers with absolute 
value 1 and will use additive notation for group product in $G$ and multiplicative notation for 
group product in $S^{1}$ and $G^{*}$. Thus, for any $\mu \in G^{*}$ 
and any $g_{1}, g_{2} \in G$ we have  $\mu(g_{1}+ g_{2}) = \mu (g_{1}) \mu(g_{2})$. 
By definition $(\mu_{1}\mu_{2})(g) = \mu_{1}(g)\mu_{2}(g)$. 
For the group $G={\mathbb R}^{p}$ one can identify $G^{*}$ with the group itself 
 by means of an exponential mapping $(x_{j}) \mapsto e^{2\pi i\alpha^{j} x_{j}} \in S^{1}$, 
$(x_{j}) \in  {\mathbb R}^{p}$, $(\alpha^{j}) \in {\mathbb R}^{p*}\cong {\mathbb R}^{p}$. 
For the group ${\mathbb Z}^{q}$ the analogous exponential mapping establishes an isomorphism 
of the corresponding dual group to a torus $T^{q}\cong {\mathbb R}^{q}/{\mathbb Z}^{q}$. 
Finally a finite abelian group can be represented as a product of cyclic groups ${\mathbb Z}_{m}$
each one of which can be identified with a subgroup of $S^{1}$ consisting of $m$-th roots of 
unity. The dual group then is  isomorphic to the finite  group itself. 
  In the most general situation 
$G={\mathbb R}^{p}\times {\mathbb Z}^{q}  \times F$ where $F$ is a finite group. Then 
$G^{*}\cong {\mathbb R}^{p}\times T^{q} \times F^{*}$.

Consider a linear space ${\cal S}(G)$ of functions on $G$ decreasing at infinity faster than any power. 
We define operators $U_{(\gamma, \tilde \gamma)}: {\cal S}(G)\to {\cal S}(G)$ labelled by a pair 
$(\gamma, \tilde \gamma)\in G\times G^{*}$ acting as follows 
\begin{equation}\label{U}
(U_{(\gamma, \tilde \gamma)}f)(x)=\tilde \gamma (x) f(x+ \gamma ) \, . 
\end{equation}
One can check that the operators $U_{(\gamma, \tilde \gamma)}$ satisfy the commutation relations 
\begin{equation} \label{nt}
U_{(\gamma, \tilde \gamma)}U_{(\mu, \tilde \mu)}=   \tilde \mu (\gamma )\tilde \gamma^{-1} (\mu ) 
U_{(\mu, \tilde \mu)}U_{(\gamma, \tilde \gamma)}   \, .
\end{equation}
If  $(\gamma, \tilde \gamma)$ run over a $d$-dimensional discrete subgroup  $\Gamma \subset G\times G^{*}$, 
$\Gamma \cong  {\mathbb Z}^{d}$         
then  formula (\ref{U}) defines a module over a $d$-dimensional noncommutative torus $T_{\theta}^{d}$ 
with 
\begin{equation}\label{cocycle}
exp(2\pi i \theta_{ij}) = \tilde \gamma_{i} (\gamma_{j} )\tilde \gamma_{j}^{-1} (\gamma_{i} )
\end{equation}
for a given basis $(\gamma_{i} , \tilde \gamma_{i})$ of the lattice $\Gamma$.  
This module is projective if $\Gamma$ is such that $G\times G^{*}/\Gamma$ is compact \cite{RieffelProj}.
(Note that in the decomposition of $G$ above one has $2p + q = d$.) 
 If that is the case then the projective   $T_{\theta}^{d}$-module at hand is called
 {\it a Heisenberg module} and denoted  $E_{\Gamma}$.

It is not hard  to see that the construction (\ref{U's}) in Example 2 above 
is precisely of this general form and corresponds to the choice $G={\mathbb R}^{1}\times {\mathbb Z}_{m}$.
The dual group $G^{*}$ can be identified with the group itself and the lattice $\Gamma \subset G\times G^{*}$ 
is generated by the basis $(\gamma_{1}, \tilde \gamma_{1}) = ((\theta - n/m, -1), (0,0)$, 
 $(\gamma_{2}, \tilde \gamma_{2}) = ((0, 0), (1,-n))$.
 
Let us give  here one more example of a Heisenberg module. 

\noindent {\bf Example 3.} If $\theta$ is rational  one 
can construct a representation of $T_{\theta}^{d}$ by means of finite-dimensional matrices. 
For a two-torus $T_{\theta}^{2}$, $\theta^{12} = m/n$ an example of such  representation  can be constructed 
via the familiar clock and shift operators on an $n$-dimensional vector space: 
\begin{equation}\label{clockandshift}
U_{jk} = e^{2\pi i jm/n} \delta_{j,k} \, , \quad V_{jk} = \delta_{j+1, k} \, .
\end{equation}
These matrices satisfy $UV = VUe^{2\pi i m/n}$ and thus represent the  $T_{\theta}^{2}$.  
From the point of view of the general construction above this module corresponds to the choice 
$G={\mathbb Z}_{n}$. It is easy to generalize this construction to any $T_{\theta}^{d}$ with a rational matrix $\theta$.

Endomorphisms of a Heisenberg module $E_{\Gamma}$ have a basis consisting 
of operators $Z_{(\nu, \tilde \nu)}= U_{(-\nu, \tilde \nu^{-1})}$ with the pair
 $(\nu, \tilde \nu)$ belonging  to the dual group $\Gamma^{*}$, i.e. satisfying 
$\tilde \nu (\mu) = \tilde \mu (\nu)$ for any $(\mu,\tilde \mu)\in \Gamma$. 
Note that in general  $\Gamma^{*}$ is not a lattice, it might have some finite order elements. 
In the case when  $\Gamma^{*}$ is a lattice the algebra $End_{T_{\theta}^{d}}E_{\Gamma}$ is naturally 
isomorphic to a $d$-dimensional noncommutative torus $T_{\tilde \theta}^{d}$ where $\tilde \theta$ 
is determined by the lattice $\Gamma^{*}$ in the same fashion as it worked for $\Gamma$. 
The module $E_{n,m}$ considered in Example 2 illustrates the above assertion. Namely when 
the integers $n$ and $m$ are relatively prime the algebra $End_{T_{\theta}^{2}}E_{n,m}$ 
is generated by operators (\ref{Z's}) that evidently are of the same general  form as (\ref{U's}).  
One can easily check that these operators correspond to a  basis in $\Gamma^{*}$ that is dual 
to the one in $\Gamma$ corresponding to (\ref{U's}).

Heisenberg modules play a special role. 
 It was shown by Rieffel (\cite{RieffelProj}) that {\it if the matrix $\theta_{ij}$ is irrational in the sense 
that at least one of its entries   is irrational then any projective 
module over $T_{\theta}^{d}$ can be represented as a direct sum of Heisenberg modules. In that sense Heisenberg 
modules can be used as  building blocks to construct an arbitrary module.}

{\it If a Heisenberg module $E$ cannot be represented as a direct sum of isomorphic modules 
$E=E' \oplus E'\oplus \dots \oplus E'$ then it is called  a basic module.} Evidently by the result stated above 
if $\theta_{ij}$ is irrational any projective module can be represented as a direct sum of basic modules. 
Moreover the algebra $End_{T_{\theta}^{d}}E$ is isomorphic to a noncommutative torus $T_{\hat \theta}^{d}$ 
for some matrix $\hat \theta_{ij}$

%%%%%%%%%%%%%%%%%%%%%%%%%%%%%%%%%%%%%%%%%%%%%%%%%%%%%%%%%%%%%%%%%%%%%%%%%%%%%%%%%%%%%%%%%%%%%%%%%%%%%%%
\subsection{Connections on noncommutative tori} \label{Consec}
In order to define a connection on  a module $E$ over  noncommutative torus $T_{\theta}^{d}$ we 
will first define a natural Lie algebra  of shifts $L_{\theta}$  acting  on  $T_{\theta}^{d}$. 
The shortest way to define this Lie algebra is by specifying a basis consisting of  derivations 
$\delta_{j}$, $j=1,\dots , d$
satisfying 
\begin{equation} \label{deltaj}
\delta_{j} (U_{\bf n}) = 2\pi i n_{j}U_{\bf n} \, . 
\end{equation}
For the multiplicative generators $U_{j}$ the above relation reads as 
\begin{equation} \label{deltaij}
\delta_{j}U_{k} = 2\pi i \delta_{jk}U_{k} \, .
\end{equation}
This derivations then span a $d$-dimensional abelian Lie algebra that we denote $L_{\theta}$. 
However this definition singles out a basis in $L_{\theta}$ and one might be interested in a 
 covariant definition of $L_{\theta}$ as well as of the corresponding Lie group $\tilde L_{\theta}$ of 
automorphisms. This covariant definition can be given in the following way. 
 The torus generators $U_{\bf n}$ we defined above are labeled by elements of a lattice ${\mathbb Z}^{d}$. 
Consider this lattice as a $d$-dimensional lattice $D\cong {\mathbb Z}^{d}$ embedded into ${\mathbb R}^{d}$. 
The dual space ${\mathbb R}^{*d}$ acts naturally on $S({\mathbb Z}^{d})$ (the space of functions decreasing 
faster than any power). Namely to every $x\in {\mathbb R}^{*d}$ we assign a map $\tau_{x}$ transforming 
a function $f(\lambda)$ into $e^{2\pi i <x, \lambda>}f(\lambda )$ where $<.,.>$ stands for a natural 
pairing of elements from  ${\mathbb R}^{*d}$ with elements form ${\mathbb R}^{d}$. For any matrix $\theta$ 
the map $\tau_{x}$ can be considered as an automorphism of $T_{\theta}^{d}$ (it preserves the  relations 
(\ref{un}) in an obvious way). This automorphism is trivial if $x\in D^{*}$ where $D$ is the lattice 
dual to $D$. Denote the group ${\mathbb R}^{*d}/D^{*}$ considered as a group of automorphisms of $T_{\theta}^{d}$ 
by $\tilde L_{\theta}$. Then it is easy to check that   the Lie algebra of $\tilde L_{\theta}$ is isomorphic to  
  $L_{\theta}$. The fact that $\tilde L_{\theta}$ is a (commutative torus) is reflected in the relations 
$$
e^{2\pi  i \delta_{j}} U_{\bf n} e^{-2\pi i \delta_{j}} = U_{\bf n}  \, . 
$$
We see that in the basis (\ref{deltaj}) the periods of the torus $\tilde L_{\theta}$ are all equal to $2\pi$. 
We call this basis standard.

A connection on a module $E$ over $T_{\theta}^{d}$ is a set of operators $\nabla_{X}:E\to E$, $X\in L_{\theta}$ 
depending linearly on $X$ and satisfying 
\begin{equation}
[\nabla_{X}, U_{\bf n} ] = \delta_{X}(U_{\bf n})  
\end{equation}
where $U_{\bf n}$ are operators $E\to E$ representing the corresponding generators of $T_{\theta}^{d}$.
In the standard basis (\ref{deltaj}) this relation reads as 
\begin{equation}
[\nabla_{j}, U_{\bf n} ] = 2\pi i n_{j}U_{\bf n} \, .
\end{equation}
Notice that this is precisely one of the relations (\ref{Td}) that emerged in   the study of  M(atrix) model 
compactifications on tori (compare also with equation ({\ref{con}) written for the two-dimensional compactification).

Let us give some examples of connections. 
First consider a free module $E = (T_{\theta}^{d})^{N}$. Its elements are $N$-tuples 
$(a_{1}, \dots , a_{N})$, $a_{i}\in T_{\theta}^{d}$. 
One constructs a connection $\nabla_{i}^{0}$ by setting 
$\nabla_{i}^{0} (a_{1}, \dots , a_{N}) = (\delta_{i}(a_{1}), \dots , \delta_{i}(a_{N}) )$.
As we already described all endomorphisms in terms of matrices with entries in $T_{\theta}^{d}$ 
acting from the right we have a description of an arbitrary connection $\nabla_{i} = \nabla_{i}^{0} + A_{i}$, 
$A_{i} \in End_{T_{\theta}^{d}}E$.

Consider now a  module $E_{m,n}$ over $T_{\theta}^{2}$ introduced in the previous section.
Its  elements are 
functions $f_{j}(x)$ of one continuous and one discrete variable. We can try  constructing  a  connection 
$\nabla_{1}$, $\nabla_{2}$ in terms of suitable linear combinations of $\frac{\partial}{\partial x}$ and 
multiplication by $x$ operators. From (\ref{U's}) we find the following commutation relations 
\begin{eqnarray*} 
&& [x, U_{1}] = \frac{n-m\theta}{m} U_{1}\, , \quad [x, U_{2}] = 0 \, , \\
&&  [\frac{\partial}{\partial x}, U_{1}] = 0 \, , \quad   [\frac{\partial}{\partial x}, U_{2}] = 2\pi i U_{2} \, .  
\end{eqnarray*}
Therefore, we can set 
\begin{equation} \label{2dconn}
\nabla_{1} = \frac{2\pi i m}{n- m \theta}x \, , \qquad \nabla_{2} = \frac{\partial}{\partial x} 
\end{equation}
(these operators satisfy the defining relations (\ref{deltaij})).
One readily calculates the curvature of this connection 
$$
F_{12} = - F_{21} = -\frac{2\pi i m}{n- m\theta} \cdot {\bf 1} 
$$
where $\bf 1$ denotes the identity endomorphism. 
In general connections whose curvature equals  the identity endomorphism times a numerical tensor are called 
 constant curvature connections. In fact {\it   on any Heisenberg module 
 there exists a constant curvature connection.} Let us sketch the construction of such a connection. 
In general a Heisenberg module is constructed on functions on a group $G={\mathbb R}^{p}\times {\mathbb Z}^{q}\times F$. 
Let $x^{1}, \dots, x^{p}$ be coordinates on the ${\mathbb R}^{p}$ factor. Then one can construct connections $\nabla_{j}$ 
in terms of suitable linear combinations of partial derivatives $\frac{\partial}{\partial x^{j} }$ and 
operators acting by multiplications by $x^{k}$. Evidently such a connection has a constant curvature. 
In the commutative gauge theory the closest analog of  a constant curvature connection over a noncommutative torus  
is a $U(N)$ connection  on a vector bundle over $T^{d}$ with a vanishing $SU(N)$ part and with a constant (coordinate 
independent) $U(1)$ part.

%%%%%%%%%%%%%%%%%%%%%%%%%%%%         K - T H E O R Y      %%%%%%%%%%%%%%%%%%%%%%%%%%%%%%%%%%%%%%%%%%%%%%%%%%%%%%%%%%%
 
\subsection{K-theory, Chern character} \label{Ksec}
Let us remind the reader the main idea of K-theory. Given a manifold $\cal M$ one can consider 
a set $Vect({\cal M})$ of all vector bundles over it. 
More precisely we need to consider bundles defined up to isomorphisms, i.e. 
equivalence classes. Since we can add bundles by means of the direct sum operation the set $Vect({\cal M})$ 
forms a commutative  semigroup 
(we have to add a formal ``zero-dimensional'' bundle $0$ that plays the role of zero element). 
There is a standard procedure called a  Grothendieck construction that turns this semigroup into a group denoted 
$K^{0}({\cal M})$. Elements of $K^{0}({\cal M})$ are equivalence classes of formal differences $E_{1} - E_{2}$ where 
$E_{1}, E_{2} \in Vect({\cal M})$. Two differences $E_{1} - E_{2}$, $E_{1}' - E_{2}'$ are set to be equivalent 
if there exists $E_{3}\in Vect({\cal M})$ such that 
$$
E_{1} +  E_{2}' + E_{3}' = E_{1}' + E_{2} + E_{3} \, .
$$
In particular for any vector bundle $E$ there corresponds  an element in  $K_{0}({\cal M})$, it is 
given by the equivalence class of $E - 0$. Note that in general two nonisomorphic bundles might 
give rise to the same element in $K^{0}({\cal M})$. Elements of $K^{0}({\cal M})$ that do not correspond 
to any vector bundle are called virtual bundles.   This definition is very natural from the physical view 
point. Formal differences $E_{1} - E_{2}$ can be considered as super vector bundles (vector bundles having 
a linear superspace as a fiber).  More precisely $E_{1} - E_{2}$ can be identified with $E_{1}\oplus\Pi E_{2}$ 
where $\Pi$ denotes the Grassmann parity inversion. The equivalence relation above now reads $E\oplus \Pi E \sim 0$.

Following the idea of algebraization we can recast the above definition of $K^{0}({\cal M})$ in 
terms of projective modules over $C^{\infty}({\cal M})$. In fact the modification is straightforward 
because a direct sum of vector bundles corresponds to the direct sum of the corresponding projective modules. 
One can simply replace everywhere  above vector bundles by projective modules.  
Consider now an associative noncommutative algebra $A$. One can first define a semigroup of 
projective modules over $A$ in a natural way. Applying the Grothendieck construction to this semigroup we 
obtain   a group $K_{0}(A)$. Again $K_{0}(A)$ can be described in terms of ${\mathbb Z}_{2}$-graded 
modules over $A$ with the equivalence relation $E\oplus \Pi E \sim 0$.  The definition of $K_{0}(A)$ 
can be applied to any associative algebra. One can define also a group $K_{i}(A)$, $i \in {\mathbb N}$ as 
$K_{i}(A) = K_{0}(A\otimes C_{0}({\mathbb R}^{i}))$ where $C_{0}({\mathbb R}^{i})$ is the algebra 
of continuous functions on ${\mathbb R}^{i}$ tending to 0 at infinity (we equip this algebra with the $sup$ norm). 
One can prove that $K_{i}(A) \cong K_{i+2}(A)$ (Bott periodicity).

The $K$-groups of a noncommutative torus can 
be computed using the technique due to Pimsner and Voiculescu \cite{PV}. The answer coincides with that for 
commutative tori: 
$$
K_{0}(T_{\theta}^{d}) \cong {\mathbb Z}^{2^{d-1}}\cong K_{1}(T_{\theta}^{d}) \, . 
$$
This result is really not surprising because $K$-groups are discrete objects and should not change under continuous 
deformations of algebra.

In the case of  standard K-theory of a manifold $\cal M$ there exists a mapping called Chern character
\begin{equation} \label{commch}
{\rm ch} : K^{0}( {\cal M}) \to H^{even}({\cal M}, {\mathbb Z}) \, . 
\end{equation}
constructed in the following way. Consider an $n$-dimensional  vector bundle  $\zeta$ specified by $p: T \to {\cal M}$
and  equipped with a connection $\nabla_{X}$. It follows from definition (\ref{comLeib}) that 
the curvature $F_{XY} = [\nabla_{X}, \nabla_{Y}]$ can be considered as a differential two-form $F$
with values in endomorphisms of  the bundle $\zeta$.  More precisely $F$ is  a section  of 
the bundle $\Lambda^{2}T^{*}({\cal M})\otimes End (\zeta ) $. Locally $F$ can be represented in terms 
of $N\times N$ matrix-valued differential 2-form. Denote $F^{k} = F\wedge ... \wedge F$ - the 
$k$-th exterior power of the form $F$. Then ${\rm tr}\, F^{k}$ are ordinary differential $2k$-forms. 
Here $\rm tr$ is a matrix trace. One can prove the following two facts about the forms ${\rm tr}\,  F^{k}$: \\
1)  The forms ${\rm tr}\,  F^{k}$ are closed. \\
2) If $F(\nabla^{1})$ and $F(\nabla^{2})$ are curvatures corresponding to two different connections 
$\nabla^{1}_{X}$ and $\nabla^{2}_{X}$ then  $ {\rm tr}\, F^{k}(\nabla^{1}) - {\rm tr}\, F^{k}(\nabla^{2})$ is 
an exact 2-form. \\
The second fact implies that the cohomology classes of ${\rm tr}\,  F^{k}$ does not depend on the choice 
of connection $\nabla_{X}$ and thus depends only on a vector bundle $\zeta$. 
Moreover the  cohomology classes of the forms $\frac{1}{(2\pi )^{k}k!}{\rm tr}\,  F^{k}$ 
turn out to be integral. Therefore, one can consider an inhomogeneous cohomology class
$$
{\rm ch}(\zeta ) =  [{\rm tr} \, exp\left(\frac{1}{2\pi } F \right)] \in H^{even}({\cal M}, {\mathbb Z})    \, .
$$
This construction  naturally gives rise to a mapping (\ref{commch}) from  the  $K^{0}$-group to the even 
integral cohomologies that is called a Chern character. In the case when $\cal M$ is a torus $T^{d}$ 
 we can choose a basis $\alpha^{i}$, $i=1,\dots , d$  of 1-forms 
in $H^{1}(T^{d}, {\mathbb Z})$. Then, the whole $H^{even}(T^{d}, {\mathbb Z})$ can be considered as 
an even part of the Grassmann algebra generated by the anticommuting variables $\alpha^{i}$ over 
the integers. The Chern character then has the form ${\rm ch}(\zeta ) = n + \frac{1}{2}m_{ij}\alpha^{i}\alpha^{j} + \dots $ 
where $n$ is the dimension of $\zeta$, $m_{ij} \in {\mathbb Z}$ are magnetic fluxes, etc.   
The Chern character is associated in physics with a collection of D-branes (for example  see \cite{Witten}). 
The integer $n$ is interpreted 
as a number of D-branes of dimension $d$, $m_{ij}$ are numbers of branes of dimension $d-2$ wrapped on 
a cycles transverse to directions $i, j$, etc.

Let us now explain how one can define a Chern character of a projective module $E$ over a 
noncommutative torus $T_{\theta}^{d}$. Let $\nabla_{X}$ be a connection on $E$ defined 
with respect to the Lie algebra  $L_{\theta}$ (Lie algebra of shifts). The curvature $F_{XY}$ is 
an exterior two-form on the adjoint vector space $L_{\theta}^{*}$ with values in $End_{T_{\theta}^{d}}E$. 
As it was already discussed in the previous sections there is 
 a canonical  trace $\rm Tr$ on the algebra $End_{T_{\theta}^{d}}E$. 
We define the {\it Chern character ${\rm ch}(E)$} as  
\begin{equation} \label{ncChern}
{\rm ch}(E) = {\rm Tr} \, exp\left(\frac{1}{2\pi i }F\right) \in \Lambda^{even}(L^{*}_{\theta})  
\end{equation}
where $\Lambda^{even}(L^{*})$ is the even part of the exterior algebra of $L_{\theta}^{*}$.
Let us choose a basis in $L_{\theta}^{*}$ in which the derivations corresponding to basis elements satisfy (\ref{deltaj}). 
Denote the exterior 1-forms corresponding to basis elements by $\alpha^{1}, \dots , \alpha^{d}$. Then an arbitrary 
element of  $\Lambda^{.}(L^{*}_{\theta})$ can be represented as a function (a polynomial) of anticommuting variables $\alpha^{i}$ 
or equivalently it corresponds to a collection of antisymmetric tensors.  
Explicitly we have for the Chern character
$$
{\rm ch}(E) = {\rm Tr} {\bf 1} + \frac{1}{2\pi i} \alpha^{j}\alpha^{k} {\rm Tr} F_{jk} + 
 \frac{1}{2\!}\frac{1}{(2\pi i)^{2}}  \alpha^{j}\alpha^{k}\alpha^{l}\alpha^{m} {\rm Tr} F_{jk}F_{lm} + \dots 
$$
Note that the first term in this expansion equals the dimension of $E$: ${\rm dim} E = {\rm Tr}{\bf 1}$.
The factor of $1/i$ entering (\ref{ncChern}) comes from  our convention that $\nabla_{j}$ are 
antihermitian operators, so that the curvature is also antihermitean.    
One can check that indeed, as the notation suggests, the element ${\rm ch}(E)$  does not depend on the choice of connection. 
The proof is very similar to the proof in the commutative case. It relies on the relation 
${\rm Tr} [\nabla_{i}, \phi] = 0$ (formula (\ref{ttt}) that holds for any connection $\nabla_{i}$ and any endomorphism $\phi$. 
Therefore, we have  a map 
\begin{equation} \label{ncch}
{\rm ch} : K_{0}(T_{\theta}^{d}) \to \Lambda^{even}(L^{*}_{\theta}) 
\end{equation}
that is an analog of (\ref{commch}). Thus defined Chern character is a homomorphism, 
i.e. ${\rm ch} (E_{1}\oplus E_{2}) = {\rm ch}(E_{1}) + {\rm ch}(E_{2})$. 
However there is no analog of 
the multiplicative property that holds in the commutative case: 
${\rm ch}(T_{1}\otimes T_{2}) = {\rm ch} (T_{1}) \cdot {\rm ch} (T_{2})$ where  $T_{1}$, $T_{2}$ are two vector 
bundles. The primary  reason for that is there is no suitable definition of a tensor product of two projective modules.   
The Chern character as defined above was first introduced and studied in \cite{Connes1}. 

A distinctive feature of the noncommutative Chern character  (\ref{ncch}) is that its image does not consist of  integral 
elements, i.e.  there is no lattice in $L_{\theta}^{*}$ that generates the image of Chern character. 
 Equivalently it can be said that the corresponding exterior forms in $\Lambda^{even}(L^{*}_{\theta})$ 
do not have integral coefficients for any choice of basis in $L^{*}_{\theta}$.   
However there is a different   integrality statement that replaces the  commutative one.
Consider a subset $\Lambda^{even} ({\mathbb Z}^{d})\subset \Lambda^{even}(L^{*}_{\theta})$ that consists 
of  polynomials  in $\alpha^{j}$ having integer coefficients.    
It was proved by Elliott (\cite{Elliott}) that  the Chern character is injective and its range on $K_{0}(T_{\theta}^{d})$ 
is given by the image of $\Lambda^{even} ({\mathbb Z}^{d})$  under the action of the  operator 
$ exp\left( 
  -\frac{1}{2}\frac{\partial}{\partial \alpha^{j}} \theta^{jk} \frac{\partial}{\partial \alpha^{k}} \right) $. For $d=2$ 
this follows from \cite{Connes1}.
 This fact implies that the K-group $K_{0}(T_{\theta}^{d})$ can be identified with the additive group  $\Lambda^{even} ({\mathbb Z}^{d})$.
Then  a K-theory class  $\mu(E)\in  \Lambda^{even} ({\mathbb Z}^{d})$ of a module $E$ can be computed 
from its Chern character  by the formula 
\begin{equation} \label{Elliott}
\mu(E) = exp\left( 
  \frac{1}{2}\frac{\partial}{\partial \alpha^{j}} \theta^{jk} \frac{\partial}{\partial \alpha^{k}} \right) 
{\rm ch}(E)  \, . 
\end{equation}
Note that the anticommuting variables $\alpha^{i}$ and the derivatives $\frac{\partial}{\partial \alpha^{j}}$ satisfy the  
anticommutation relation $\{ \alpha^{i}, \frac{\partial}{\partial \alpha^{j}} \} = \delta^{i}_{j}$.
The coefficients of $\mu(E)$ standing at monomials in $\alpha^{i}$ are integer numbers to which we will refer 
to as the topological numbers of module $E$. This numbers also can be interpreted as numbers of D-branes of a definite kind 
although in  noncommutative geometry it is difficult to talk about branes as geometrical objects wrapped on torus cycles.

Let us consider here an  example of application of  formula (\ref{Elliott}). In section \ref{Consec} we found  a connection on a 
module $E_{m, n}$ over a noncommutative two-torus $T_{\theta}^{d}$. Its curvature was calculated to be equal to 
$F_{12} = - F_{21} = -\frac{2\pi i m}{n- m\theta}$ where $\theta = \theta^{12}$. 
 One readily finds that the Chern character is 
$$
{\rm ch}(E_{m, n}) = {\rm dim}(E) -m \frac{{\rm dim}(E)}{n- m\theta} \alpha^{1}\alpha^{2} \, . 
$$  
Applying formula (\ref{Elliott}) to this expression we obtain 
$$
\mu(E) = \frac{{\rm dim}(E)}{n- m\theta} ( n - m\alpha^{1}\alpha^{2} ) \, .
$$ 
We see that for this expression to be integral we need the fraction $ \frac{{\rm dim}(E)}{n- m\theta}$ to be at least 
rational. In fact a direct calculation  via the corresponding projector \cite{RieffelC*} shows that 
${\rm dim}(E)= |n-m\theta|$ so that the above fraction equals $\pm 1$ \cite{Connes1}. 
Thus, the corresponding K-theory class is $\mu (E) = \tilde  n - \tilde m\alpha^{1}\alpha^{2}$. 
where $\tilde n = n\cdot {\rm sgn}(n-m\theta)$, $\tilde m = m\cdot {\rm sgn}(n-m\theta)$. 
 One could alternatively label modules $E_{m,n}$ by pairs of integers $(\tilde n, \tilde m)$ such that 
$\tilde n - \tilde m \theta > 0$ \footnote{ When ${\rm dim}(E) = |n - m\theta| = \tilde n - \tilde m \theta= 0$ 
(that can happen only for rational $\theta$) formulas (\ref{U's}) still define some module, however it is not projective.}.

Formula (\ref{Elliott}) to which we will sometimes refer as Elliott's formula 
is probably the most important formula as far as the applications are concerned. 
As another example of how it works let us discuss a  characterization of  modules  that 
admit a constant curvature connection in terms of their $K$-theory class. 
If a module $E$ can be equipped  with a constant curvature connection 
$[\nabla_{j}, \nabla_{k}] =  2\pi i f_{jk}$ the Chern character (\ref{ncChern}) of module $E$ 
is a quadratic exponent 
$$
{\rm ch}(E) = {\rm dim}(E)\cdot e^{\alpha^{i}f_{ij}\alpha^{j}} \, . 
$$ 
In order to calculate $\mu(E)$ by formula (\ref{Elliott}) it is convenient to apply a Fourier 
transform for the odd Grassmann variables $\alpha^{i}$. This allows us to write $\mu(E)$ as 
$$
\mu(E) = {\rm dim}(E) (-1)^{d(d-1)/2} \cdot \int d\beta  \int d\gamma \, 
exp(\gamma_{i}(\alpha^{i} - \beta^{i}) + \beta_{i}f^{ij}\beta_{j} + \gamma_{i}\theta_{ij}\gamma_{j} ) \, . 
$$
We see that $\mu(E)$ can be expressed as a Gaussian integral in $2d$ anticommuting variables. 
If the corresponding quadratic form is nondegenerate the result of integration is a quadratic exponent. 
Noting that a degenerate quadratic form can be considered as a limit of nondegenerate ones 
we see that $\mu(E)$  can be represented as  a limit of quadratic exponents; we can call such a function a generalized   
 quadratic exponent. Thus we see that if a module admits a constant curvature connection its K-theory class 
has to be a generalized quadratic exponent. One can  prove \cite{AstSchw}, \cite{KS} that the opposite statement 
is also true. In the case when $\theta$ is irrational (see more below) this means that  whenever $\mu(E)$ is a generalized quadratic exponent 
there exists  a constant curvature connection on $E$.

As it was already noted above not every element in $K_{0}(T_{\theta}^{d})$ corresponds to an equivalence class 
of some projective module. The set of elements  of $K_{0}(T_{\theta}^{d})$  that are K-theory 
classes of projective modules is called a positive cone of $K_{0}(T_{\theta}^{d})$. 
A projective module over $T_{\theta}^{d}$ can be specified by a hermitian projector $P=P^{2}$, $P=P^{*}$ that is an element 
of the matrix algebra $Mat_{N}(T_{-\theta})$.  From formulas (\ref{dim}) and (\ref{trace=int}) it follows then 
$$
{\rm dim}(E) = \tau_{N} P = \tau_{N} (P\cdot P^{*}) \ge 0  
$$
(the last inequality  follows from the positivity of trace). 
Thus, the dimension of  module  is always a  positive number. 
Moreover it was proved in 
\cite{RieffelProj} that {\it if $\theta$ is irrational the positive cone consists exactly of elements with positive  dimension.}
More precisely given an element $\mu \in  \Lambda^{even} ({\mathbb Z}^{d})$ one can compute the corresponding Chern 
character  inverting formula (\ref{Elliott}): 
\begin{equation}\label{Elliott^-1}
{\rm ch} =  exp\left( 
  -\frac{1}{2}\frac{\partial}{\partial \alpha^{j}} \theta^{jk} \frac{\partial}{\partial \alpha^{k}} \right) \mu \, .
\end{equation}
Then $\mu$ is a K-theory class of some projective module $E$, i.e. $\mu = \mu(E)$, if and only if the zeroth order term 
${\rm ch}_{0}$ in (\ref{Elliott^-1}) is strictly positive.

There is another question that arises naturally in regard with the K-group. K-theory classes  might not 
distinguish the modules completely, i.e. two nonisomorphic modules can lie in the same K-theory class. 
This is what happens in general for vector bundles  over commutative tori. However {\it for noncommutative tori 
$T_{\theta}^{d}$ with irrational matrix $\theta$ any two projective modules which represent the same element 
of $K_{0}(T_{\theta}^{d})$ are isomorphic} (\cite{RieffelProj}).  In terms of the definition of $K_{0}$ group 
as equivalence classes of differences we can equivalently say that for any three projective modules 
$E_{1}$, $E_{2}$, $F$ the isomorphism $E_{1}\oplus F \cong E_{2}\oplus F$ implies $E_{1}\cong E_{2}$. 
This is the reason the above fact can be called a cancellation property.

It is noteworthy that for irrational $\theta$ there are projective modules whose dimension is arbitrarily small. 
In particular this implies that 
any projective module $E$ over a noncommutative torus with irrational $\theta$ 
can always be split into a direct sum of projective modules. To see how it follows from the results quoted above
 take a module $E'$ such that ${\rm dim}(E') < {\rm dim}(E) $. Consider the element in the $K$-group 
corresponding to the difference $E - E'$. Since  ${\rm dim}(E - E' ) =  {\rm dim}(E) - {\rm dim}(E')$ is positive
there is a module $E''$ specifying the same element in the $K$-group as  $ E - E'$. Therefore by cancellation property 
$E\cong E' \oplus E''$. 
We see that although the $K_{0}(T_{\theta})$ is a finitely generated group there is no finite set of generators for 
the positive cone (modules themselves).  (By taking direct sums of modules from a given finite set of modules we can never obtain 
a module whose dimension is smaller than the smallest dimension of  modules from the given set.)       

%%%%%%%%%%%%%%%%%%%%%%%%%%%             (ccc modules)                     %%%%%%%%%%%%%%%%%%%%%%%%%%%%%%%%

\subsection{ Modules with nondegenerate constant curvature connection} \label{cccsec}
Any Heisenberg module has a constant curvature connection, i.e. a connection $\nabla_{j}$ that 
satisfies 
\begin{equation} \label{fjk}
[\nabla_{j}, \nabla_{k}] = 2\pi i f_{jk} {\bf 1} 
\end{equation}
where $f_{jk}$ is a real-valued antisymmetric tensor and $\bf 1$ is a unit endomorphism. 
In this section we are going to describe a particular class of modules 
admitting a constant curvature connection, those  that have a nondegenerate curvature tensor $f_{jk}$ 
(of course this  is possible only if the dimension $d$ of the torus is even). 
One can construct 
examples of such modules taking commutation relations (\ref{fjk})
as a starting point. If $f_{jk}$ is nondegenerate  then the operators 
$\nabla_{j}$ define a representation of Heisenberg algebra. It is well known that there is a unique irreducible representation 
$\cal F$ of this algebra. Suppose that a representation space  $E$ can be decomposed into a direct sum of a finite number of 
irreducible components: $E\cong {\cal F}^{N}\cong {\cal F}\otimes {\mathbb C}^{N}$.

We fix the representation $\cal F$ as follows. First let us bring the matrix $f_{ij}$ to a canonical block-diagonal form 
\begin{equation} \label{cform}
(f_{ij})= \left( \begin{array}{cccc}
f_{1} {\bf \epsilon} & 0 & \dots & 0 \\
0 & f_{2} {\bf \epsilon} & \dots & 0 \\
0 & 0 & \ddots & 0 \\
0 & 0 & \dots  & f_{g}  {\bf \epsilon} 
\end{array} \right) 
\end{equation}
where 
\begin{equation} \label{epsilon}
{\bf \epsilon} = \left(
\begin{array}{cc} 0 & 1 \\ 
-1 & 0 \end{array} \right) 
\end{equation}
is a $2\times 2$ matrix 
and $f_{i}$ are positive numbers. 

Then we can define a representation space as $L_{2}({\mathbb R}^{g})$ 
and the operators $\nabla_{i}$  as 
\begin{equation} \label{nabla} 
\nabla_{j} = \sqrt{f_{(j+ 1)/2}}\partial_{j} \, , j - \mbox{odd} \, , \quad \nabla_{j} = 2\pi i\sqrt{f_{j/2}}x_{j-1}\, , j - \mbox{even} 
\end{equation} 
where $\partial_{j}$, $x_{k}$, $j,k=1,\dots , g$  are derivative and multiplication by $x^{k}$ operators acting 
on smooth functions $f(x)\in L_{2}({\mathbb R}^{g})$.
An arbitrary representation of the torus generators $U_{i}$, $i=1, \dots, d$ has the form 
$
U_{i} = U_{i}^{st}\cdot u_{i} 
$  
where $U_{i}^{st}$ is some standard representation satisfying 
$$
[\nabla_{j}, U_{k}^{st}]= 2\pi i \delta_{jk} U_{k}^{st}
$$ 
and $u_{i}$ is an $N\times N$  unitary matrix. This form of representation of $U_{i}$ follows from the irreducibility 
of representation $\cal F$. 
A straightforward calculation shows that one can take $U_{i}^{st}$ to be   
\begin{equation}\label{Ust}
U_{k}^{st} = e^{-(f^{-1})^{kl}\nabla_{l}} \, .
\end{equation}
These operators satisfy
\begin{equation} \label{stcom}
U_{j}^{st}U_{k}^{st} = e^{-2\pi i (f^{-1})^{jk}} U_{k}^{st}U_{j}^{st} \, .
\end{equation}
Since $U_{i}= U_{i}^{st}\cdot u_{i}$ must give a representation of a noncommutative torus it follows from (\ref{stcom}) that 
so must do  the operators $u_{i}$. But the last ones are finite-dimensional matrices so they can only represent a 
noncommutative torus whose noncommutativity matrix has rational entries, i.e. $u_{i}$'s have to satisfy  
\begin{equation} \label{u}
u_{i}u_{j} = e^{2\pi i n^{ij}/N} u_{j}u_{i}
\end{equation}
where $N$ is a positive  integer and $n^{ij}$ is an integer valued antisymmetric matrix. 
Putting the formulas (\ref{stcom}) and (\ref{u}) together one finds that $U_{i}$'s give a representation of a noncommutative torus 
$T_{\theta}$ with 
\begin{equation} \label{theta}
\theta^{ij} = -(f^{-1})^{ij} + n^{ij}/N \, . 
\end{equation}

 It follows from the results obtained by M.~Rieffel (\cite{RieffelProj}) that 
for finite $N$ (i.e. when $E$ decomposes into a finite number of irreducible components) 
 the module $E$ endowed with $U_{i}=U^{st}_{i}\cdot u_{i}$ as above is a finitely generated projective module over $T_{\theta}$ with 
$\theta$ given in (\ref{theta}).  Conversely one can show that the finiteness of $N$ is required by the condition of 
$E$ to be finitely generated and projective. (See \cite{AstSchw} for a detailed discussion of modules admitting a 
constant curvature connection.)

\noindent {\bf Topological numbers}. Let us calculate here the topological numbers of the modules constructed above.
We assume here that the matrix $\theta_{ij}$ given in  (\ref{theta}) has irrational entries. Then a  
projective module $E$ is uniquely characterized by an integral element $\mu(E)$ of the even part of  Grassmann algebra 
$\Lambda^{even}({\mathbb R}^{d})$.  
In order to calculate $\mu(E)$ we can use  the Elliot's formula (\ref{Elliott}) together with the following expression for  the Chern character 
\begin{equation}\label{ch}
ch(E) = {\rm dim}(E) \cdot  exp( \alpha^{i} f_{ij} \alpha^{j}  ) \, .
\end{equation}
 Substituting (\ref{ch}) into (\ref{Elliott}) and applying a Fourier transform in Grassmann variables we obtain
\begin{eqnarray} \label{mu}
\mu(E) &=& {\rm dim}(E)\cdot {\rm Pfaff}(f) \cdot \int d\beta \, 
exp( \frac{1}{2}\beta_{i} ((f^{-1})^{ij} + \theta^{ij})\beta_{j} + \alpha^{i}\beta_{i}) = \nonumber \\
&=&  {\rm dim}(E) \cdot {\rm Pfaff}(f)  \cdot \int d\beta \, exp(\frac{1}{2}\beta_{i} n^{ij} \beta_{j} /N  + \alpha^{i}\beta_{i}) \, .
\end{eqnarray}
At this point it is convenient to assume that the matrix $n_{ij}$ is brought 
to a canonical block-diagonal 
form similar to (\ref{cform}) with integers $n_{i}$, $i=1,\dots, g$ on the diagonal  by means of an $SL(d, {\mathbb Z})$ transformation 
(this is always possible, see \cite{Igusa}). 
Then we can explicitly do the integration in (\ref{mu}) and obtain 
\begin{equation}\label{mu2}
\mu(E) = C\cdot \left(\frac{n_{1}}{N} + \alpha^{1}\alpha^{2}\right) \left(\frac{n_{2}}{N} + \alpha^{3}\alpha^{4}\right)\cdot \dots 
\cdot \left(\frac{n_{g}}{N} + \alpha^{2g-1}\alpha^{2g}\right)
\end{equation}
where $C={\rm dim}(E) \cdot {\rm Pfaff}(f) $ is a constant that can be determined by the requirement that $\mu(E)$ is an integral element 
of Grassmann algebra $\Lambda({\mathbb R}^{d})$  (i.e. each coefficient  is an integer). By looking at the term of the highest order in 
  $\alpha$  in (\ref{mu2}) we immediately realize that $C$ must be an integer.   In fact     $C=N$. We will give a partial proof 
of this below.

Let us introduce  numbers
$$
N_{i} = \frac{N}{g.c.d. (N, n_{i})} \, , \qquad \tilde N_{i} = \frac{n_{i}}{g.c.d. (N, n_{i})}
$$
so that for each $i=1, \dots , g$ the pair   $N_{i}$, $\tilde N_{i}$ is relatively prime.
Then we can rewrite (\ref{mu2}) as 
\begin{equation} \label{mu3} 
\mu(E) = \frac{C}{N_{1}N_{2}\cdot \dots \cdot N_{g}} \prod_{i=1}^{g} (\tilde N_{i} + N_{i}\alpha^{2i-1}\alpha^{2i}) \equiv 
\frac{C}{N_{1}N_{2}\cdot  \dots \cdot N_{g}}\mu_{0}(E) \, .
\end{equation}
For any integral element $\nu \in \Lambda^{even}({\mathbb R}^{d})$  let us introduce a number $g.c.d. (\nu)$ which is 
defined to be  the largest integer $k$ such that  $\nu = k\nu'$ where 
$\nu'$ is also integral. Or, in other words  $g.c.d. (\nu)$ is simply the greatest common divisor of the coefficients of $\nu$. 
It is a simple task to prove by induction in $g$ that  $g.c.d. (\mu_{0}(E))=1$. 
Hence, $C$ must be an integer divisible by the product $N_{1}N_{2}\cdot \dots \cdot N_{g}$. Moreover 
 $C = g.c.d. (\mu(E) )N_{1}N_{2}\cdot \dots \cdot N_{g}$.  
It is known (for example see \cite{vanBaal})  that the dimension of  an irreducible representation of the algebra (\ref{u}) is equal to 
the product $N_{1}\cdot N_{2} \cdot \dots \cdot N_{g}$. 
Thus,  necessarily this product divides $N$, i.e. $N =  N_{1}\cdot N_{2} \cdot \dots \cdot N_{g}\cdot N_{0}$ 
where $N_{0}$ is an integer equal to the  number of  irreducible components in the  
representation ${\mathbb C}^{N}$ of the algebra (\ref{u}). 
Evidently $N_{0}$ divides $g.c.d.(\mu(E))$. We will  show below that   $g.c.d.(\mu(E))$ cannot be bigger than $N_{0}$.
This implies that $C=N$.

Let us look at some particular examples of formula (\ref{mu2}). 
 If the matrix $n^{ij}$ is nondegenerate then $\mu(E)$ is a quadratic exponent:
\begin{equation}
\mu(E) = p\cdot exp( \frac{1}{2}\alpha^{i}(n^{-1})_{ij} \alpha^{j} N) \, , \quad p=N_{0}\tilde N_{1}\cdot \dots \cdot \tilde N_{g} 
\end{equation} 
where $p=N\cdot {\rm Pfaff}(n/N)$ is written in a form where it is manifestly an integer.  
If $n^{ij}$ is degenerate then $\mu(E)$ is a so called generalized quadratic exponent (see \cite{AstSchw} and \cite{KS}, Appendix D). 
For example if $n^{ij}=0$ for all $i$ and $j$ then we obtain from (\ref{mu}) 
\begin{equation}
\mu(E) = N \alpha^{1}\alpha^{2}\cdot \dots \cdot \alpha^{d} \, .
\end{equation}

\noindent {\bf Moduli space of constant curvature connections}.
We showed above that modules endowed with a constant curvature connection correspond to  representations of 
matrix algebra (\ref{u}). The residual gauge transformations preserving (\ref{nabla}) correspond to $N\times N$ 
unitary transformations acting on the ${\mathbb C}^{N}$ factor  of $E$. Thus, we see that the moduli space of 
 constant curvature connections on a module with fixed $(N, n^{ij})$ (or fixed $\mu(E)$, 
which is the same)  can be described as a space of inequivalent representations of the matrix algebra (\ref{u}).
 The center of algebra  (\ref{u}) is spanned by elements $u_{\bf k}$ with ${\bf k}\in D\cong {\mathbb Z}^{d}$ satisfying 
$k_{i}n^{ij}m_{j}/N \in {\mathbb Z}$ for any ${\bf m}\in D$. Such elements correspond to a sublattice of $D$ that we denote $\tilde D$. 
To describe the center in a more explicit way it is convenient to choose the basis we used above, in which the matrix  $n^{ij}$ 
is brought to a block-diagonal canonical form. 
In this basis  generators of the center can be chosen to be elements  $(u_{i})^{M_{i}}$ where we set 
 $M_{i} = N_{(i+1)/2}$, $i$-odd and  $M_{i} = N_{i/2}$, $i$-even.
Thus, in an irreducible 
representation $(u_{i})^{M_{i}}=\lambda_{i} \in {\mathbb C}^{\times}$ are constants of absolute value 1. 

Using the substitution
\begin{equation}\label{subst}
u_{i} \mapsto c_{i} u_{i}
\end{equation}
where $c_{i}$ are constants, $|c_{i}|=1$,  we obtain an irreducible representation  
with values of center generators  $\lambda_{i}'=\lambda_{i}c_{i}^{M_{i}}$. 
 By means of this substitution one can transform any irrep into the one with $\lambda_{i} =1$. 
The last one corresponds to a representation of the 
algebra specified by relations (\ref{u}) along with the relations $(u_{i})^{M_{i}} =1$.
 This algebra has a
 unique irreducible representation of dimension  $N_{1}\cdot N_{2} \cdot \dots \cdot N_{g}$ (for example see \cite{vanBaal}). 
Therefore, the space of irreducible representations of algebra (\ref{u}) is described by means of $d$
complex numbers $\lambda_{i}$ with absolute value 1, i.e. is isomorphic to a (commutative) torus $\tilde T^{d}\cong {\mathbb R}^{*d}/\tilde D^{*}$. 
We denote the corresponding 
irreps by $E_{\Lambda}$, $\Lambda=(\lambda_{1}, \dots , \lambda_{d})$.
In general for any noncommutative torus $T_{\theta}$ one can construct a group $L_{\theta}$ of automorphisms isomorphic to 
a commutative torus of the same dimension by means of (\ref{subst}). This torus acts naturally on the space of 
unitary representations of $T_{\theta}$. If $\theta$ is rational we obtain a transitive action of this automorphism 
group on the space of irreducible representations. In this case one can consider $L_{\theta}$ as a finite covering 
of $\tilde T^{d}$.

Let us assume now that the space   ${\mathbb C}^{N}$ is decomposed into irreducible representations of algebra (\ref{u})
\begin{equation} \label{dec}
{\mathbb C}^{N} = E_{\Lambda_{1}}\oplus \dots \oplus E_{\Lambda_{N_{0}}}\, .
\end{equation}
Note that in the picture we are working with,  gauge transformations are given by unitary linear operators acting on $E$ that 
commute with all $\nabla_{i}$'s that is by unitary $N\times N$ matrices.  
 The matrices representing central elements are diagonalized in the basis specified by  decomposition (\ref{dec}). 
There are residual gauge transformations corresponding to permutations of diagonal entries. Thus, we see that in  general 
 the moduli space is isomorphic to $(\tilde T^{d})^{N_{0}}/S_{N_{0}}$. 
As it was noted   in the previous subsection $N_{0}$ divides $g.c.d. (\mu(E))$.
On the other hand as we know from \cite{AstSchw}, \cite{KS} any module $E$ over a noncommutative torus $T_{\theta}$ 
admitting a constant curvature connection $\nabla_{i}$ 
can be represented as a direct sum of $k$ identical modules  $ E=E'\oplus \dots \oplus E' $ with 
$k=g.c.d.(\mu(E))$. This implies that the moduli space of constant curvature connections necessarily contains 
a subset isomorphic to $(\tilde T^{d})^{k}/S_{k}$. Thus, on dimensional grounds we conclude that $k=g.c.d.(\mu(E))=N_{0}$. 

The considerations  above were made in the assumption that we have a constant curvature connection whose 
curvature tensor is nondegenerate. By using the technique of Morita equivalence one can show that a generic case 
can be always   reduced to the nondegenerate one \cite{moduli}. 
Therefore, {\it the moduli space of constant curvature connections on a module $E$ is isomorphic to 
$(\tilde T^{d})^{g.c.d.(\mu(E) )}/S_{N_{0}}$.} For two-dimensional tori this result was first proved in 
\cite{ConnesRieffel}.

%%%%%%%%%%%%%%%%%%%%%%%%%%%%%%%%%%%%%%%%%%%%%%%%%%%%%%%%%%%%%%%%%%%%%%%%%%%%%
\subsection{Heisenberg modules  as deformations of vector bundles} \label{Deformsec}
Taking a closer look at the example of a Heisenberg module over $T_{\theta}^{2}$ 
(\ref{U's}) we notice that 
it actually gives us a one-parameter family of modules depending on $\theta$. 
Moreover this dependence is continuous. An analogous fact is   true for a general 
Heisenberg module. A Heisenberg module is constructed via a lattice 
$\Gamma \in G\times G^{*}$ where $G$ is an abelian finitely generated group. 
This group in general is isomorphic to ${\mathbb R}^{p}\times {\mathbb Z}^{q}\times F$ 
where $F$ is a finite group. Thus, $G$ has a continuous factor ${\mathbb R}^{p}$ and 
a discrete factor ${\mathbb Z}^{q}\times F$. The dual group has the form 
$G^{*}={\mathbb R}^{p}\times T^{q}\times F^{*}$ and its continuous and discrete parts are 
${\mathbb R}^{p}\times T^{q}$ and $F^{*}$ respectively. We can vary the lattice $\Gamma$ by 
varying the continuous components of vectors only. The idea is that deforming $\Gamma$   
this way  will change $\theta^{jk}$ without changing the topological numbers $\mu(E)$.  
We might be able to deform  continuously 
 a projective module $E$ into a projective module over a commutative torus with $\theta_{ij}=0$ 
or more generally $\theta^{ij}$ can have integer entries.  But the last one has a description in terms 
of sections of a vector bundle over a (commutative) torus $T^{d}$.  
    One has to make sure that $\Gamma$ remains  a $d$-dimensional lattice through 
the deformation. For example if we consider a module over $T_{\theta}^{2}$ specified by  (\ref{U's}) 
with $n=0$  it does not make sense to go to the $\theta\to 0$ limit because we do not 
end up with a Heisenberg module ($\Gamma$ does not stay a 2-dimensional lattice in the sense that the factor 
$G\times G^{*}/{\Gamma}$ does not stay isomorphic to $T^{2}$). The result of such deformation is still a module 
over functions on a commutative torus but it is not projective and thus does not correspond to any vector bundle.
However in this case one can take instead a limit $\theta \to n$ for some nonzero integer $n$.
It turns out that choosing an appropriate end point specified by $\theta^{jk}$ with integral coefficients one can always 
deform a Heisenberg module into a Heisenberg module over a commutative torus.
We conclude from this discussion  that it should be possible to go into the opposite direction and 
 describe Heisenberg modules as deformations of spaces of sections of vector bundles over commutative tori. 
The appropriate vector bundles can be described via twisted boundary conditions.

Let us illustrate this ideas on two-dimensional tori. Consider a nontrivial (twisted) $U(n)$ gauge 
bundle  over $T^{2}$.
 A section  
of this bundle  bundle  can be represented by a vector function $\phi_{j}(\sigma_{1}, \sigma_{2})$ 
on $T^{2}$ (or more precisely on the universal covering of $T^{2}$  satisfying twisted boundary conditions:
\begin{eqnarray} \label{twist}
&& \phi_{j}(\sigma_{1} + 2\pi, \sigma_{2}) = (\Omega_{1})_{j}^{k}(\sigma_{2}) \phi(\sigma_{1}, \sigma_{2}) \, , 
\nonumber \\
&& \phi_{j}(\sigma_{1} , \sigma_{2}+ 2\pi ) = (\Omega_{2})_{j}^{k}(\sigma_{1}) \phi(\sigma_{1}, \sigma_{2}) 
\end{eqnarray}
where $(\Omega_{1})_{j}^{k}(\sigma_{2})$ and  $(\Omega_{2})_{j}^{k}(\sigma_{1})$ are $U(n)$-matrix valued 
functions specifying the twists. These matrices must satisfy the consistency condition 
$$
\Omega_{1}(\sigma_{2} + 2\pi) \Omega_{2}(\sigma ) = \Omega_{2}(\sigma_{1} + 2\pi)\Omega_{1}(\sigma_{2} )
$$ 
called the cocycle condition. (Here for brevity we omitted the matrix indices.) 
This condition   can be solved as 
$$
\Omega_{1}(\sigma_{2}) = e^{im\sigma_{2}/n}U \, , \quad \Omega_{2}(\sigma_{1}) = V
$$
where $U$ and $V$ are $n\times n$ clock and shift matrices (\ref{clockandshift}) and $m$ is some integer 
relatively prime with $n$.

One can find then a general solution to the  twisted boundary conditions (\ref{twist}) (\cite{GRTaylor}). 
A section $\phi_{j}(\sigma_{1}, \sigma_{2})$ is specified by $m$ functions $\hat \phi_{j}(x)$ on ${\mathbb R}^{1}$ 
(smooth and fast decreasing at infinity) by means of the following formula 
$$
\phi_{j}(\sigma_{1}, \sigma_{2}) = \sum_{s\in {\mathbb Z}}\sum_{k=1}^{m} 
exp\Bigl[ i \Bigl( \frac{m}{n}(\sigma_{2}/2\pi + j + ns) + k \Bigr)\sigma_{1}\Bigr] \hat 
\phi(\sigma_{2}/2\pi + j + ns + \frac{nk}{m}) \, . 
$$ 
One can check now that the multiplication of a section  $\phi_{j}(\sigma_{1}, \sigma_{2})$ by  $e^{i\sigma_{1}}$,  $e^{i\sigma_{2}}$ 
results in the action on the functions $\hat \phi_{j}(x)$ on ${\mathbb R}^{1}$ that is given by operators $U_{1}$, $U_{2}$ defined in 
(\ref{U's}) for $\theta = 0$. The action of operators $U_{i}$ as we know can be deformed to give  (\ref{U's}) with an arbitrary $\theta$. 
We see thus that, indeed, a Heisenberg module over $T_{\theta}^{d}$ can be described as a deformation of the space of sections 
of a  vector bundle over a commutative torus. In this review we will primarily work with the Hilbert space picture. 
For the details of  deformation construction we refer the reader to papers \cite{Ho}, \cite{MorZum}, \cite{BrMorZum}, \cite{HofVerII}.

%%%%%%%%%%%%%%%%%%%%%%%%%%%%%%%%%%%%%%%%%%%%%%%%%%%%%%%%%%%%%%%%%%%%%%%%%
\section{Noncommutative Yang-Mills and super Yang-Mills theories} 
\subsection{YM and SYM on free modules} \label{YMsec}
Let us take a break now from developing the math formalism and describe an object of 
a more immediate interest in physics - Yang-Mills theory on noncommutative tori.

Let $E$ be a projective module over an associative noncommutative algebra $A$. And 
let  $L$ be a Lie algebra acting on $A$ with respect to which we define connections. 
Then  one can introduce a noncommutative Yang-Mills action functional 
more or less in the same way as usual. The action functional is defined on the space of connections 
on $E$  and can be written in the form 
\begin{equation} \label{S_YM}
S_{YM}= \frac{V}{4g^{2}}{\rm Tr}F^{jk}F_{jk}
\end{equation}
where $F_{jk}$ are components of a curvature tensor in some fixed basis in $L$. 
Here we assume that   the Lie algebra $L$ is equipped with some metric $g_{ij}$ and  
we raise and lower indices by means of this metric, $V=\sqrt{\pm det(g_{ij})}$ is the corresponding 
volume and $g$ is a coupling constant. Note that when $g_{ij}$ has the Minkowski signature 
the overall sign in (\ref{S_YM}) is plus in contrast with the conventional minus in the 
commutative gauge theory. This is 
due to our conventions discussed in section \ref{Invsec} in which the curvature $F_{ij}$ is an 
antihermitian operator.

If we consider connections on a free module over ${\mathbb R}^{d}_{\theta}$ or $T_{\theta}^{d}$ 
this formula can be written in terms of a star product (\ref{qplane}) or (\ref{torus_star}) respectively.
If $\sigma_{j}$ are coordinates on the underlying commutative space ${\mathbb R}^{d}$ or $T^{d}$ (in the last case 
we assume that $0\le \sigma_{j} \le 2\pi$)
a  connection on a free module of rank $N$ can be written as 
 $\nabla_{j} =   2\pi  \frac{\partial}{\partial \sigma_{j}} + A_{j}^{ab}(\sigma_{1}, \dots , \sigma_{d})$
where  $A_{j}^{ab}(\sigma_{1}, \dots , \sigma_{d})$, $a,b=1, \dots , N$ are  matrix valued functions 
specifying an endomorphism. The corresponding curvature tensor reads 
\begin{eqnarray} \label{Fjk}
&& F_{jk} = F_{jk}^{ab}(\sigma_{1}, \dots , \sigma_{d}) = [\nabla_{j}, \nabla_{k}]^{ab}(\sigma ) = \nonumber \\
&&  2\pi  \partial_{j} A_{k}^{ab}(\sigma ) - 
 2\pi  \partial_{k}A^{ab}_{j}(\sigma ) + \sum_{c}(A^{ac}_{j}\ast A^{cb}_{k} - A^{ac}_{k}\ast A^{cb}_{j} )(\sigma) \, .      
\end{eqnarray}
It remains to plug in this expression into the general formula (\ref{S_YM}) bearing in mind that one should take 
for the trace $\rm Tr$  a composition of the matrix trace and the canonical trace on ${\mathbb R}_{\theta}^{d}$ or 
$T_{\theta}^{d}$ respectively that were defined above.

A gauge transformation in the noncommutative Yang-Mills theory is specified by a unitary endomorphism $Z\in End_{A}E$. 
Let us remind that if $A$ is equipped with an involution and $E$ is projective there is an induced involution on 
the algebra  $Z\in End_{A}E$. For example in the case of a free module when an endomorphism  is specified by a matrix 
with entries in $A$ the corresponding involution is just a composition of the matrix transposition and an involution in $A$. 
 The gauge transformation on connection reads  
$$
\nabla_{j} \mapsto Z\nabla_{j} Z^{*} \, . 
$$
In the example of Yang-Mills theory on a free module over ${\mathbb R}^{d}_{\theta}$ ($T_{\theta}^{d}$) 
considered above a unitary endomorphism is specified by means of a matrix-valued function 
$Z^{ab}(\sigma_{1}, \dots , \sigma_{d})$ satisfying 
$$
Z\cdot Z^{*} \equiv  \sum_{c} Z^{ac}(\sigma) \ast \bar Z^{bc} (\sigma ) = \delta^{ab} \cdot 1 \equiv {\bf 1} \, .
$$
The corresponding gauge transformation acts on the gauge field $A_{j}^{ab}$ as  
$$
A_{j}^{ab} \mapsto Z^{ac}(\sigma) \ast A^{cd}_{j}(\sigma)  \ast (Z^{*})^{ db} (\sigma) +  2\pi  Z^{ac}(\sigma) \ast 
\partial_{j} (Z^{*})^{cb}(\sigma) 
$$
where $Z^{* cd} (\sigma ) = \bar Z^{dc}(\sigma )$ (a bar stands for the usual conjugation of complex numbers),  
and   summation over repeated matrix indices is assumed.

 Consider now a compactification of M(atrix) theory on a $d$-dimensional noncommutative torus $T_{\theta}^{d}$.
 Let us look at the solutions to the system of equations (\ref{Td}) corresponding to connections and 
endomorphisms of a free module. The fields  $X_{i}$, $i=0,\dots , d-1$ up to a factor of $i$ are  covariant derivatives: 
 $X_{j} = i\nabla_{j}$ and are specified 
as above by means of a gauge field $A_{j}^{ab}(\sigma)$.  The scalar and spinor fields  $X_{I}$, $I=d, \dots , 9$, 
 $\psi^{\alpha}$ are 
endomorphisms specified  by  matrix valued functions of the corresponding Grassmann parity 
on commutative $T^{d}$: $X_{I}^{ab}(\sigma_{1}, \dots , \sigma_{d})$, 
$\psi^{\alpha ab}(\sigma_{1}, \dots , \sigma_{d})$  where $a, b = 1, \dots N$. 
A commutator of $X_{I}$ and a covariant derivative is also an endomorphism and is given by a function 
\begin{equation} \label{com}
[\nabla_{j}, X_{I}]^{ab}(\sigma ) = 2\pi  \partial_{j}X_{I}^{ab}(\sigma ) + \sum_{c}(A_{j}^{ac}\ast X^{cb}_{I} - 
X_{I}^{ac}\ast A_{j}^{bc} )(\sigma) \, .
\end{equation} 
The same formula also works for the covariant derivative of $\psi^{\alpha}$. 
A commutator of two scalar fields $X_{I}$  is 
\begin{equation} \label{comend}
[X_{I}, X_{J}]^{ab}(\sigma ) = \sum_{c} ( X_{I}^{ac}\ast X_{J}^{cb} - X_{J}^{ac}\ast X_{I}^{cb})(\sigma ) \, .
\end{equation}

One can derive an action of the compactified theory from 
the M(atrix) theory action (\ref{BFSS}) or (\ref{IKKT}) following 
a  procedure similar to the one described in section \ref{Comp1sec}. 
If we start from a solution corresponding to a free module of rank $N$  
 we  get an action functional defined on the set of $N\times N$ matrix valued 
fields $A_{j}^{ab}(\sigma )$, $X_{I}^{ab}(\sigma )$, $\psi^{\alpha ab}(\sigma )$ introduced above, and  reads as 
\begin{eqnarray} \label{ncYM}
&& S = \frac{V}{4g^{2}}\int d^{d}\sigma \sum_{ab} \Bigl( F^{ab}_{jk}\ast F^{jkba} + 
[\nabla_{i}, X_{J}]^{ab}\ast [\nabla^{i}, X^{J}]^{ba} + \nonumber \\
&&  [X_{I}, X_{J}]^{ac}\ast [X^{I}, X^{J}]^{cb} - 
2  \psi^{\alpha ab} \sigma^{j}_{\alpha \beta} \ast [\nabla_{j}, \psi^{\beta}]^{ba} - 
2  \psi^{\alpha ab} \sigma^{J}_{\alpha \beta} \ast [X_{J}, \psi^{\beta}]^{ba} \Bigr)
\end{eqnarray}
where $F^{ab}_{jk}$, $[\nabla_{j}, X_{I}]$ and $[\nabla_{j}, \psi^{\alpha} ]^{ab}$, $[X_{i}, X_{J}]^{ab}$ 
are defined as in (\ref{Fjk}), (\ref{com}), (\ref{comend});  $g$ is a coupling constant, the indices are being raised and 
lowered by means of some fixed metric on $L_{\theta}\cong {\mathbb R}^{d}$.  
Note that it follows from the definition of Moyal product that 
$\int d^{d}\sigma (f\ast g)(\sigma) = \int d^{d}\sigma (f\cdot g)(\sigma)$ so one can omit stars from (\ref{ncYM}); 
however  Moyal products are still  present in   (\ref{Fjk}), (\ref{com}). 
The action functional (\ref{ncYM}) is a functional defining a maximally supersymmetric Yang-Mills 
theory on a noncommutative torus $T_{\theta}^{d}$ written explicitly for a free module. 
We wrote down this functional in such detail  mainly for illustrative purposes.  

\subsection{SYM on arbitrary projective modules} \label{SYMsec}
Writing out the action functional (\ref{ncYM}) on a free module we could use the star product combined with the 
matrix multiplication and 
integration of ordinary functions on a commutative torus. This was possible due to a simple structure of 
the algebra $End_{T_{\theta}^{d}}E$  which our fields are elements of. As it was noted in section \ref{Projsec}
when we consider a Heisenberg module $E$ the algebra  $End_{T_{\theta}^{d}}E$ is isomorphic to a matrix algebra 
over some other (dual) noncommutative torus: $Mat_{N}(T_{\hat \theta})$. In that case one could also represent 
the fields $A_{i}$, $\psi^{\alpha}$, $\phi_{I}$ in terms of functions of the corresponding Grassmann parity 
on a commutative torus. Then one can write a noncommutative SYM action functional on this module 
using the Moyal product corresponding to $\hat \theta$ and integration over the commutative torus that corresponds to 
taking the trace. 
However for  an arbitrary projective module over $T_{\theta}^{d}$ there is no such explicit description of 
the algebra  $End_{T_{\theta}^{d}}E$ and we should proceed in a more abstract way.

Formally in order to  define a super Yang-Mills 
action functional on an arbitrary  projective module over  $T_{\theta}^{d}$ 
 one only needs a suitable trace operation. As we already discussed, the last one always 
exists for   any projective module.  Below ${\rm Tr}$ stands for the canonical  
trace on the algebra $End_{T_{\theta}^{d}}E$. With this notation in mind we can write an action 
functional of noncommutative supersymmetric Yang-Mills theory on   $T_{\theta}^{d}$  for any projective 
module $E$ as 
\begin{eqnarray} \label{ncsYM}
S &=& \frac{V}{4g^{2}} {\rm Tr}  \Bigl( F_{jk}F^{jk} + [\nabla_{i}, X_{J}][\nabla^{i}, X^{J}] + \nonumber \\
&& [X_{I}, X_{J}][X^{I}, X^{J}] - 2 \psi^{\alpha} \sigma_{\alpha \beta}^{j}[\nabla_{j}, \psi^{\beta}] -
2 \psi^{\alpha}\sigma^{J}_{\alpha \beta} [X_{J}, \psi^{\beta}] \Bigr)
 \end{eqnarray}
where products and commutators are those of operators acting in $E$. 

On an arbitrary projective module one can also define an analog of the  rank of the gauge group in 
the commutative case. Namely, {\it we say that $N$ is the rank of the  gauge group  if the module $E$ can be represented 
as a direct sum of $N$ isomorphic modules $E=E' \oplus \dots \oplus E'$}. More precisely we should take 
the largest number $N$ with this property. It is easy to see then that the algebra of endomorphisms
$End_{T_{\theta}^{d}}E$ is isomorphic to the matrix algebra $Mat_{N}(End_{T_{\theta}^{d}}E')$. 
This means that the  group of gauge transformations is a subalgebra in this matrix algebra 
that consists of unitary endomorphisms.

The are two kinds of supersymmetry transformations of action (\ref{ncsYM}) denoted  $\delta_{\epsilon}$,  $\tilde \delta_{\epsilon}$  
and defined as   
\begin{eqnarray} \label{ncSUSY}
&& \delta_{\epsilon} \psi = \frac{1}{2}(\sigma^{jk} F_{jk} \epsilon + \sigma^{jI}[\nabla_{j}, X_{I}]\epsilon + 
\sigma^{IJ}[X_{I}, X_{J}] \epsilon )  \, , \nonumber \\
&& \delta_{\epsilon}  \nabla_{j} =  \epsilon \sigma_{j} \psi \, , \quad   
 \delta_{\epsilon} X_{J} =   \epsilon \sigma_{J} \psi \, , \nonumber \\
&& \tilde \delta_{\epsilon} \psi = \epsilon \, , \quad \tilde \delta_{\epsilon}  \nabla_{j} = 0 \, , \quad 
 \tilde \delta_{\epsilon} X_{J} = 0 \, .
\end{eqnarray}
where $\epsilon$  is a  constant $16$-component Majorana-Weyl spinor. 
To check that the action (\ref{ncsYM}) is invariant under these transformations one needs to use 
the Fierz identities for 10-dimensional Gamma matrices and the identities 
$$
{\rm Tr} [A, B] = 0 \, , \qquad {\rm Tr}[\nabla_{i}, A] = 0 
$$
that hold for any endomorphisms $A$, $B$. The first of these identities is obvious while the last one 
can be easily proved by noting that it suffices to prove it for a any single connection  $\nabla_{i}^{0}$ 
and then explicitly checking that it holds for the Levi-Civita connection defined in section \ref{Endsec}.

Solutions to the equations of motions of (\ref{ncYM}) invariant under half of the supersymmetries are 
called 1/2 BPS solutions. One finds from (\ref{ncSUSY})  that such solutions satisfy 
\begin{equation} \label{ccc}
[\nabla_{j}, \nabla_{k}] = 2\pi i f_{jk}{\bf 1} \, , 
\end{equation} 
\begin{equation} 
\psi^{\alpha} = 0 \, , \qquad   [\nabla_{j}, X_{J}] = 0 \, , \qquad [X_{J}, X_{K}] = 0
\end{equation}
where $\bf 1$ is the identity operator and $f_{jk}$ is a constant antisymmetric matrix. 
Thus we see that, as far as gauge fields are concerned,  1/2 BPS configurations  
 correspond to constant curvature connections (\ref{ccc}).  

%%%%%%%%%%%%%%%%%%%%%%%%%%%%%%%%%%%%%%%%%%%%%%%%%%%%%%%%%%%%%%%%%%%%%%%%%%
\subsection{BPS states on $T_{\theta}^{2}$} \label{BPSsec}
 In this section we will explicitly calculate the energies of BPS states of 
the   SYM theory on a noncommutative spatial two-torus.  We will use   a semiclassical 
approximation the exactness of which for BPS states is ensured by  the  supersymmetry. 
 Instead of working with the 
full supersymmetric action (\ref{ncsYM}) we will consider  the $1+2$-dimensional  YM  action 
functional and constant curvature connections which, by abuse of 
terminology, we will call  BPS fields (they satisfy BPS condition 
in the supersymmetric theory). It is a valid thing to do because the 
calculation  leads to the same result;  
 the sole role of supersymmetry is to ensure the exactness of the 
semiclassical approximation.  We will consider a noncommutative YM theory on 
${\mathbb R}^{1}\times T_{\vartheta}$ where ${\mathbb R}^{1}$ is the time direction that is 
assumed to be commutative and $T_{\vartheta}\equiv T_{\theta}^{2}$ is a noncommutative two-torus with 
$$
\theta = \left( \begin{array}{cc} 
0&\vartheta\\
-\vartheta & 0\\
\end{array} \right) \, .
$$ 

We start with the following  action functional 
\begin{equation}\label{action} 
S=\frac{V}{4g^{2}}{\rm Tr}\left( 
\sum_{\rho \sigma} 
(F_{\rho \sigma}+ \phi_{\rho\sigma}\cdot {\bf 1})
g^{\rho \mu} g^{\sigma \nu} (F_{\mu \nu}+ \phi_{\mu\nu}\cdot {\bf 1}) 
\right) \, . 
\end{equation}
Here we explicitly wrote the metric tensor $g_{\mu \nu}$; $g$ is the YM coupling constant, $\phi_{\rho\sigma}$ 
plays the role of a background field,  
$V=R_{1} R_{2}$ is the volume of the  torus,   
the indices $\rho$, $\sigma$, $\mu$, $\nu$  take values $0, 1, 2$.  
Note that adding the field $\phi_{\rho\sigma}$ essentially means adding a term 
$\phi_{\mu \nu} {\rm Tr} F^{\mu \nu}$ which being proportional to Chern numbers is topological.
 This means that  it does not depend 
on the choice of connection (it  depends only on topological numbers of the module at hand and 
on $\theta$). We added such a term to the action because it will be important in the discussion of Morita 
equivalence in sections \ref{Moritasec}, \ref{GMoritasec}.

\noindent {\bf Hamiltonian formalism}. We would like to calculate the energies of 
BPS states in the Hamiltonian formalism. For the sake of completeness let us carefully go 
over the phase space structure.
In the gauge $\nabla_{0}=\frac{\partial}{\partial t}$ we obtain  
a Hamiltonian  
\begin{equation}\label{hamiltonian}
H= \frac{g^{2}}{2V} {\rm Tr} 
P^{i}g_{ij}P^{j}  + 
 \frac{V}{4g^{2}} {\rm Tr}(F_{ij} + 
\phi_{ij}\cdot {\bf 1})g^{ik}g^{il}(F_{kl} + 
\phi_{kl}\cdot {\bf 1}) \, .  
\end{equation}
Here $F_{ij}$ is the curvature of connection $\nabla_{i}$  on a 
$T_{\vartheta}$-module $E$, $P^{i} \in End_{T_{\vartheta}} E$, and 
${\rm Tr}$ denotes the trace in $End_{T_{\vartheta}} E$. Thus, $H$ is 
defined on the space $Conn \times (End_{T_{\vartheta}}E)^{2}$.
In the derivation of (\ref{hamiltonian}) we assumed
 that the metric $g_{\mu \nu}$ obeys
$g_{0i} = 0$, $g_{00} = -1$,  
$g_{ij}= \delta_{ij} R_{i}^{2}$ and the antisymmetric  
tensor $\phi_{\alpha \beta}$ has  only spatial nonzero components 
$\phi_{ij}$ (in this section the  Greek indices $\mu, \nu, \dots$ 
run from $0$ to $d$ and Latin indices $i, j, \dots$ run from $1$ to 
$d$). 
The Hamiltonian (\ref{hamiltonian}) should be restricted to a 
subspace $\cal N$ where the constraint 
\begin{equation} \label{constr} 
[\nabla_{i}, P^{i}] = 0 
\end{equation}
is satisfied. More precisely, one should consider $H$ as a function 
on the space ${\cal N}/G = {\cal P}$ where $G$ is a group of unitary 
elements of $End_{T_{\vartheta}}E$ (the group of spatial gauge 
transformations). The symplectic form on the space 
$Conn \times (End_{T_{\vartheta}}E)^{d}$ can be written as 
\begin{equation}
\omega = {\rm Tr}\delta P^{i}\wedge \delta \nabla_{i} \, . 
\end{equation}
The restriction of this form to $\cal N$ is degenerate , but it 
descends to a nondegenerate form on ${\cal P} = {\cal N}/G$ 
(on the phase space of our theory). The phase space $\cal P$ is 
not simply connected. Its fundamental group is the group of connected 
components of the gauge group $G$. In other words, 
$\pi_{1}({\cal N}/G) = G/G_{0}\equiv G^{large}$ where $G_{0}$ is the 
group of ``small'' gauge transformations (connected component of $G$). 
One can say that $\pi_{1}({\cal N}/G)$ is a group of ``large'' gauge 
transformations. It is useful to consider the phase space $\cal P$ 
as a quotient $\tilde {\cal P}/G^{large}$ where 
$\tilde {\cal P} = {\cal N}/G_{0}$ is a symplectic manifold obtained 
from $\cal N$ by means of factorization with respect to  small 
gauge transformations.

Modules over $T_{\vartheta}$ 
are labeled by pairs of integers $(n,m)$. It is convenient to assume that 
these numbers are such that $n - \vartheta m> 0$.  
We gave an  explicit description of these modules $E_{n,m}$  in section \ref{Projsec}. 
Let us remind that  ${\rm dim} (E_{n,m}) = n - m\vartheta$, 
$\mu(E_{n,m}) = n - m\alpha_{1}\alpha_{2}$, 
${\rm ch}(E_{n,m}) = {\rm dim} (E_{n,m})  - m\alpha_{1}\alpha_{2}$, and the curvature of a 
constant curvature connection is 
$F_{12}=-2\pi i \frac{m}{n-m\vartheta }\cdot {\bf 1}$ in the standard basis.

 Let us consider  
modules $E_{n,m}$ where $n$ and $m$ are relatively prime. These 
modules are basic (see section \ref{Projsec})  because every module 
$E_{n,m}$ can be represented as a direct sum of $D$ copies of 
identical basic modules $E_{n', m'}$. (Here $D= g.c.d.(n,m)$ 
and $Dn'=n$, $Dm'=m$.) Let us fix a constant curvature connection 
$\nabla^{0}$. It follows from the results of \cite{ConnesRieffel} that for 
a basic module any other constant curvature connection can be 
transformed to the form $\nabla^{0}_{j} + iq_{j}\cdot {\bf 1}$ 
by means of small gauge transformations. Using large gauge transformations 
one can prove that  $\nabla^{0}_{j} + iq_{j}\cdot {\bf 1}$  is 
gauge equivalent to  
$\nabla_{j}^{0} + i(q_{j} - \frac{ 2\pi n_{j}}{{\rm dim}(E_{n,m})})\cdot {\bf 1}$    
where $n_{j} \in {\mathbb Z}$. Therefore, the space of gauge 
classes of constant curvature connections is a two-dimensional torus.
From now on we fix $n$, $m$ and omit subscripts in the notation of the module 
$E_{n,m}$.

We will say that a set $(\nabla_{i}, P^{j})$ is a BPS field if 
$\nabla_{i}$ is a constant curvature connection and 
$P^{j} = p^{j}\cdot {\bf 1}$. (After supersymmetrization these fields satisfy the BPS 
condition). We will obtain the energies of quantum BPS fields 
restricting our Hamiltonian  to the neighborhood of the space of BPS 
states and quantizing the restricted Hamiltonian. Let us consider 
the fields of the form 
\begin{equation} \label{con'} 
\nabla_{j}=\nabla_{j}^{0} + iq_{j}\cdot {\bf 1} + x_{j} \, ,  
\end{equation} 
\begin{equation} \label{p}
P^{i} = p^{i}\cdot {\bf 1} + \pi^{i}
\end{equation}
where  $x_{i}, \pi^{j} \in End_{T_{\vartheta}}E$, 
${\rm Tr}x_{i} = {\rm Tr}\pi^{j} = 0$. 
For a basic module we can identify $End_{T_{\vartheta}}E$ 
with a noncommutative torus $T_{\tilde \vartheta}$ where 
\begin{equation} \label{varteta}
\tilde \vartheta = (b-a\vartheta)(dim E)^{-1} \,  , \enspace \mbox{and} \enspace 
 a, b \, \mbox{satisfy} \enspace  mb - an = 1 
\end{equation}
( \cite{ConnesRieffel}). (In the explicit construction of the module $E_{n, m}$ (\ref{U's}) 
these operators were defined in (\ref{Z's}).) 
 We can 
consider $x_{i}$ and $\pi^{j}$ as functions on a lattice: 
$x_{i} = \sum_{\bf k} x_{i}({\bf k})Z_{\bf k}$, 
$\pi^{j} = \sum_{\bf k} \pi^{j}({\bf k})Z_{\bf k}$ where 
$Z_{\bf k}$ are elements of $T_{\tilde \vartheta}$ satisfying 
\begin{equation} \label{rel1}
Z_{\bf k}Z_{\bf n} = 
exp(2\pi i \tilde \vartheta (k_{2}n_{1} - k_{1}n_{2})) 
Z_{\bf n}Z_{\bf k} \, .  
\end{equation}  
Substituting  expressions (\ref{con'}) and (\ref{p}) 
into the Hamiltonian (\ref{hamiltonian}), 
 keeping the terms up to the second order in fluctuations $x_{i}$, 
$p^{i}$  we obtain 
\begin{eqnarray} \label{hfluct} 
&&H_{fluct} =  \frac{g^{2}{\rm dim}(E)}{2R_{1}R_{2}} (p^{i})^{2}R_{i}^{2} +
  \frac{1}{2g^{2}R_{1}R_{2}dimE} ( \phi {\rm dim}(E) - \pi m)^{2} + 
\nonumber \\ 
&& +  \frac{g^{2}{\rm dim}(E)}{2R_{1}R_{2}} \sum_{{\bf k}} 
\pi^{i}({\bf k})\pi^{i}(-{\bf k}) R_{i}^{2} + \nonumber \\
&& +  \frac{2\pi }{2g^{2}R_{1}R_{2}{\rm dim}(E)}
\sum_{{\bf k}} (k_{1}x_{2}({\bf k}) - k_{2}x_{1}({\bf k}))
(k_{1}x_{2}(-{\bf k}) - k_{2}x_{1}(-{\bf k})) \, . \nonumber \\
&&
\end{eqnarray} 
In the derivation of this formula we used the relation 
\begin{equation} \label{rel2}
[\nabla_{j}, x_{l}] (k_{1}, k_{2}) = \frac{2\pi i k_{j}}{dim E} 
x_{l}(k_{1}, k_{2}) \, .  
\end{equation} 
The constraint (\ref{constr}) in the approximation at hand 
now takes the form 
\begin{equation} \label{constr2}
k_{j}\pi^{j}({\bf k}) = 0 \, . 
\end{equation}
In a neighborhood of the space of BPS fields every  field 
satisfying (\ref{constr2}) 
can be transformed by means of a small gauge transformation 
into a field obeying 
\begin{equation} \label{transver}
k_{i}x_{i}({\bf k})R_{i}^{-2} = 0 \, . 
\end{equation}
This means that in our approximation the conditions (\ref{constr2}), 
(\ref{transver}) single out a symplectic manifold $\check {\cal P}$ 
that can be identified with $\tilde {\cal P}={\cal N}/G_{0}$. 
It remains to factorize with respect to large gauge transformations 
to obtain the phase space $\cal P$. The group 
$G^{large}=G/G_{0}$ of large gauge transformations can be identified 
with the subgroup $G^{mon}$ of $G$ consisting of the elements 
$Z_{\bf k}$ (more precisely, every coset in $G/G_{0}$ has a unique 
representative of the form $Z_{\bf k}$). It is easy to check that 
$\check {\cal P}$ is invariant under the action of $G^{mon}$. 
This observation permits us to identify 
${\cal P}= \tilde {\cal P}/G^{large}$ with $\check {\cal P}/G^{mon}$. 
The Hamiltonian $H$ on $\tilde P$ describes a free motion on a plane and 
an infinite system of harmonic oscillators with frequencies 
 $$
\omega({\bf k}) = \frac{2\pi}{{\rm dim}(E)}
\sqrt{ \frac{k_{1}^{2}}{R_{1}^{2}} + \frac{k_{2}^{2}}{R_{2}^{2}} }
\, . 
$$
More precisely, the fields under consideration can be 
represented in the form 
\begin{equation}
\nabla_{j} = \nabla_{j}^{0} + iq_{j}\cdot {\bf 1} + \sum_{\bf k} 
\mu({\bf k}) x^{\perp }_{j}({\bf k}) 2^{-1/2} 
( a({\bf k}) + a^{*}(\bf{-k}) ) Z_{\bf k}
\end{equation}
\begin{equation}
P^{j} = p^{j}\cdot {\bf 1} + \sum_{\bf k} \pi^{\perp j}({\bf k}) 
\mu({\bf k})^{-1} ({\rm dim}(E))^{-1}2^{-1/2} (a(-{\bf k}) - a^{*}({\bf k}))
Z_{\bf k}
\end{equation}
where $a^{*}({\bf k})$, $a({\bf k})$ are classical counterparts of 
creation and annihilation 
operators obeying the canonical commutation relations, 
$x^{\perp }_{j}({\bf k})$ is a unit vector satisfying  (\ref{transver}), 
$ \pi^{\perp j}({\bf k}) $ is a unit vector satisfying (\ref{constr2}), 
and 
$$
\mu({\bf k}) = \left( \frac{ g^{2}}{R_{1}R_{2}{\rm dim}(E)\omega({\bf k})} 
\right)^{1/2} \, . 
$$

 The Hamiltonian now reads as  
\begin{eqnarray} \label{hosc}
H& =&  \frac{g^{2}{\rm dim}(E)}{2R_{1}R_{2}} (p^{i})^{2}R_{i}^{2} +
  \frac{1}{2g^{2}R_{1}R_{2}{\rm dim}(E)} ( \phi {\rm dim}(E) - \pi m)^{2} + 
\nonumber \\ 
&+& \sum_{\bf k} \omega({\bf k}) a^{*}({\bf k}) a({\bf k}) \, . 
\end{eqnarray}
The action of the group $G^{mon}$ on the coordinates 
$q_{j}$, $p^{j}$, $a^{\dagger}({\bf k})$, $a({\bf k})$ can 
be expressed by the formulas
\begin{equation} \label{gaugetr} 
\begin{array}{c} 
q_{j} \mapsto  q_{j} - \frac{2\pi n_{j}}{{\rm dim}( E)} \\
p^{j} \mapsto p^{j} \\
a({\bf k}) \mapsto exp( 2\pi i\tilde \vartheta (n_{2}k_{1} - n_{1}k_{2})) a({\bf k})
\\
a^{*} ({\bf k}) \mapsto 
exp(- 2\pi i\tilde \vartheta (n_{2}k_{1} - n_{1}k_{2})) a^{*}({\bf k})
\end{array}
\end{equation}
These formulas follow immediately from the relations 
(\ref{rel1}), (\ref{rel2}).

\noindent {\bf Quantization}. The quantization of the system with Hamiltonian  (\ref{hosc}) 
is straightforward. The corresponding space of states is 
 spanned by the wave functions   
\begin{equation} \label{psi}
\Psi_{p_{m}; {\bf k}^{1}, N_{1}; \dots ;{\bf k}^{l}, N_{l}} = 
exp(ip^{m}q_{m}{\rm dim}(E)) 
\prod_{j=1}^{l} (a^{\dagger}({\bf k}^{j}))^{N_{j}}|0\rangle
\end{equation} 
where $|0\rangle $ is the oscillators ground state.
Here $a^{\dagger}({\bf k})$ are creation operators and  $p^{i}$ are 
eigenvalues of the quantum operator corresponding to the coordinate $p^{i}$
(which by abuse of notation we denote by the same letter). 
The group $G^{mon}$ acts on the space of states. Under this action the state 
(\ref{psi})   gets multiplied by the 
exponential  factor 
$$
exp\left( 2\pi i (-n_{j}p^{j} +n_{j}\lambda^{j}+ \tilde \vartheta 
\sum_{j=1}^{l} N_{j}(k^{j}_{1}n_{2} - k^{j}_{2}n_{1}) \right) \, . 
$$ 
where the parameters $\lambda^{j}$ have the meaning of topological 
``theta-angles''.  
Thus, the invariance of state vectors under the gauge 
transformations leads to the following quantization law of the 
$p^{i}$ values : 
\begin{eqnarray} \label{pquant} 
p^{1} &=&  e^{1} + \lambda^{1} + 
\tilde \vartheta \sum_{j=1}^{l} N_{j}k_{2}^{j}  
\, , \nonumber \\
p^{2} &=&  e^{2} + \lambda^{2} - 
\tilde \vartheta \sum_{j=1}^{l} N_{j}k_{1}^{j}  
\end{eqnarray}
where $e^{1}$ and $e^{2}$ are integers. 
Substituting this quantization condition into the Hamiltonian (\ref{hosc}) 
we get the energy spectrum 
%%%%%%%%%%%%%%%%%%%%%%%%%%%%%%%%%%%%%%%%%%%%%%%%%%%%%%%%%%%%%%%%%%%
%%%%%%%%%%%%%%%%%%%%%%              SPECTRUM             %%%%%%%%%%
%%%%%%%%%%%%%%%%%%%%%%%%%%%%%%%%%%%%%%%%%%%%%%%%%%%%%%%%%%%%%%%%%%%
\begin{eqnarray}\label{BPSspec} 
&&  E = \frac{g^{2}{\rm dim}(E)}{2R_{1}R_{2}}  
(e^{1} + \lambda^{1} + (b-a\vartheta )({\rm dim}(E))^{-1}\sum_{j=1}^{l} N_{j}k_{2}^{j} 
 )^{2}R_{1}^{2} + \nonumber \\
&& + \frac{g^{2}{\rm dim}(E)}{2R_{1}R_{2}}   (e^{2} + \lambda^{2} -  
 (b-a\vartheta )({\rm dim}(E))^{-1} \sum_{j=1}^{l} N_{j}k_{1}^{j})^{2} R_{2}^{2} 
  + \nonumber \\ + 
&&\frac{1}{2g^{2}R_{1}R_{2}{\rm dim}(E)} ( \phi {\rm dim}(E) - \pi m)^{2} 
+  \frac{2\pi }{{\rm dim}(E)} \sum_{j=1}^{l} N_{j}
\sqrt{\frac{(k^{j}_{1})^{2}}{R_{1}^{2}} + \frac{(k^{j}_{2})^{2}}{R_{2}^{2}} } 
\nonumber \\
&& 
\end{eqnarray}

The method used above to obtain the energy spectrum can be 
applied to calculate eigenvalues of a gauge invariant translation operator (momentum operator). 
A classical  functional defining this operator  has the form 
${\rm \bf P}_{i}={\rm Tr}F_{ij}P^{j}$. In the vicinity of a BPS field it 
takes the form
$$
{\rm \bf P}_{i} = m\epsilon_{ij}p^{j} + \sum_{\bf k}\frac{k_{i}}{{\rm dim}(E)}
a^{*}({\bf k})a({\bf k}) \, . 
$$ 
The corresponding operator has the following eigenvalues
\begin{equation} \label{momspec}
m_{i} = m\epsilon_{ij}e^{j}  
- a\sum_{j}k_{i}^{j}N_{j}  
\end{equation}
where $a\in {\mathbb Z}$ is the  integer that enters the 
expression (\ref{varteta}) for $\tilde \theta$ and the parameters $\lambda^{i}$ are assumed 
to be equal to zero.  
Thus, we see that the total momentum is quantized in the usual way 
 (provided $\lambda^{i}=0$). This is not surprising, the integrality 
of eigenvalues is related to the periodicity of the torus. 
One can rewrite the first two terms  of the spectrum (\ref{BPSspec}) (contribution of 
``electric charges'')  using the numbers (\ref{momspec}):
\begin{eqnarray} \label{compare}
&&  E = \frac{g^{2}}{2R_{1}R_{2}{\rm dim}(E)}  
(n^{1} + n\lambda^{1} + \vartheta (m_{2}-m\lambda^{1})
 )^{2}R_{1}^{2} + \nonumber \\
&& + \frac{g^{2}}{2R_{1}R_{2}{\rm dim}(E)}   (n^{2} + n\lambda^{2} -\vartheta 
(m_{1}+ m\lambda^{2}) )^{2} R_{2}^{2} 
 \end{eqnarray}
where 
\begin{equation} \label{n}
n^{1} = e^{1}n + b\sum_{j} N_{j}k_{2}^{j} \, , \enspace 
n^{2} = e^{2}n - b\sum_{j} N_{j}k_{1}^{j} \, . 
\end{equation}
(In (\ref{compare}), for simplicity, we set $\lambda^{i}=0$.)
  When some of the fluctuations are in the excited state 
they contribute to the energy and the integers $n^{i}$ can not be considered separately 
of the quantum numbers related to fluctuations. When all oscillators representing the 
fluctuations are in the ground state the numbers $n^{j}$ and $m_{i}$ are related by 
the formula $m_{i}n = m\epsilon_{ij}n^{j}$ and thus are not independent. 
One can fix the numbers $m_{i}$ and $n^{j}$ and minimize (\ref{BPSspec}) over all 
$N_{j}$ obeying (\ref{momspec}), (\ref{n}). We obtain 
\begin{eqnarray}\label{1/4BPS} 
&&   E = \frac{g^{2}}{2R_{1}R_{2}{\rm dim}(E)}  
(n^{1} +n\lambda^{1} + \vartheta (m_{2} - m\lambda^{1}) )^{2}R_{1}^{2} + \nonumber \\
&& + \frac{g^{2}}{2R_{1}R_{2}{\rm dim}(E)}(n^{2} +n\lambda^{2} - \vartheta (m_{1} + 
m\lambda^{2}) )^{2} R_{2}^{2} 
  + \nonumber \\ 
&& + \frac{1}{2g^{2}R_{1}R_{2}{\rm dim}(E)} ( \phi {\rm dim}(E)- \pi m )^{2} + \nonumber \\ 
&& + \frac{2\pi}{{\rm dim}(E)}  
\sqrt{\frac{(m_{1}n - n^{2}m)^{2}}{R_{1}^{2}} + \frac{(m_{2}n + n^{1})^{2}m}{R_{2}^{2}} } 
\end{eqnarray}
(When minimizing it is convenient to use the simple fact that a norm of a sum of vectors 
is always larger or equal then the corresponding sum of norms.) 
It is easy to check that (\ref{1/4BPS}) gives energies of 1/4 BPS states; we obtain 
1/2 BPS states when all oscillators are in the ground states 
(i.e.  $m_{i}n = m\epsilon_{ij}n^{j}$).

%%%%%%%%%%%%%%%%%%%%%%%%%%%%%%%%%%%%%%%%%%%%%%%%%%%%%%%%%%%%%%%%%%%%%%%%%%%%%%%%%%%%%%%%%%%%%%%%%
%%%%%%%%%%%%%%%%%%%%%               SUSY       ALGEBRA                      %%%%%%%%%%%%%%%%%%%%%
%%%%%%%%%%%%%%%%%%%%%%%%%%%%%%%%%%%%%%%%%%%%%%%%%%%%%%%%%%%%%%%%%%%%%%%%%%%%%%%%%%%%%%%%%%%%%%%%%

\subsection{Supersymmetry algebra} \label{Salgsec}
In this section we will derive the supersymmetry algebra of SYM on an arbitrary projective module 
over $T_{\theta}^{d}$. It is convenient to adopt the following conventions  about indices.
The Greek indices from the middle of the alphabet
 $\mu,\nu, \dots$  run from 0 to 9, 
  the small Latin indices $i,j, \dots$  take values
from 1 to $d$, and the capital  Latin indices $I, J, \dots$ 
take values from $d+1$ to $9$.
We fix a standard basis   of the Lie algebra $L$ such that 
 the metric tensor 
 $g_{\mu \nu}$ in this basis satisfies   $g_{00} = -1$, $g_{\mu 0} = 0$  if
$\mu \ne 0$, $g_{iJ}=0$, $g_{IJ} =
\delta_{IJ}$.
In other words  $g_{\mu \nu}$  can be written as the following block matrix
\begin{equation}
g = \left(
\begin{array}{ccccc}
-1 & 0 & 0 & \dots & 0 \\
0 & (g_{ij} ) & \dots &  0 & 0 \\
0& \dots & 1 & \dots & 0 \\
0& \dots & \dots& \ddots & 0 \\
0& 0 & \dots & \dots & 1\\
\end{array} \right)
\end{equation}
where $g_{ij}$ stands for a $d\times d$ matrix that defines a metric
on the spatial torus.
The Greek indices from the beginning of the alphabet will be used as spinor indices.
 It is convenient to denote the scalar fields $X_{I}\in End_{T_{\theta}^{d}}E$ as $\nabla_{I}$
then all of the formulas can be written in a more uniform way.

In the previous section  for simplicity we considered only the bosonic part of the whole SYM theory 
(= compactified M(atrix) theory) and   discussed the quantization  of  Yang-Mills theory on a
noncommutative torus in the $\nabla_{0} = \partial_{t}$ gauge.
Those considerations can be easily generalized to the supersymmetric case. 
 The Minkowski
action functional (\ref{action}) is defined on the configuration space 
$ConnE\times (\Pi End_{T_{\theta}}E)^{16}$
where $ConnE$ denotes the space of connections on $E$, $\Pi$ denotes the
parity reversion operator. To describe the Hamiltonian formulation we
first restrict ourselves
to the space
${\cal M} = Conn' E\times (End_{T_{\theta}}E)^{9} \times (\Pi
End_{T_{\theta}}E)^{16}$
where $Conn'E$ stands for the space of connections satisfying
$\nabla_{0}=\partial_{t}$, the second factor
corresponds to a cotangent space to $Conn'E$. We denote coordinates on that cotangent 
space by $P^{\mu}$. 
Let ${\cal N}\subset {\cal M}$ be a subspace where the Gauss constraint 
$[\nabla_{\mu},P^{\mu}] +  \psi \sigma^{0}  \psi=0$
is satisfied.  
Then the  phase space of the theory is the quotient ${\cal P}={\cal N}/G$ where $G$
is the  group of spatial gauge transformations.
The presymplectic form (i.e. a degenerate closed 2-form)  $\omega$ on ${\cal M}$ is defined as
\begin{equation} \label{sform}
\omega = {\rm Tr} \delta P^{\mu}\wedge \delta \nabla_{\mu}
+ \frac{V}{2g^{2}}{\rm Tr}\delta  \psi^{\alpha}  \sigma^{0}_{\alpha \beta} \delta \psi^{\beta} \, .
\end{equation}
It descends to a symplectic form on the phase space $\cal P$ which determines Poisson brackets
$\{. , . \}_{PB}$.

The Hamiltonian corresponding to (\ref{ncsYM}) (with the topological $\phi$-term added) reads
\begin{eqnarray}\label{Hamiltonian}
&&H= \frac{g^{2}}{2V} {\rm Tr} 
P^{\mu}g_{\mu \nu }P^{\nu}   +  \frac{V}{4g^{2}}{\rm Tr}(F_{\mu \nu } +
\phi_{\mu \nu }\cdot {\bf 1})g^{\nu \rho} g^{\mu \sigma}(F_{\rho \sigma} +
\phi_{\rho \sigma}\cdot {\bf 1}) +  \nonumber \\
&& + \frac{1}{2} \sum_{\mu \ne 0} {\rm Tr} \psi^{\alpha}\sigma^{\mu}_{\alpha \beta}[\nabla_{\mu}, \psi^{\beta}]  \, .
\end{eqnarray}

The  action (\ref{ncsYM}) is invariant
under the  supersymmetry
transformations $\delta_{\epsilon}$ and $\tilde \delta_{\epsilon}$ defined in (\ref{ncSUSY}).
The corresponding supercharges  are given by expressions
\begin{equation} \label{sch1}
 Q_{\alpha }  = \frac{1}{2}{\rm Tr} P^{\mu} (\sigma_{\mu })_{\alpha \beta}\psi^{\beta}  +
\frac{V}{4g^{2}}{\rm Tr}F_{\mu \nu}  (\sigma^{ [0 \mu \nu]})_{\alpha \beta}\psi^{\beta} \, ,
\end{equation}
\begin{equation}\label{sch2}
 \tilde Q_{\alpha} = -\frac{V}{g^{2}}\sigma^{0}_{\alpha \beta}{\rm Tr}\psi^{\beta} \, .
\end{equation}
(Supersymmetry transformations are odd vector fields preserving the
symplectic form and therefore are generated by odd functions on the phase
space - supercharges.)

As one can readily calculate using (\ref{ncSUSY}) and (\ref{sch1}), (\ref{sch2}) the supersymmetry
algebra has the form
\begin{eqnarray} \label{salgebra}
&&\{Q_{\alpha}, Q_{\beta} \}_{PB} = \frac{1}{2} (H - \Phi ) (\sigma_{0})_{\alpha \beta}
+  \frac{1}{2}(\sigma_{\mu})_{\alpha \beta}{\cal P}^{\mu} -\frac{V}{16g^{2}}
(\sigma^{[0ijkl]})_{\alpha \beta} C_{ijkl}  \nonumber \\
&&\{\tilde Q_{\alpha}, Q_{\beta} \}_{PB} = -\frac{1}{2} (\sigma_{\mu})_{\alpha \beta} p^{\mu}
+ \frac{V}{4g^{2}}(\sigma^{[0ij]})_{\alpha \beta} C_{ij} \nonumber \\
&& \{ \tilde Q_{\alpha}, \tilde Q_{\beta} \}_{PB} = \frac{V}{g^{2}}(\sigma_{0})_{\alpha \beta}{\rm dim}(E)
\end{eqnarray}
where $H$ is the Hamiltonian (\ref{Hamiltonian}),
$\Phi = \frac{V}{4g^{2}}{\rm Tr}(\phi_{ij}\phi^{ij}{\bf 1} + 2F_{ij}\phi^{ij})$,
$p^{\mu} = {\rm Tr}P^{\mu}$,
 ${\cal P}_{\mu} = {\rm Tr}F_{\mu \nu}P^{\nu} + \mbox{fermionic term}$,
$C_{ij} = {\rm Tr} F_{ij}$, $C_{ijkl}= {\rm Tr} F_{ij} F_{kl}$.
 Note that in (\ref{salgebra}) we assume
that the index $\mu$ does not take the zero value. When calculating Poisson brackets (\ref{salgebra})
it is convenient to use the identity $\{ A, B \}_{PB} = \frac{1}{2}({\bf a}(B) - {\bf b}(A))$ where
$\bf a$, $\bf b$ denote the Hamiltonian vector fields corresponding to functions $A$ and $B$
respectively.

After quantization the supersymmetry algebra (\ref{salgebra}) preserves its form 
 provided $Q_{\alpha}$, $\tilde Q_{\alpha}$, $H$, ${\cal P}_{\mu}$,
 and $p^{\mu}$  are now considered as self-adjoint operators in a Hilbert space. The quantities $C_{ij}$,
$C_{ijkl}$, $\Phi$, ${\rm dim}(E)$ are central charges.
As usual we define BPS states as states annihilated by a part of supersymmetry operators.
The energy $H$ eigenvalues of BPS states can be expressed via
values of central charges and eigenvalues of operators ${\cal P}_{\mu}$, $p^{\mu}$ (the operators
$H$, ${\cal P}_{\mu}$, and $p^{\mu}$ all commute with each other).
%%%%%%%%%%%%%%%%%%%%%%%%%%%%%%%%%%%%%%%%%%%%%%%%%%%%%%%%%%%%%%%%%%%%%%%%%%%%%%%%%%%%%%%%%%%%%%
%%%%%%%%%%%%%%%%%%            GENERAL FORMULA FOR THE ENERGY                     %%%%%%%%%%%%%
%%%%%%%%%%%%%%%%%%%%%%%%%%%%%%%%%%%%%%%%%%%%%%%%%%%%%%%%%%%%%%%%%%%%%%%%%%%%%%%%%%%%%%%%%%%%%%
The energy of  a 1/4-BPS state on a $d$-dimensional torus for $d\le 4$ is given by the formula
\begin{eqnarray} \label{energy}
&&E=\frac{g^{2}}{2V{\rm dim}(E)}p^{\mu}g_{\mu \nu}p^{\nu} + 
 \frac{V}{4g^{2}{\rm dim}(E)}(C_{ij}+ {\rm dim}(E)\phi_{ij})(C^{ij} + {\rm dim}(E)\phi^{ij}) + \nonumber \\
&& \sqrt{\|{\bf v}\|^{2} + {\cal P}_{I}^{2}
+ (\pi /g)^{4}(C_{2})^{2} } \, .
\end{eqnarray}
where $\|{\bf v}\|^{2}= v_{i}g^{ij}v_{j}$ stands for the norm squared of  a $d$-dimensional vector
  ${\bf v} = ({\cal P}_{i} - ({\rm dim}(E))^{-1}C_{ij}p^{j})$, and
$$
C_{2}=\frac{1}{8\pi^{2}} \epsilon^{ijkl}(C_{ijkl} - ({\rm dim}(E))^{-1}C_{ij}C_{kl}) \, .
$$ In formula (\ref{energy}) we
use the same notations ${\cal P}_{\mu}$, $p^{\mu}$ for the eigenvalues of the corresponding operators.
Note that  when the term with the square root
vanishes we get a 1/2-BPS state.
Here we would like to make few remarks on how to obtain formula (\ref{energy}).
First one notices
that commutators of supercharges (\ref{salgebra}) form a block
matrix $M_{\alpha \beta}$. The BPS condition means that this matrix has a zero eigenvector. As the block
$\{\tilde Q_{\alpha}, \tilde Q_{\beta}\}$ is non-degenerate one can reformulate the BPS  condition
as the  degeneracy condition on some square matrix of dimension twice smaller than  that of
$M^{\alpha \beta}$. At that point one can apply the
standard technique of finding zero eigenvalues of matrices expressed in terms of Gamma matrices
(for example see \cite{OP} Appendix B).

Note that operators ${\cal P}_{I}$, $p^{J}$ have a continuum spectrum.
  We will  restrict ourselves to the zero eigenvalue subspace for these operators. Moreover, we will not 
consider any effects of scalar fields on the spectrum.  Below 
${\cal P}_{i}$ denotes the operator  $ {\rm Tr} F_{ij}P^{j}$. 

%%%%%%%%%%%%%%%%%%%%%%%%%%%%%%%%  TOP. TERMS 

\subsection{Topological terms from geometric quantization} \label{Topsec}
In this section we would like to discuss how topological terms similar to the theta angle term in the Yang-Mills 
theory can be incorporated  into the  Hamiltonian formalism. To this end we use the framework of geometric quantization. 
This approach is not new and belongs to a kind of mathematical physics folklore (however  see \cite{Gawedzki}). 
Still by some reason it is not  widely known. In this section we will briefly explain the main idea.
The material in this section  is not essential for  understanding of the forthcoming sections
 and can be skipped at the first reading.

The phase space $\cal P$ of the theory we consider  is not simply connected. It is well known that in such a 
case there is some freedom in  quantization. Namely, in the framework of geometric quantization 
 we can assign  to every function $F$ on a phase space an operator $\check F$ defined
 by the formula
\begin{equation} \label{preq}
\check F \phi = F\phi + \omega^{ij}\frac{\partial F}{\partial x^{j}} 
\nabla_{i} \phi 
\end{equation} 
where  $\nabla_{i} \phi = \hbar \partial_{i} + \alpha_{i}$ is a 
covariant derivative with respect to $U(1)$-gauge field having the 
curvature $\omega_{ij}$. (Such a gauge field exists if the cohomology class of $\omega$ is integral.) 
Here $\omega_{ij}$ is a matrix of symplectic form $\omega$ and $\omega^{ij}$ 
is the inverse matrix. The operator $\check F$ acts on the space of sections of a line bundle over 
the phase space. The transition from $F$ to $\check F$ is called prequantization. It satisfies 
$$ 
[\check F, \check G ] = i\hbar (\{ F, G \} )^{\vee}
$$
where $\{ F, G \} = \frac{\partial F}{\partial x^{i}}\omega^{ij} 
\frac{\partial G}{\partial x^{j}} $ is the Poisson bracket.
Notice that replacing $\alpha$ with $\alpha + \delta \alpha$ we change $\check F$ in the following 
way
\begin{equation} \label{F}
\check F = \check F_{old} + \delta \alpha (\xi_{F}) = 
\check F_{old} + \delta \alpha_{i}\xi^{i}_{F}
\end{equation}  
where $\xi_{F}$ stands for the Hamiltonian vector field corresponding to the function $F$. 
To define a quantization we should introduce the so called polarization, i.e. exclude half of the 
variables. Then for every function $F$ on the phase space we obtain a quantum operator $\hat F$.

If the gauge field in (\ref{preq}) is replaced by a gauge equivalent field we obtain an equivalent 
quantization procedure. However, in the case when the phase space is not simply connected the gauge 
class of $\nabla_{i}$ in (\ref{preq}) is not specified uniquely. The simplest choice of the 1-form 
$\alpha$ in the system we consider is 
\begin{equation} \label{alpha}
\alpha = {\rm Tr}  P^{\mu} \delta \nabla_{\mu}
\frac{V}{2g^{2}}{\rm Tr}  \psi \sigma^{0}  \delta \psi \, .
\end{equation}
But we can also add to this $\alpha$ any closed 1-form $\delta \alpha$, for example 
\begin{equation} \label{dalpha}
\delta \alpha = \lambda^{i}{\rm Tr} \delta \nabla_{i} + \lambda^{ijk}{\rm Tr}\delta \nabla_{i}
\cdot F_{jk} 
\end{equation}
where $\lambda^{ijk}$ is an antisymmetric 3-tensor and $\lambda^{i}$ is a 1-tensor.
One can check that for $d\le 4$  we obtain all gauge classes by adding (\ref{dalpha}) 
to (\ref{alpha}) (for $d>4$ one 
should include additional terms labeled by antisymmetric tensors of odd rank $\ge 5$).
In the Lagrangian formalism  one can include topological terms into the action functional.
The addition of all topological terms to the Lagrangian corresponds to the consideration 
of all possible ways of (pre)quantization in the Hamiltonian formalism.
The topological terms can be interpreted as Ramond-Ramond backgrounds from the string theory 
point of view (see \cite{HofVerII}, \cite{BrMorZum2} for more details).
The additional terms  in the action can be expressed in terms of the Chern character.
In particular, the consideration of the form $\alpha + \delta \alpha$ given by  
(\ref{alpha}), (\ref{dalpha}) corresponds to adding the following
 terms to the action
\begin{equation}\label{Stop}
S_{top} = \lambda^{i}{\rm Tr} F_{0i} + \lambda^{ijk}{\rm Tr} F_{0i}F_{jk}  \, .
\end{equation}

%%%%%%%%%%%%%%%%%%%%%%%%%%%%%%%%%%%%%%%%%%%%%%%%%%%%%%%%%%%%%%%%%%%%%%%%%%%%%%%%%%%%%%%%%%%%%%

\subsection{Spectrum of translation operators}  \label{Translsec}
Let us make one  general remark first. Assume that a first order linear differential operator $L_{A}$
defined by $L_{A}F = \{A,F\}_{PB}$ obeys $e^{2\pi  L_{A}}=1$ (in other words the Hamiltonian vector field 
corresponding to $A$ generates the group $U(1)={\mathbb R}/{\mathbb Z}$). Then it is easy to 
check that 
\begin{equation}\label{A'}
e^{2\pi  \check A} \check F e^{-2\pi  \check A} = \check F
\end{equation} 
for any function $F$ on the phase space. 
It follows from (\ref{A'}) that after quantization we should expect 
$$
e^{2\pi i \hat A}=const. 
$$
We see that the eigenvalues of $\hat A$ are quantized; they have the form $m + \mu$ 
where $m\in {\rm \bf Z}$ and $\mu$ is a fixed constant.

Now we would like to discuss the spectrum of the operators  $\hat {\cal P}_{i}$, $\hat p^{i}$.
Consider a Hamiltonian vector field $\xi_{{\cal P}_{i}}$ corresponding to the gauge invariant momentum 
functional ${\cal P}_{i} = {\rm Tr}F_{ij}P^{j}$. The variation of $\nabla_{j}$ under the action 
of $\xi_{{\cal P}_{i}}$  is equal to 
$F_{ij}$. Let us fix a connection $\nabla^{0}_{i}$ on the module $E$. 
Then an arbitrary connection has the form $\nabla_{i} = \nabla_{i}^{0} + X_{i}$ where 
$X_{i}\in End_{T_{\theta}}E$. Hence, we have
\begin{equation} \label{P'}
\xi_{{\cal P}_{i}}(\nabla_{j}) = F_{ij} = -[\nabla_{j}, X_{i}] +  [\nabla_{i}^{0},\nabla_{j}] \, . 
\end{equation} 
Here the first term corresponds to infinitesimal gauge transformation. Therefore, up to a 
gauge transformation, ${\cal P}_{i}$ generates a vector field acting as 
$\delta \nabla_{j} =  [\nabla_{i}^{0},\nabla_{j}]$. Now one can check using 
 the identity $e^{ \delta_{j}}=1$ 
that the operator $exp( \nabla^{0}_{j})$ is an endomorphism, i.e. 
$$
exp(  \nabla^{0}_{j}) U_{i} exp(-  \nabla^{0}_{j}) = U_{i}
$$
and since it is  unitary it  represents a global gauge transformation.
This means that on the phase space of our theory (after taking a quotient with respect to the 
gauge group) the identity $exp(  L_{{\cal P}_{j}})=1$ is satisfied. Here $L_{{\cal P}_{j}}$
stands for the differential operator corresponding to ${\cal P}_{j}$. 
As we already mentioned this identity leads to a quantization of $\hat {\cal P}_{j}$ eigenvalues. 
Namely, for some fixed constant $\mu_{j}$ the eigenvalues of $\hat {\cal P}_{j}$ have the form 
\begin{equation} \label{qc}
{\cal P}_{j} = 2\pi( \mu_{j} + m_{j}) 
\end{equation}
where $m_{j}$ is an integer. 
The quantization condition (\ref{qc}) is valid for every choice of quantization procedure
(for every $\alpha + \delta \alpha$). However, the constants $\mu_{j}$ depend on 
this choice. Using the freedom that we have in the quantization we can take $\mu_{j}=0$ 
when $\delta \alpha=0$. In the case of non-vanishing $\alpha$ we obtain using (\ref{F}) and (\ref{P'})
\begin{equation}
 \label{momentum}
 {\cal P}_{j} = 2\pi m_{j} +  C_{jk}\lambda^{k} + C_{ijkl}\lambda^{jkl} \, .
\end{equation}

 Quantization of the electric field zero mode ${\rm Tr}P^{i}$ comes from periodicity conditions on
the space $Conn' E/G$. The operator  ${\rm Tr}P^{i}$ can be considered as a generator of translations 
on that space. 
As it was first noted in \cite{HofVerII} 
the operator $U_{j}exp(-  \theta^{jk}\nabla_{k})$  commutes with all $U_{i}$ and thus is
an endomorphism. Being unitary this operator determines a gauge transformation.  One can easily 
check that this gauge transformation can be equivalently written as an action of the 
operator 
\begin{equation} \label{op}
exp((2\pi \delta_{jk} - \theta^{jl}F_{lk})\frac{\delta}{\delta\nabla_{k}})
\end{equation}
on the space $Conn'E$. Therefore, on the space $Conn' E/G$ this operator descends to the  identity. 
The Hamiltonian vector field defined by the exponential of (\ref{op}) corresponds to the functional 
$2\pi p^{k} - \theta^{ki}{\cal P}_{i}$. Hence, as we explained above the quantum operator 
$\hat p^{k} - (2\pi)^{-1}\theta^{ki}\hat {\cal P}_{i}$ has eigenvalues of the form 
$ n^{i} + \nu^{i}$ 
where $n^{i} \in {\rm \bf Z}$ and $\nu^{i}$ is a fixed number. 
  Proceeding as above we obtain the following quantization law for eigenvalues of $Tr\hat P^{i}$
\begin{equation} \label{p'}
p^{i} =  n^{i} + \theta^{ij}m_{j} +  \lambda^{i}dimE +
\lambda^{ijk} C_{jk}
\end{equation}
where $n^{i}$ is an integer and $m_{j}$ is the integer specifying the eigenvalue of total momentum 
operator (\ref{momentum}).

%%%%%%%%%%%%%%%%%%%%%%%%%%%%%%%%%%%%%%%%%%%%%%%%%%%%%%%%%%%%%%%%%%%%%%%%%%%%%%%%%%%%%%%%%%%%
\subsection{Energies of  BPS states in d= 2, 3, 4} \label{234sec}

Now we are ready to write explicit answers for $d=2,3,4$. For
$d=2$ the Chern character can be written as
$$
{\rm ch}(E) = (n - m\vartheta)  -m \alpha^{1}\alpha^{2}
$$
where $n$ and $m$ are integers such that $n-m\vartheta > 0$.
 Hence $dimE=n-m\vartheta$, $C_{ij} = -\pi \epsilon_{ij} m$, $C_{2}=0$ and we get the
following answer for the energies of BPS states
\begin{eqnarray} \label{d=2}
&&E= \frac{g^{2}}{2V{\rm dim}(E)}(n^{i} +\theta^{ik}m_{k}+  \lambda^{i}{\rm dim}(E) )g_{ij}\cdot
\nonumber \\
&&\cdot (n^{j} + \theta^{jl}m_{l}+  \lambda^{j}{\rm dim}(E)) +  \nonumber \\
&& + \frac{1}{2Vg^{2}{\rm dim}(E)}(\phi dimE - \pi m)^{2}  
+  \frac{2\pi}{{\rm dim}(E)}\| { \bf v }\|
\end{eqnarray}
where ${\bf v} = (m_{i}n- m\epsilon_{ij}n^{j} )$.
It is easy to see that this formula coincides with formula (\ref{1/4BPS}) that we derived in section 
\ref{BPSsec} by a different method.

For a three-dimensional  torus the Chern character has the form
$$
{\rm ch}(E) = n + \frac{1}{2}{\rm tr}(\theta q)  + \frac{1}{2}q_{ij}\alpha^{i}\alpha^{j}
$$
where $n$ is an integer and $q_{ij}$ is an antisymmetric matrix with integral entries.
Substituting the expressions ${\rm dim}(E) = n + \frac{1}{2}{\rm tr}(\theta q)$,
$C_{ij} = \pi q_{ij}$, $C_{2}=0$ into the main formula (\ref{energy}) and using (\ref{p'}), (\ref{momentum})
we obtain
\begin{eqnarray} \label{d=3}
&&E= \frac{g^{2}}{2V{\rm dim}(E)}(n^{i} + \theta^{ik}m_{k}+ \lambda^{i}{\rm dim}(E) + 
\pi \lambda^{ikl}q_{kl})g_{ij}\cdot \nonumber \\
&& \cdot (n^{j} + \theta^{jr}m_{r} + \lambda^{j}{\rm dim}(E ) + \pi \lambda^{jrs}q_{rs}) +  \nonumber \\
&& + \frac{V}{4g^{2}{\rm dim}(E)}(\pi q_{ij} + {\rm dim}(E) \phi_{ij})g^{ik}g^{jl}(\pi q_{kl} + {\rm dim}(E) \phi_{kl})
+ \nonumber \\
&& + \frac{\pi }{dimE}\| {\bf v}\|
\end{eqnarray}
where ${\bf v} = (m_{i}{\rm dim}(E) - q_{ij}(n^{j} + \theta^{jk}m_{k}))$.

Finally let us consider the case $d=4$. The Chern character now reads as
$$
{\rm ch}(E) = n + \frac{1}{2}{\rm tr}(\theta m) + q{\rm Pfaff}(\theta ) +
\frac{1}{2}(m + q\cdot \tilde \theta )_{ij}\alpha^{i}\alpha^{j} + q\alpha^{1}\alpha^{2}\alpha^{3}\alpha^{4}
$$
where $n$ and $q$ are integers, $m_{ij}$ is an antisymmetric $4\times 4$ matrix with integral
entries, $(\tilde \theta)_{ij}=\frac{1}{2}\epsilon_{ijkl}\theta^{kl}$, and 
${\rm Pfaff}(\theta )=\frac{1}{8}\theta^{ij}\epsilon_{ijkl}\theta^{kl} = \sqrt{det \theta_{ij}}$ 
stands for the Pfaffian of $\theta$. From this expression for ${\rm ch}(E)$   we see that 
$$
{\rm dim}(E)=n + \frac{1}{2}{\rm tr}(\theta m) + q{\rm Pfaff}(\theta ) \, , \quad
C_{ij} = \pi (m + \tilde \theta q)_{ij}
$$
$$
C_{[ijkl]} = \pi^{2}\epsilon_{ijkl} q \, , \quad C_{2} = ({\rm dim}(E))^{-1}(nq - {\rm Pfaff}(m)) \, .
$$
Substituting the above expressions  into (\ref{energy}) we obtain the following answer for the
BPS spectrum
\begin{eqnarray}\label{d=4}
&&E= \frac{g^{2}{\rm dim}(E)}{2V}(n^{i} + \lambda^{i}dimE + \pi \lambda^{ikl}(m+\tilde \theta q)_{kl} + \theta^{ik}m_{k})g_{ij}\cdot \nonumber \\
&& \cdot (n^{j} + \lambda^{j}{\rm dim}(E) + \pi\lambda^{jrt}(m+\tilde \theta q)_{rt} + \theta^{jr}m_{r}) + 
   \nonumber \\ 
&& \frac{V\pi^{2}}{4g^{2}{\rm dim}(E)}((m+ \tilde \theta q)_{ij} + 
{\rm dim}(E) \frac{\phi_{ij}}{\pi})g^{ik} g^{jl}((m+ \tilde \theta q)_{kl} + {\rm dim}(E) \frac{\phi_{kl}}{\pi})
+ \nonumber \\
&& + \frac{\pi }{{\rm dim}(E)}\sqrt{ \|{\bf v}\|^{2} + (\pi /g^{2})^{2}(nq- {\rm Pfaff}(m))^{2} } \, .
\end{eqnarray}
Here
$$
{\bf v} = (v_{i}) = (m_{i}{\rm dim}(E) - (m + \tilde \theta q)_{ij}(n^{j} + \theta^{jk}m_{k}) +
2\pi (\tilde \lambda_{3})_{i}(nq- {\rm Pfaff}(m) ))\, ,
$$ and $(\tilde \lambda_{3})_{i}=\frac{1}{3!}\epsilon_{ijkl}\lambda^{jkl}$.

%%%%%%%%%%%%%%%%%%%%%%%%%%%%%%%%%%%%%%%%%%%%%%%%%%%%%%%%%%%%%%%%%%%%%%%%%%%
\section{Morita equivalence}
\subsection{Morita equivalence of associative algebras} \label{Moritasec}
One says that two associative algebras $A$ and $\hat A$ are Morita equivalent if there exists 
a natural identification of modules over these algebras (in mathematical terms this means that the category 
of $A$-modules is equivalent to the category of $\hat A$-modules). In other words to every $A$-module $E$ 
there should correspond an $\hat A$-module $\hat E$ and to every $A$-linear map 
$\phi: E \to E'$ there should correspond a map $\hat \phi : \hat E \to \hat E'$. 
It is required that $\widehat{ \phi_{1}\circ \phi_{2}} = \hat \phi_{1} \circ \hat \phi_{2} $ and 
that the above correspondences are bijections (i.e. one-to-one and onto). 
More precisely we assume that there exists a map assigning to every $\hat A$-module $F$ an $A$-module 
$\tilde F$ in such a way that $\widetilde{\hat E}$ is canonically isomorphic $E$.  
We also assume that there exists a similar mapping transforming $\hat A$-linear maps into $A$-linear ones. 
In particular {\it we have an isomorphism between $End_{A}E$ and $End_{\hat A} \hat E$}.
Let us introduce a notation $P= \widehat{A^{1}}$. In other words $P$ denotes the $\hat A$-module 
corresponding to the one-dimensional free $A$-module $A^{1}$. Notice that $P$ can be also considered as 
an $(\hat A , A)$-bimodule because $End_{\hat A}P = End_{A}(A^{1}) = A^{op}$. 
Similarly $Q  = \widetilde{\hat A^{1}}$ can be considered as an $(A, \hat A )$-bimodule. 
If we know the bimodule $P$ we can describe the correspondence $E\mapsto \hat E$ by means of the formula 
\begin{equation} \label{P}
\hat E  = P\otimes_{A} E  \, .
\end{equation}
Here we use the notion of a tensor product over the algebra $A$ that can be obtained from the usual tensor product 
$\otimes_{\mathbb C}$ (tensor product over $\mathbb C$) by means of identification 
$ (pa)\otimes e \sim p\otimes (ae)$ for any $p\in P$, $e\in E$, $a\in A$.
Note that in this definition we use the fact that $P$ is a right $A$-module and $E$ is a left module.
We omit the proof of (\ref{P}). Instead we will use (\ref{P}) as a constructive definition of Morita 
equivalence. In other words we will take an arbitrary $(\hat A, A)$-bimodule $P$ as a starting point and 
use it to define a correspondence between $A$-modules and $\hat A$-modules by  formula  (\ref{P}).
This construction immediately extends to give a correspondence between $A$-linear and $\hat A$-linear 
maps. Namely, every $A$-linear map $\phi: E \to E'$ induces an $\hat A$-linear map 
$\hat \phi = 1\otimes \phi  : P\otimes_{A}E \to  P\otimes_{A}E'$. We say that $P$ generates a Morita equivalence 
of algebras $A$ and $\hat A$ if there exists an $(A, \hat A)$-bimodule $Q$ generating the inverse correspondence 
$\hat E \mapsto E$. Then $P$ is called an $(\hat A, A)$ Morita equivalence bimodule.

One can prove that Morita equivalence bimodules can be characterized as follows. 
Consider a projective $A$-module $P$ that has the property that $A^{1}$ is isomorphic to a direct summand 
in $P^{N}=P\oplus \dots \oplus P$ for some $N$. Then $P$ is an $(A, \hat A)$ Morita equivalence bimodule 
where  $\hat A = End_{A}P$. Every Morita equivalence bimodule can be obtained by means of this construction. 
We are not going to give a complete proof of this statement. Let us remark only that the necessity of the 
conditions we imposed on $P$ is obvious (the second condition is equivalent  to projectivity of the 
$\hat A$-module $\widehat{A^{1}}$).

Let us give some examples. First of all for any algebra $A$ we can take $P=A^{N}$. We obtain that any 
algebra is Morita equivalent to the algebra of $N\times N$ matrices $Mat_{N}(A)$. 
If $A$ is a two-dimensional noncommutative torus $T_{\theta}^{2}$ we can take as $P$ the module 
defined by formulas (\ref{R1}). We deduce that the algebra $T_{\theta}^{2}$ is Morita equivalent 
to the algebra $T^{2}_{1/\theta}$. Taking as $P$ the module $E_{m,n}$ with $m$ and $n$ relatively prime 
we obtain that $T_{\theta}^{2}$ is Morita equivalent to $T_{\hat \theta}^{2}$ where 
$$
\hat \theta = \frac{b- a\theta }{n - m\theta} \, .
$$ 
Here $a$ and $b$ are integers satisfying $an - mb =1$. 
This fact follows from the description of endomorphisms of  $E_{m,n}$, they are generated by 
the operators (\ref{Z's}) that satisfy the commutation relation defining $T_{\hat \theta}^{2}$.

Consider now the module described in  section \ref{cccsec} and defined by formula (\ref{Ust}). 
It is not hard to check that its endomorphisms are generated by operators $e^{\nabla_{j}}$ that 
satisfy the commutation relations of a noncommutative torus $T_{\theta^{-1}}^{d}$.
This means that $d$-dimensional noncommutative tori $T_{\theta}^{d}$ and $T_{\theta^{-1}}^{d}$ are Morita 
equivalent (we assume that $\theta$ is nondegenerate).

Another set of examples  of Morita equivalence is provided by automorphisms of algebra $A$. 
In general for any associative algebra $A$ and any automorphism $\phi : A \to A$ one can consider 
a bimodule $P$ that consists of elements of $A$ itself with the right action defined by the multiplication 
by $a\in A$ from the right and the left multiplication defined as a multiplication by $\phi(a)$ from the left. 
Clearly this defines an $(A, A)$ Morita equivalence bimodule. It defines some mapping of projective modules. 
 In the case when $\phi$ is an inner automorphism the corresponding Morita equivalence is  always trivial. 
 Outer automorphisms as the examples below  show may result in a nontrivial relabeling of modules.

Noncommutative  tori parametrized by the matrices $\theta^{ij}$ and 
$\theta^{ij} + N^{ij}$ that differ by an antisymmetric matrix $N^{ij}$ with integer entries, are isomorphic. 
However trivial this isomorphism may look it generates a nontrivial transformation on modules due to the 
Elliott formula (\ref{Elliott}) that explicitly depend on the fixed $\theta^{ij}$. The resulting  transformation 
shuffles the topological numbers  that can be considered as a mere relabelling of modules. 
Another kind of Morita equivalence  of this sort has to do with geometric $SL(d, {\mathbb Z})$ symmetry of the torus.
If $A\in SL(d, {\mathbb Z})$ then the tori with  $\theta^{ij}$ and $\hat \theta^{ij} = (A^{t}\theta A)^{ij}$ are 
isomorphic. The corresponding Morita equivalence results in a  change of basis of the Grassmann algebra $\Lambda(L^{*})$: 
$\alpha^{i} \to (A^{t})^{i}_{j}\alpha^{j}$ 
that changes the topological numbers in the corresponding way.

In general we can take for $P$ any basic $T_{\theta}^{d}$-module. Then the algebra $End_{T_{\theta}^{d}}P$ is 
again a noncommutative torus $T^{d}_{\hat \theta}$ (see section \ref{Projsec}).  
One can prove that  the matrix $\hat \theta$ is related to $\theta$ 
by means of a fractional linear transformation 
\begin{equation} \label{fr_tr}
\hat \theta = (M\theta + N)(R\theta + S)^{-1}
\end{equation}
where 
\begin{equation} \label{matr}
\left(
 \begin{array}{cc} 
M&N\\
R&S 
\end{array}
\right)
\end{equation}
is a $2d\times 2d$ matrix belonging to the group $SO(d,d|{\mathbb Z})$. (This result will be proved 
in the next section.) Moreover  given an element of $SO(d,d|{\mathbb Z})$ 
specified by a matrix (\ref{matr}) the torus $T_{\theta}^{d}$ is Morita equivalent to the torus $T_{\hat \theta}^{d}$ 
where $\hat \theta$ is related to $\theta$ by means of the fractional transformation (\ref{fr_tr}) (provided 
it is well defined, i.e. the denominator is invertible). Let us sketch a proof of this fact here.
First note that the group  $SO(d,d|{\mathbb Z})$ acting on $\theta$ is generated by three kinds transformations:
the shifts  $\theta\to \theta + N$ that embed in the $SO(d,d|{\mathbb Z})$ as matrices with vanishing
blocks $M$, $R$, $S$, the  $SL(d, {\mathbb Z})$ rotations $\theta \to A^{t}\theta A$   corresponding  to 
$M = A^{t}$, $R= (A^{t})^{-1}$, $N=S=0$, a single transformation $\sigma$ called a flip that inverts any given $2\times 2$  block 
in matrix $\theta$. More precisely the last transformation act on $\theta$ as follows. Without loss 
of generality we can assume that $\theta$ has a block form
$$
\theta = \left( 
\begin{array}{cc}
\theta_{11}&\theta_{12}\\
\theta_{21}&\theta_{22}
\end{array}
\right) \, .
$$  
where $\theta_{11}$ is a $2\times 2$ nondegenerate matrix. Then a flip $\sigma$ sends $\theta$ into 
\begin{equation} \label{flip}
\sigma_{2}(\theta ) =  \left( 
\begin{array}{cc}
\theta_{11}^{-1}&-\theta_{11}^{-1}\theta_{12}\\
\theta_{21}\theta_{11}^{-1}&\theta_{22} - \theta_{21}\theta_{11}^{-1}\theta_{12}
\end{array}
\right) \, .
\end{equation}
For the particular case of a two-dimensional torus the flip simply inverts the whole matrix $\theta$. 
One can check that $SL(d, {\mathbb Z})$ rotations, shifts and the flip $\sigma$ generate the whole $SO(d,d|{\mathbb Z})$.
The construction of Morita equivalence bimodules corresponding to  $SL(d, {\mathbb Z})$ transformations and shifts is 
obvious. As about the flip transformation one can also explicitly construct the corresponding bimodule. However the construction 
is a bit technical and we skip it here (see \cite{RS}). Let us only note that for two-dimensional tori this 
bimodule  is defined by (\ref{R1}), (\ref{Z1}).

%%%%%%%%%%%%%%%%%%%%%%%%%%%%%%%%%%%%%%%%%%%%%%%%%%%%%%%%%%%%%%%%%%%%%%%%%%%%%%%
\subsection{Gauge Morita equivalence} \label{GMoritasec}
For the case of noncommutative tori one can define a notion of gauge Morita equivalence \cite{ASMorita}\footnote{In 
\cite{ASMorita} it was called ``complete Morita equivalence''. 
We adopt the term ``gauge Morita equivalence'' from \cite{SeibWitt} that seems to be more illuminating.}
that allows one to transport connections between modules $E$ and $\hat E$. Let $L$ be a $d$-dimensional 
commutative Lie algebra. 
We say that $(T_{\hat \theta}^{d},T_{ \theta}^{d})$ Morita equivalence bimodule $P$ establishes a gauge Morita equivalence if it
is endowed with operators $\nabla^{P}_{X}$, $X\in L$ that determine a constant curvature connection 
simultaneously with respect to $T_{\theta}^{d}$ and $T_{\hat \theta}^{d}$, i.e. satisfy  
\begin{eqnarray}\label{biconnect}
&&\nabla^{P}_{X}(ea)=(\nabla^{P}_{X}e)a + e(\delta_{X}a) \, , \nonumber \\
&&\nabla^{P}_{X}(\hat ae )=\hat a(\nabla^{P}_{X}e) + (\hat \delta_{X} \hat a)e \, , \nonumber \\
&&[\nabla^{P}_{X},\nabla^{P}_{Y}]=2\pi i \sigma_{XY}\cdot {\bf 1} \, .
\end{eqnarray} 
Here  $\delta_{X}$ and $\hat \delta_{X}$ are  standard derivations on $A_{\theta}$ and $A_{\hat \theta}$ respectively. 
In other words we have two Lie algebra homomorphisms
\begin{equation} \label{deltas}
\delta : L \to L_{\theta} \, , \qquad \hat \delta : L \to L_{\hat \theta} \, . 
\end{equation}
It is important to have covariant (independent of the choice of basis in $L$) formulas because it does  not happen in general that 
a standard basis in $L$ defined with respect to  $L_{\theta}$, in which derivations $\delta_{i}$ satisfy (\ref{deltaij}), has 
the same property with respect to $L_{\hat \theta}$. Putting it differently we can say that homomorphisms (\ref{deltas}) 
set some particular isomorphism  between $L_{\theta}$ and $L_{\hat \theta}$. This isomorphism is just some linear transformation.

 Any basic module $E$ is equipped with a standard constant curvature connection $\nabla_{i}$ the construction of which was 
described at the end of section \ref{Consec}. One can check that this standard connection satisfies (\ref{biconnect}) that 
gives us an example of gauge Morita equivalence bimodule.

Sometimes for brevity we will omit the word Morita in the term (gauge) Morita equivalence bimodule.
If a pair $(P, \nabla^{P}_{X})$ specifies a gauge  $(A_{\theta},A_{\hat \theta})$ equivalence  bimodule then there exists a correspondence 
between connections on $E$ and connections on $\hat E$. 
A connection $\hat \nabla_{X}$ on $\hat E$ corresponding to a given connection $\nabla_{X}$ on $E$  is defined as 
$$
\nabla_{X} \mapsto \hat \nabla_{X} = 1\otimes \nabla_{X} +  \nabla_{X}^{P}\otimes 1
$$
More precisely, an operator 
$1\otimes \nabla_{X} +  \nabla_{X}^{P}\otimes 1$ on $E\otimes_{\mathbb C}P$ descends to a 
connection $\hat \nabla_{X}$ on $\hat E = P\otimes_{A_{\theta}}E  $. 
It is straightforward  to check that under this mapping gauge equivalent connections 
go to gauge equivalent ones
$$
\widehat{Z^{\dagger}\nabla_{X} Z} = \hat Z^{\dagger} \hat \nabla_{X} \hat Z
$$
where $\hat Z = 1\otimes Z$ is the endomorphism of $\hat E= P\otimes_{A_{\theta}}E$ corresponding
to $Z\in End_{T_{\theta}^{d}}E$.

The curvatures of 
$\hat \nabla_{X}$ and $\nabla_{X}$ are connected by the formula 
\begin{equation} \label{curv_shift}
F_{XY}^{\hat \nabla}=\hat F_{XY}^{\nabla} + {\bf 1}\sigma_{XY} 
\end{equation}
that in particular shows that constant curvature connections go to constant curvature ones.

As we now have a mapping of connections it is natural to ask whether this mapping preserves the Yang-Mills equations of motion. 
It turns out it does. Moreover, one can define a slight generalization of Yang-Mills action functional in such a way 
that the values of functionals on $\nabla_{X}$ and $\hat \nabla_{X}$ coincide up to an appropriate rescaling of coupling 
constants. By a slight modification we mean an action functional of the form
\begin{equation} \label{modifYM}
S_{YM} = \frac{V}{4g^{2}}{\rm Tr} (F_{jk} + \Phi_{jk}\cdot {\bf 1}) (F^{jk} + \Phi^{jk}\cdot {\bf 1})
\end{equation}  
where $\Phi^{jk}$ is a number valued tensor that can be thought of as some background field. 
Adding this term  corresponds to adding the terms $ \Phi^{2} {\rm Tr}{\bf 1}$ and $\Phi^{jk}{\rm Tr} F_{jk}$ 
that are both topological in nature (they are proportional to ${\rm ch}_{0}$ and $({\rm ch}_{1})_{jk}$ and thus do not affect 
the Yang-Mills equation of motion. On the other hand adding this term will allow us to compensate the shift 
(\ref{curv_shift}) by adopting the transformation rule 
\begin{equation} \nonumber 
\Phi_{XY} \mapsto \Phi_{XY} - \sigma_{XY} \, .
\end{equation} 
To show that thus defined $S_{YM}$ is invariant under gauge Morita equivalence one has to take into account two more effects. 
Firstly, the values of  trace  change by a factor $c = {\rm dim}(\hat E) ({\rm dim}(E ))^{-1}$  
as  $ \hat {\rm Tr} \hat A = c {\rm Tr} A $. Secondly, the identification of $L_{\theta}$ and $L_{\hat \theta}$ 
is established by means of some linear transformation $A_{j}^{k}$ the  determinant of which will 
rescale the volume $V$. Both effects can be absorbed into an appropriate rescaling of the coupling constant. 
Note that in this consideration it is  not important that we consider the Yang-Mills action. It will go 
through for any gauge invariant action depending only on the combination $F_{ij} + \Phi_{ij}$. In particular 
one could consider  Born-Infeld type  action functionals.

Let us turn now to the derivation of precise transformation rules and the group governing the above duality.
Note that  formula (\ref{curv_shift}) implies the following relation between the Chern characters of the modules 
$E$ and $\hat E$ related by a gauge Morita equivalence 
\begin{equation} \label{Cherns}
{\rm ch}(\hat E ) = \frac{{\rm dim}(\hat E )}{{\rm dim}( E )} e^{\alpha^{j}\sigma_{jk}\alpha^{k}} {\rm ch}( E) \, .
\end{equation}
(Here $\alpha^{j}$ corresponds to some basis in $L_{\theta}$, the particular choice is not essential here.) 
This formula is a straightforward consequence of the definition (\ref{ncChern}) and the last formula in (\ref{biconnect}). 
This relation between Chern characters induces by (\ref{Elliott}) a relation between K-theory classes $\mu (E)$ and 
$\mu (\hat E)$. The last ones are integral elements of the  Grasmann algebras 
$ \Lambda^{.}(L_{\theta}^{*})$ and $ \Lambda^{.}(L_{\hat \theta}^{*})$ respectively. This integrality puts a strong 
restriction on the form of the transformation $\theta \mapsto \hat \theta$ induced by (gauge) Morita equivalence.
Eventually it will allow us to determine the group governing the Morita equivalences of noncommutative tori. 
But first we need to develop some useful machinery.

 The Grassmann algebra $\Lambda \equiv \Lambda(L^{*})$ 
can be considered as a fermionic Fock space carrying two irreducible representations 
$\Lambda= \Lambda^{even}\oplus\Lambda^{odd} $ of the group $O(d,d|{\mathbb C})$. Namely,  we 
have operators $a^{k}$ of multiplication by $\alpha^{k}$ and operators 
$b_{k}=\frac{\partial}{\partial \alpha^{k}}$ acting on $\Lambda$ and satisfying canonical  anticommutation 
relations 
\begin{equation} \label{CAR}
\{a^{k}, b_{l} \}_{+} = \delta_{l}^{k} \, , \enspace 
\{a^{k}, a^{l} \}_{+} = 0 \, , \enspace 
\{b_{k}, b_{l} \}_{+} = 0 
\end{equation}
that correspond to a Clifford algebra specified by a metric in ${\mathbb R}^{2d}$  
$$
\left( \begin{array}{cccc} 
0&\dots & 0&1\\
0&\dots & 1&0\\
\vdots & \ddots & \vdots & \vdots\\
1 & \dots & 0& 0 \end{array} \right) \, .
$$
This metric is equivalent to a metric with signature $(d,d)$.
One can construct a spinor representation of the group  $O(d,d|{\mathbb C})$ in the Fock space $\Lambda$ 
very much in the same way as one constructs a spinor representation of the Lorentz group starting with a Clifford
algebra specified by  Dirac gamma matrices.
More formally  the group $O(d,d|{\mathbb C})$ can be regarded as a group of automorphisms 
of the  Clifford algebra (\ref{CAR}) (a group of linear canonical transformations).  
A transformation  $W_{g}$ given by the formulas 
\begin{eqnarray}\label{tildeab}
&& W_{g}: a^{k} \mapsto \tilde  a^{k} = M_{l}^{k} a^{l} + N^{kl}b_{l} \, , \nonumber \\ 
&& W_{g}: b_{k} \mapsto \tilde b_{k} = R_{kl}a^{l} + S_{k}^{l}b_{l} 
\end{eqnarray}
preserves  the canonical anticommutation relations (\ref{CAR})  if (and only if)  the matrix 
\begin{equation} \label{g1}
g= \left( \begin{array}{cc}
M&N\\
R&S\\
\end{array} \right) 
\end{equation}
belongs to the group $O(d,d|{\mathbb C})$. For the inverse matrix we have 
\begin{equation} \label{g2}
g^{-1}= \left( \begin{array}{cc}
S^{t}&N^{t}\\
R^{t}&M^{t}\\
\end{array} \right) \, . 
\end{equation}
 One can define a projective action of $O(d,d|{\mathbb C})$ on 
$\Lambda(L^{*})$ assigning to every $g\in O(d,d|{\mathbb C})$ an operator $V_{g}: \Lambda(L^{*}) \to \Lambda(L^{*})$ 
that satisfies 
\begin{equation} \label{Vg}
 V_{g}a^{k}V_{g}^{-1} = W_{g^{-1}}(a^{k}) \, , \enspace 
 V_{g}b_{k}V_{g}^{-1} = W_{g^{-1}} (b_{k})  
\end{equation}
where $W_{g}$ is defined  by formula (\ref{tildeab}). 
The projectivity of this action means that the operators operators $V_{g}$ are defined only up to a constant factor. 
It is possible however to define a double-valued spinor representation of $SO(d,d|{\mathbb C})$ on the Fock space $\Lambda(L^{*})$. 
This can be done by choosing a bilinear form on $\Lambda(L^{*})$ and imposing a requirement that operators $V_{g}$ should preserve 
this form (see Appendix B in \cite{KS} for details). The operators $V_{g}$  fixed this way specify a spinor representation 
of $SO(d,d|{\mathbb C})$.

We also define an action  $\theta \mapsto g\theta = \hat \theta$   of $O(d,d|{\mathbb C})$ on 
the space of  antisymmetric matrices by the formula 
\begin{equation} \nonumber 
\hat \theta = (M\theta + N)(R\theta + S)^{-1}
\end{equation}
where $d\times d$ matrices $M$, $N$, $R$, $S$ correspond to an element $g\in O(d,d|{\mathbb C})$ 
by formula (\ref{g1}).
More precisely, this action is defined on a subset of the space of all antisymmetric matrices 
where the matrix $R\theta + S$ is invertible.

Consider now two noncommutative tori $T_{\theta}$ and $T_{\hat \theta}$ related by a gauge Morita 
equivalence. If $E$ is a $T_{\theta}$-module, $\hat E$ is the corresponding 
$T_{\hat \theta}$-module  related by Morita equivalence  then it follows from (\ref{Cherns}), (\ref{Elliott}) that  
\begin{equation}\label{mues}
\mu(\hat E) = V\mu(E) 
\end{equation} 
where $V_{g}  = V_{1}V_{2}V_{3}V_{4}$ with 
\begin{eqnarray}
&&V_{1}f = exp (-\frac{1}{2} b_{k}\hat \theta^{kj} b_{j})f \nonumber \\
&& V_{2}f = exp(  a^{k}\sigma_{kj}a^{j}) f \nonumber \\
&& V_{3}f = \frac{dim\hat E}{dimE} f(A^{t}\alpha )\nonumber \\
&& V_{4}f = exp(\frac{1}{2} b_{k}\theta^{kj}b_{j})f 
\end{eqnarray}
where $f\in \Lambda$.
The operator $V_{1}$ relates $ \mu(\hat E)$ and $ch (\hat E)$, the operator $V_{4}$ relates 
$\mu(E)$ and $ch(E)$. The operator $V_{2}V_{3}$ relates $ch(\hat E)$ and 
$ch(E)$. The last relation follows from (\ref{Cherns}) if we take into account that we should identify 
 $L_{\theta}$ and $L_{\hat \theta}$ by means of 
 some linear operator  $A$. (Note that  formula (\ref{Elliott}) holds only 
in a standard basis.) 
It is clear from the formulas above that the operators $V_{1}, V_{2}, V_{3}, V_{4}$, and 
hence their product too, are linear canonical transformations. 
We know that $ \mu (\hat E)$ and $\mu (E)$ are integral elements of $\Lambda(L^{*})$. Therefore, 
the operator $V$ transforms integral elements of $\Lambda(L^{*})$ into integral elements 
(i.e. $V$ is an integrality preserving operator). The same is true for the inverse operator $V^{-1}$ 
because $\mu(E)$ and $ \mu(\hat E)$ are on equal footing. Thus, we can say that the linear 
 canonical transformation $V$ corresponds to an element of  $SO(d,d|{\mathbb Z})$. 
We proved that $ \mu(\hat E)$ and $\mu(E)$ are related by a linear canonical transformation 
corresponding to an element of $SO(d,d|{\mathbb Z})$. We denote this element by
$$
g = \left( \begin{array}{cc} 
M&N\\
R & S\\
\end{array} \right) \, .
$$ 
 This transformation, as well as transformations $V_{1}$, $V_{2}$, $V_{4}$, 
preserves the  bilinear form on $\Lambda(L^{*})$  (for $V$ this follows from integrality of $V$ and 
$V^{-1}$, and for   $V_{1}$, $V_{2}$, $V_{4}$ it can be checked directly). This means that 
$V_{3}$ also preserves the form  and therefore 
\begin{equation} \label{dim/dim}
\frac{dim \hat E}{dimE} = |det(A)|^{-1/2} \, .   
\end{equation}
Going from $V_{1}$, $V_{2}$, $V_{4}$ to the corresponding elements of 
$SO(d,d|{\mathbb C})$ we obtain 
\begin{eqnarray*} 
&& 
\left( \begin{array}{cc} 
M&N\\
R & S\\
\end{array} \right) = 
\left( \begin{array}{cc} 
1&\hat \theta\\
0 & 1\\
\end{array} \right) 
\left( \begin{array}{cc} 
1&0\\
-\sigma & 1\\
\end{array} \right)  
\left( \begin{array}{cc} 
(A^{t})^{-1}&0\\
0 & A\\
\end{array} \right)
 \left( \begin{array}{cc} 
1&-\theta\\
0 & 1\\
\end{array} \right) = \nonumber \\ 
&& = \left( \begin{array}{cc} 
(A^{t})^{-1} - \hat \theta \sigma (A^{t})^{-1}&-(A^{t})^{-1}\theta - \hat \theta
( \sigma(A^{t})^{-1} - A)\\
-\sigma(A^{t})^{-1} & \sigma  (A^{t})^{-1}\theta + A \\
\end{array} \right) \, .   
\end{eqnarray*} 
From the last formula one readily obtains 
\begin{equation} \label{newteta} 
\hat \theta = (M\theta + N)(R\theta + S)^{-1} \, , 
\end{equation}
\begin{equation} \label{A}
A = R\theta + S \, , 
\end{equation} 
\begin{equation} \label{sigma}
\sigma = -RA^{t} = -R( R\theta + S)^{t} \, . 
\end{equation}
This proves that whenever two tori $T_{\theta}$ and 
$T_{\hat \theta}$ are gauge Morita 
equivalent the matrices $\theta$ and $\hat \theta$ are connected by a fractional transformation (\ref{newteta})  
corresponding to a subgroup $SO(d,d|{\mathbb Z})\subset SO(d,d|{\mathbb C})$- 
the group of automorphisms of the Clifford algebra $\Lambda(L^{*})$. 
Conversely one can also prove \cite{RS} that if $\theta$ and $\hat \theta$ are connected by a fractional transformation 
(\ref{newteta}) corresponding to some element of  $SO(d,d|{\mathbb Z})$ then the tori $T_{\theta}$ and 
$T_{\hat \theta}$ are gauge Morita equivalent. In the previous section we sketched the proof of the analogous  result for 
Morita equivalences. It follows from the explicit constructions of equivalence bimodules that the ordinary Morita equivalences 
can be always promoted to the gauge Morita equivalences.  One can  choose  a suitable constant curvature connection satisfying 
(\ref{biconnect}) on a Heisenberg module  specifying the equivalence bimodule. 

Summarizing our results we see that {\it gauge Morita equivalence of $d$-dimensional noncommutative 
tori is governed by the group $SO(d,d|{\mathbb Z})$. 
This group acts on matrices $\theta$ by means of 
fractional transformations (\ref{newteta}) and on topological numbers of modules by means of 
a spinor representation.} 

It follows from the above formulas that the curvature tensor, metric tensor, background field $\Phi_{ij}$  and the volume transform according to 
\begin{eqnarray} \label{tr_rules}
&& F_{ij}^{\hat \nabla} = A^{k}_{i}F_{kl}^{\nabla} A^{l}_{j} + \sigma_{ij} \, , \qquad \hat g_{ij} =  A^{k}_{i}g_{kl}A^{l}_{j}  \, , \nonumber \\ 
&& \hat \Phi_{ij} =  A^{k}_{i}\Phi_{kl}  A^{l}_{j} - \sigma_{ij} \, , \qquad        \hat V = V|{\rm det}\, A| \, 
\end{eqnarray}
where $\sigma_{ij}$ and $A^{i}_{j}$ are given in (\ref{sigma}), (\ref{A}).

%%%%%%%%%%%%%%%%%%%%%%%%%%%%%%%%%%%%%%%%%%%%%%%%%%%%%%%%%%%%%%%%%%%%%%%%%%%%%%%%%%%%%%%%%%
\subsection{Invariance of BPS spectrum} \label{BPSMoritasec}
It is straightforward to check that the transformation rules (\ref{tr_rules}) along with  formula  (\ref{dim/dim}) 
imply that {\it the action functional $S_{YM}$ given by (\ref{modifYM}) is invariant under gauge Morita equivalence 
provided the coupling constant changes according to} 
\begin{equation} \label{gYM}
\hat g^{2} = g^{2}\cdot |{\rm det}A|^{1/2} \, .
\end{equation}
This result  extends straightforwardly to the SYM action functional (\ref{ncsYM}).

As one can see directly  from formula (\ref{curv_shift}) the constant curvature connections go into constant curvature ones. 
Thus 1/2 BPS states are mapped again into 1/2 BPS states. Furthermore it follows from the invariance of the 
Yang-Mills action functional and the fact that BPS solutions always minimize the Yang-Mills action for a given set of 
topological numbers that any BPS solution should go to a BPS solution under a gauge Morita equivalence. 
In particular the BPS spectrum should be invariant. In sections \ref{BPSsec}, \ref{Salgsec}, \ref{234sec} we derived explicit formulas 
for BPS energies. One can check directly that they are invariant under transformations  induced by (\ref{newteta}). Namely we can use 
the fact that topological numbers transform according to the spinor representation of $SO(d,d|{\mathbb Z})$ and formulas  
(\ref{tr_rules}), (\ref{gYM}) for transformations of the  continuous moduli $g_{ij}$, $\Phi_{ij}$, $g$. 
To derive the transformation of quantum numbers $m_{j}$, $n^{j}$ one first notices that the transformation rule 
for the translation operators ${\cal P}_{j}$, $p^{j}$ is 
\begin{equation} \label{Ptr}
\hat {\cal P}_{j} = A_{j}^{k}{\cal P}_{k} - 2\pi R_{jk}p^{k}\, , \quad       
\hat p^{j} = B^{j}_{k}p^{k}
\end{equation}
where $A= (R\theta + S)$, $B=(A^{t})^{-1}$. This can be derived directly using the expressions $p^{j} = {\rm Tr} P^{j}$, 
${\cal P}_{j} = {\rm Tr} F_{jk}P^{k}$ and formulas (\ref{tr_rules}). Taking into account the quantization laws 
(\ref{momentum}), (\ref{p'}) one finds from  (\ref{Ptr}) that the numbers $(-n^{j}, m_{i})$ transform in the vector representation 
of $SO(d,d|{\mathbb Z})$: 
\begin{equation} \label {nmtr}
\left( 
\begin{array}{c} 
-\hat n \\
\hat m 
\end{array}
\right) = 
\left(
\begin{array}{cc}
M & N\\
R & S
\end{array}
\right) 
\left( 
\begin{array}{c} 
-n \\
m 
\end{array}
\right) \, . 
\end{equation}

The transformation laws for the topological tensors $\lambda^{i}$, $\lambda^{ijk}$ can be derived from (\ref{Stop}) and (\ref{tr_rules})
\begin{eqnarray} \label{lambda_tr}
&& \hat \lambda^{i} = |det(A)|^{1/2}B^{i}_{j}\lambda^{j} + \lambda^{ijk}\sigma_{jk} \, , \nonumber \\
&& \hat \lambda^{ijk} =  |det(A)|^{1/2} B^{i}_{l}B^{j}_{m}B^{k}_{n}\lambda^{lmn}
\end{eqnarray}
where $\sigma_{jk}$ is given in (\ref{sigma}).

For example for the case of a noncommutative two-torus $T_{\vartheta}^{2}$ and the Morita equivalence corresponding to 
$\vartheta \mapsto -1/\vartheta$ we have in the notations of section \ref{BPSsec} 
\begin{equation}\label{mor}
\begin{array}{c}
m \mapsto n  \, , \quad   n\mapsto - m   \, , \\    
n^{1} \mapsto -m_{1}\, , \quad n^{2} \mapsto - m_{2} \, , \quad 
m_{1} \mapsto -n^{1} \, , \quad m_{2} \mapsto -n^{2} \, , \\ 
R_{2} \mapsto R_{1}\vartheta  \,, \quad   R_{1} \mapsto R_{2}\vartheta  \, , \\ 
\phi \mapsto \phi \vartheta^{2} -  \vartheta  \, , \quad  
\lambda^{1} \mapsto \lambda^{2}  \, , \quad  \lambda^{2} \mapsto -\lambda^{1} \, , \quad 
g^{2} \mapsto g^{2} \vartheta 
\end{array}
\end{equation} 
One readily checks that  formula (\ref{1/4BPS}) for the energy of 1/4 BPS states on  $T_{\vartheta}^{2}$ 
is invariant under this transformation.

In the general case the invariance of BPS spectrum can be checked most easily using the general formula (\ref{energy}) and 
the transformation rules (\ref{tr_rules}), (\ref{Ptr}), (\ref{lambda_tr}).

%%%%%%%%%%%%%%%%%%%%%%%%%%%%%%%%%%%%%%%%%%%%%%%%%%%%%%%%%%%%%%%%%%%%%%%%%%%%%%%%%%%%%%%%%%%%%%%%%%%%%%%%%%%%%%%%%%%%%%%%%% 
\section{Noncommutative instantons} 
\subsection{Instantons on $T_{\theta}^{4}$. Definition and a simple example.} \label{Inst1sec}
Instantons in noncommutative gauge theories were first considered in \cite{NS} for noncommutative Euclidean space
${\mathbb R}^{4}_{\theta}$. In \cite{NS} the ADHM construction for commutative instantons was generalized. Remarkably it 
turned out that the presence of nonzero noncommutativity parameter $\theta^{ij}$ leads to a Nakajima compactification 
of the instantons moduli space. Despite this interesting results the theory of istantons on ${\mathbb R}^{4}_{\theta}$ 
has a mathematical subtlety of dealing with an algebra without a unit element. For this reason we prefer to discuss 
instantons on a noncommutative four-torus $T_{\theta}^{4}$.

Throughout this section we will assume that the Lie algebra $L$ of shifts on the torus is equipped 
with the standard Euclidean metric $g_{ij} = \delta_{ij}$. 
In the commutative gauge theory an instanton on Euclidean ${\mathbb R}^{4}$ (effectively compactified to a 
four-sphere by imposing the boundary conditions at infinity) 
is a connection for which the corresponding curvature has a vanishing self-dual 
part
$$
F^{+}_{jk} \equiv \frac{1}{2}(F_{jk} + \tilde F_{jk}) = 0 \, , \enspace \tilde F_{jk}\equiv \frac{1}{2}\epsilon_{jkmn}F^{mn} \, . 
$$
Respectively an antiinstanton has a vanishing anti-self-dual part $F^{-}_{jk} \equiv \frac{1}{2}(F_{jk} - \tilde F_{jk})$. 
An istanton or antiinstanton solution minimizes the Yang-Mills action for the given Pontryagin number
$$
 q = \frac{1}{32\pi^{2}}\int d^{4}x\, \langle F_{jk}, \tilde F^{jk} \rangle \, . 
$$

For $U(N)$ gauge theories this number is nothing but the appropriately normalized second Chern-number.

When considering $U(N)$ Yang-Mills fields on a torus a vector bundle in addition to the Pontryagin number 
is characterized by the first Chern numbers that are equal to magnetic fluxes $\int dx^{i}dx^{j} {\rm tr} F_{ij}$ and 
the (anti)instanton solutions in general do not minimize the action. One should also  consider solutions 
whose (anti)self dual part of the curvature lies in the $U(1)$ part and is constant. (For example see 
\cite{GurRamg} for a discussion of this kind of solutions). We will show in the next section 
how the bound on action generalizes in the presence of nonzero first Chern numbers.
However the $U(1)$ part  decouples  and one can concentrate on $SU(N)$ instanton solutions.

In noncommutative gauge theory one cannot decouple the $U(1)$ part. Let us illustrate this on a simple example.
Consider a connection $\nabla_{j}$ 
 field on a free module of rank two over $T_{\theta}^{d}$. This connection can be written as 
$ \nabla_{j} = \partial_{j} + i A_{j}^{ab}(\sigma ) $ where $A_{j}^{ab}(\sigma )$, $a, b = 1,2$ 
functions on the underlying  commutative torus $T^{2}$ with values in $2\times 2$ matrices. 
Assume that  $A_{j}^{ab}(\sigma )$  has a form 
$$
A_{j}^{ab}(\sigma ) = \left( \begin{array}{cc} 
a_{j}(\sigma )& 0\\
0 & - a_{j}(\sigma )\\
\end{array} \right) 
$$
that would be analogous to having the vanishing $U(1)$ part in the commutative theory. 
Let us make an infinitesimal gauge transformation parameterized by a $2\times 2$ matrix valued function 
of the same form with $\pm \lambda(\sigma ) $ on the diagonal. In the commutative theory such a transformation 
would be trivial. But in the noncommutative theory we have
$$
\delta A_{j} = {\bf 1}_{2\times 2} \cdot (a_{j}\ast \lambda - \lambda \ast a_{j} )(\sigma )
$$
that is the $U(1)$ part was generated as a result of   the gauge transformation.

In view of the above remarks one can give the following definition of an instanton on a noncommutative space.
{\it An instanton in a noncommutative gauge theory is a connection such that the selfdual part of the 
corresponding curvature tensor is proportional to the identity  operator, i.e.  
$$
F^{+}_{jk} = i \omega_{jk}\cdot {\bf 1} 
$$ 
where $\omega_{jk}$ is a constant matrix with real entries.} 
The antiinstanton is defined the same way by replacing the selfdual part by antiselfdual.

Obviously  by such a definition a constant curvature connection is an instanton and an  antiinstanton 
simultaneously. It is easy however to construct an example of an instanton solution on 
a module that does not admit a constant curvature connection. 
To this end assume that the matrix $\theta_{ij}$ specifying $T_{\theta}^{4}$ is of particular form
$$
(\theta_{ij})  = \vartheta \left( \begin{array}{cccc}
0 &1 & 0 & 0 \\
-1 & 0 & 0 & 0 \\
0& 0 & 0 & -1\\
0& 0 & 1 & 0 
\end{array}  \right) 
$$
where $\vartheta$ is a number. Evidently  $\theta_{ij}$ is an antiselfdual matrix.
Consider a module $E=E_{1}\oplus E_{2}$ that is a direct sum of a rank one free module $E_{1}\cong (T_{\theta}^{4})^{1}$ 
and a module $E_{2}$ considered in section \ref{Projsec}, specified by formulas (\ref{nabla}) (\ref{Ust}). 
For the matrix $\theta_{ij}$ at hand we have the following representation on functions $\phi \in {\cal S}({\mathbb R}^{2})$
\begin{eqnarray*}
&& U_{1} \phi( x_{1}, x_{2}) = \phi(x_{1} + \vartheta, x_{2}) \, \quad U_{2}\phi(x_{1}, x_{2}) = \phi(x_{1}, x_{2}) e^{2\pi i x_{1}} \, , \\
&& U_{3} \phi( x_{1}, x_{2}) = \phi(x_{1} , x_{2} - \vartheta) \, \quad U_{4}\phi(x_{1}, x_{2}) = \phi(x_{1}, x_{2}) e^{2\pi i x_{2}} \, . \\
\end{eqnarray*}
Let $\nabla^{1}_{j}=\delta_{j}$ be the standard connection on $E_{1}$ and let  $\nabla^{2}_{j}:E_{2}\to E_{2}$
be a connection defined as  
$$
\nabla^{2}_{1} = -\frac{2\pi i }{\vartheta }x_{1}\, , \quad \nabla^{2}_{2} = \partial_{1}\, , \quad 
\nabla^{2}_{3} = \frac{2\pi i }{\vartheta }x_{2}\, , \quad \nabla^{2}_{4} = \partial_{2}\, .
$$ 
This connection has 
a constant curvature $[\nabla^{2}_{j} ,\nabla^{2}_{k}  ] = 2\pi i (\theta^{-1})_{jk}$ that is also an antiselfdual tensor. 
Evidently the connection 
$$
\nabla_{j} = \left( \begin{array}{cc} 
\nabla^{1}_{j}& 0 \\
0&\nabla^{2}_{j} \end{array} 
\right) : E \to E
$$
has a vanishing selfdual part of the curvature and hence is an instanton. On the other hand the module $E$ does not admit a 
constant curvature connection because its topological numbers are $\mu (E) = 1 + \alpha^{1}\alpha^{2}\alpha^{3}\alpha^{4}$.

One can generalize this example as follows. Consider a module $E$ over $T_{\theta}^{4}$ (where $\theta$ need not be in general of 
the particular form above) that splits into a direct sum of modules $E=E_{1}\oplus \dots \oplus E_{n}$ such that 
each $E_{a}$ is equipped with a constant curvature connection $\nabla^{a}_{j}$. Consider a block diagonal connection 
$\nabla_{j}$ that acts as $\nabla_{j}^{a}$ on each block $E_{a}$. The curvature of this connection has the form 
$$
F_{ij} = 2\pi i \left( \begin{array}{cccc} 
f^{1}_{ij} \cdot {\bf 1}& 0 & \dots & 0\\
0 & f^{2}_{ij} \cdot {\bf 1}& \dots & 0 \\
\vdots & \vdots & \ddots & \vdots \\
0 &  0& \dots & f^{n}_{ij} \cdot {\bf 1} 
\end{array} \right) 
$$ 
where $f^{a}_{ij}$ are constant tensors. The connection at hand will be an (anti)instanton 
if the tensors $f^{a}_{ij}$  all have the same (anti)selfdual part.

\subsection{Instanton action} \label{Inst2sec}
Instanton and antiinstanton solutions minimize the Euclidean Yang-Mills action functional 
$$
S_{YM} (E, \nabla) = -\frac{1}{4g^{2}}{\rm Tr} F_{jk}F^{jk} 
$$
on an arbitrary projective module $E$ over  $T_{\theta}^{4}$.

A bound on the value of Yang-Mills action can be derived starting with an obvious inequality 
$$
{\rm Tr} (F_{jk} \pm \tilde F_{jk} + i\omega_{jk}\cdot {\bf 1})^{2} \le 0
$$ 
that follows from antihermiticity of $F_{jk}$. Here $\omega_{jk}$ is an arbitrary constant antisymmetric matrix. 
Expanding the product and using the definition of the Chern character we obtain 
$$
S_{YM}  \ge \frac{1}{g^{2}}\left( \mp 2\pi^{2} {\rm ch}_{2} -\frac{1}{8}\omega^{2}{\rm ch}_{0} -\pi \omega^{jk}({\rm ch}_{1})^{\pm}_{jk}  \right) \, .  
$$
Maximizing the right hand side over $\omega_{jk}$ we obtain 
\begin{equation} \label{bound}
S_{YM} \ge S_{\pm}  
\end{equation}
where 
\begin{equation} \label{Spm}
S_{\pm} = \frac{2\pi^{2}}{g^{2}} ( \mp {\rm ch}_{2} +  \frac{ ({\rm ch}_{1})^{\pm}_{jk}({\rm ch}_{1})^{\pm jk}}{2ch_{0}} ) \, .
\end{equation} 
Evidently the bound (\ref{bound}) is saturated on  instanton or antiinstanton solutions.
 The instanton action equals  $S_{+}$ and the antiinstanton one is $S_{-}$.
Note that deriving the above bound we have not really used the fact that we are dealing with a noncommutative torus. Any other 
space for which the expressions ${\rm Tr} {\bf 1}$, ${\rm Tr}F_{ij}$, ${\rm Tr}F_{ij}F_{jk}$ make sense 
would also do.  

For a noncommutative four-dimensional torus $T_{\theta}^{4}$
 the $K$-theory class of a projective module $E$ is given by an element $\mu(E) \in \Lambda^{even}(L^{*})$ that can be written 
explicitly as  
\begin{equation}
\mu(E) = n + \frac{1}{2}m_{jk}\alpha^{j}\alpha^{k} + q\alpha^{1}\alpha^{2}\alpha^{3}\alpha^{4} \, . 
\end{equation}
The corresponding Chern character can be calculated using Elliott's formula and is equal to 
\begin{equation}
{\rm ch}(E) = n + \frac{1}{2}{\rm tr}(\theta m) + q\cdot {\rm Pfaff}(\theta)  + 
\frac{1}{2}(m + q\cdot \tilde \theta)_{jk}\alpha^{j}\alpha^{k} + q \alpha^{1}\alpha^{2}\alpha^{3}\alpha^{4}  
\end{equation}
 where $\tilde \theta_{jk} = \frac{1}{2}\epsilon_{jklm}\theta^{lm}$ and 
${\rm Pfaff}(\theta) = \frac{1}{8} \theta^{jk}\epsilon_{jklm}\theta^{lm}$. 
One can substitute now these expressions  into (\ref{Spm}) to obtain  explicitly  the values  of  instanton/antiinstanton 
actions.  

%%%%%%%%%%%%%%%%%%%%%%%%%%%%%%%%%%%%%%%%%%%%%%%%%%%%%%%%%%%%%%%%%%%%%%%%%%%%%%%%%%%%%%%%%%%

A saturation of bounds of the type (\ref{bound}) is characteristic of BPS solutions in supersymmetric theories. 
We have discussed  this phenomenon in sections \ref{Salgsec}, \ref{234sec}. 
Indeed we can extend an instanton or 
antiinstanton to a solution in a maximally supersymmetric Yang-Mills theory on $T_{\theta}^{4}\times {\mathbb R}^{1}$ 
where the ${\mathbb R}^{1}$ component stands for the time direction. This theory can be obtained by a compactification 
of BFSS matrix model on $T_{\theta}^{4}$. The $10$-dimensional Majorana-Weyl spinors can be decomposed 
with respect to representations of the subgroups $SO(4)\times SO(5,1)\subset SO(9,1)$ corresponding to 
Lorentz transformations in the  compactified and transverse directions:
\begin{equation} \label{spin_dec}
\psi = \left( \begin{array}{c} 1 \\
0 \end{array} \right) \otimes 
\left( \begin{array}{c} \lambda^{\alpha , A}  \\
0 \end{array} \right) +
 \left( \begin{array}{c} 0 \\
1 \end{array} \right) \otimes 
\left( \begin{array}{c} 0 \\
\bar \lambda_{\bar \alpha , A}  \end{array} \right)  
\end{equation}
where $\lambda^{\alpha , A}$ ($\alpha =1,2$) transforms only under the first $SU(2)$ in $SO(4)=SU(2)\times SU(2)$, while 
$\bar \lambda_{\bar \alpha , A}$ changes only under the second $SU(2)$. Or, equivalently,  $\lambda^{\alpha , A}$  
($\bar \lambda_{\bar \alpha , A}$) is a Weyl spinor of left (right) chirality for each $A=1, \dots, 4$. 
Furthermore,  $\bar \lambda_{\bar \alpha , A}$ transforms in a 
 complex conjugate of the $SO(5,1)$  representation 
 $\lambda^{\alpha , A}$ while together  they furnish a  pesudoreal  two-spinor representation (for each $\alpha$ that 
is assumed to be fixed).
The particular details 
of this decomposition will not be important for us here (they can be found for example in  paper \cite{D_lect}). 
We assume that the Dirac $\Gamma$-matrices $\Gamma_{\mu}$ are chosen now according to this decomposition. 
In particular the $SO(4)$ chirality  is defined  
with respect to the matrix 
$$
\tilde \gamma_{5} \equiv \Gamma_{1}\Gamma_{2}\Gamma_{3}\Gamma_{4} \, . 
$$
 
The   supersymmetry transformation $\delta_{\epsilon}$ (\ref{ncSUSY}) in the presence of decomposition (\ref{spin_dec})
takes the form  
 $$
\delta_{\epsilon}\psi = \frac{1}{2}(  F^{-}_{jk}\Gamma^{jk}\mu + F^{+}_{jk}\Gamma^{jk}\bar \mu + 
\Gamma^{jI}[\nabla_{j}, X_{I}]\epsilon + \sigma^{IJ}[X_{I}, X_{J}]\epsilon ) 
$$
where $\mu$ and $\bar \mu$ are components of $\epsilon$ corresponding to the first and second terms in the decomposition 
(\ref{spin_dec}) so that $\tilde \gamma_{5} \mu =\mu$, $\tilde \gamma_{5} \bar \mu = -\bar \mu$. 
From this formula  we see that the solution  
$$
F^{+}_{jk}  = i\omega_{jk}\cdot {\bf 1}\, , \quad \psi^{\alpha} = 0 \, , \quad X_{I} = 0 
$$
is preserved by  supersymmetry transformations $\delta_{\epsilon_{1}} + \tilde \delta_{\epsilon_{2}}$ 
where 
$$
\epsilon = (0, \bar \mu) \, , \qquad \tilde \epsilon = -i\omega_{ij}\Gamma^{ij}\bar \mu \, . 
$$  
An analogous formula with $\epsilon= (\mu, 0)$ gives supersymmetry transformations preserving an 
antiinstanton solution.  
This means that in our conventions (anti)instantons are $1/4$-BPS fields. 
 
%%%%%%%%%%%%%%%%%%%%%%%%%%%%%%%%%%%%%%%%%%%%%%%%%%%%%%%%%%%%%%%%%%%%%%%%%%%%%%%%%%%%%%%%%%%%%%%%%%%%%%%%%%%%%%%%%%%
\section{Noncommutative orbifolds}
\subsection{Noncommutative toroidal orbifolds} \label{Orb1sec}
In this section we will consider  compactifications of M(atrix) Theory on toroidal orbifolds, including 
the case when the underlying torus is noncommutative.

Let $D\subset {\mathbb R}^{d}$ be a $d$-dimensional lattice  embedded in ${\mathbb R}^{d}$ and let $G$ be a finite 
group acting on   ${\mathbb R}^{d}$ by linear transformations mapping the lattice $D$ to itself. 
For an element $g\in G$ we will denote the corresponding representation matrix $R_{i}^{j}(g)$.
One can write down constraints describing compactification of M(atrix) theory on the orbifold $T^{d}/G$, 
where $T^{d}={\mathbb R}^{d}/D$: 
\begin{equation} \label{e1}
X_{j} + \delta_{ij}2\pi \cdot {\bf 1} = U_{i}^{-1}X_{j}U_{i} \, , 
\end{equation}
\begin{equation} \label{eq1'}
  X_{I} = U_{i}^{-1}X_{I}U_{i} \, \qquad \psi_{\alpha} = U_{i}^{-1}\psi_{\alpha}U_{i} \, ,
\end{equation}
\begin{equation} \label{e2}
R_{i}^{j}(g)X_{j} = W^{-1}(g)X_{i}W(g) \, , 
\end{equation}
\begin{equation} \label{eq2'}
 \Lambda_{\alpha \beta}(g)\psi_{\beta} = W^{-1}(g)\psi_{\alpha}W(g) \, , \quad X_{I} = W^{-1}(g) X_{I} W(g) \, . 
\end{equation}
Here  $i,j =1, \dots , d$ are indices for directions along the torus , $I= d+1,\dots , 9$ is an index corresponding to  the transverse 
directions, $\alpha$ is a spinor index; $\Lambda_{\alpha \beta}(g)$ is the matrix of spinor representation of $G$ obeying 
$\Lambda^{\dagger}(g)\Gamma^{i}\Lambda(g)=R_{ij}(g)\Gamma_{j}$; $ U_{i}$, $W(g)$ - unitary operators.  
One can check that the quantities $U_{i}U_{j}U_{i}^{-1}U_{j}^{-1}$ 
commute with all $X_{i}$, $X_{I}$, and $\psi_{\alpha}$. It is natural to set them to be proportional to the identity operator. 
This gives us defining relations of  a noncommutative torus
$$
U_{j}U_{k} = e^{2\pi i\theta_{jk}}U_{k}U_{j} \, .  
$$
It is convenient to work with linear generators $U_{\bf n}$ that can be expressed in terms of products of $U_{i}$.  
One can further check that expressions $W(gh)W^{-1}(g)W^{-1}(h)$ and $W^{-1}(g)U_{\bf n}W(g)U^{-1}_{R^{-1}(g){\bf n}}$
also commute with all  fields $X_{i}$, $X_{I}$, $\psi_{\alpha}$. We assume that these expressions are proportional 
to the identity operator. This leads us to the following relations  
\begin{eqnarray}
&&W(g)W(h) = W(gh) e^{i\phi(g,h)} \, , \nonumber\\ 
&& W^{-1}(g)U_{\bf n}W(g) = U_{R^{-1}(g){\bf n}} e^{i \chi({\bf n},g)}
\end{eqnarray}
where $\phi(g,h)$, $\chi({\bf n},g)$ are constants.
The first equation means that operators $W(g)$ furnish a projective representation of $G$. 
It follows from these equations that the matrix $\theta$ is  invariant under the group action $R(g)$.  
In this review we will confine ourselves to the case  of vanishing  cocycles $\phi$ and $\chi$. (See \cite{HoWu}, 
\cite{Doug_discr} for a discussion of cases when  cocycle $\phi$ does not vanish.
(Note that for cyclic groups both cocycles are always trivial. This means that they can be absorbed into 
redefined  generators.)

One can define an algebra of functions on a  noncommutative orbifold as an algebra generated by 
the operators $U_{\bf n}$ and  $W(g)$ satisfying  (\ref{un}) and 
\begin{equation} \label{WU}
 W^{-1}(g)U_{\bf n}W(g) = U_{R^{-1}(g){\bf n}} \, , 
\end{equation}
\begin{equation}
  W(g)W(h) = W(gh) \, .
\end{equation}
This construction in mathematics is called a crossed product. Namely, the algebra generated 
by $U_{\bf n}$, $W(g)$ is a crossed product of the algebra $T_{\theta}^{d}$ and the finite group 
$G$  whose action  on $T_{\theta}^{d}$ is specified by means of representation $R$. We will denote 
this algebra as $T_{\theta}\rtimes_{R} G$. Again we remark here that allowing central extensions leads to  
a more general case of so called twisted crossed products.

The algebra $T_{\theta}\rtimes_{R} G$ can be equipped with an involution $*$ by setting 
$U_{\bf n}^{*} = U_{-{\bf n}}$, $W^{*}(g) = W(g)$. This makes it possible to embed these algebras 
into the general theory of $C^{*}$ algebras. 
A projective module over  an orbifold can be considered as a projective module $E$
over $T_{\theta}$ equipped with operators $W(g)$, $g\in G$ satisfying (\ref{WU}).
The equations (\ref{e1}), ({\ref{e2})  mean that $X_{i}$ specifies
a $G$- equivariant connection  on $E$, i.e. $X_{j} = i\nabla_{j}$ where $\nabla_{j}$ is  a $T_{\theta}$-connection  satisfying 
\begin{equation}\label{eqcon}
R_{i}^{j}(g)\nabla_{j} = W^{-1}(g)\nabla_{i}W(g) \, .
\end{equation}
The fields $X_{I}$ are endomorphisms of $E$, commuting both with $U_{\bf n}$ and $W(g)$ and the spinor fields 
 $\psi_{\alpha}$ can be called equivariant spinors.

Let us comment here  on the supersymmetry of these compactifications. 
The surviving supersymmetry transformations are transformations (\ref{dynsusy}), (\ref{kinsusy})   
corresponding to invariant spinors $\epsilon$, i.e. the ones satisfying 
$\Lambda(g)\epsilon = \epsilon$. For $d=4$, $6$  this equation has a nontrivial solution 
provided the representation $R(g)$  lies  within  an $SU(2)$ or $SU(4)$ subgroup respectively.
 The possible finite groups $G$ that can 
be embedded in this way are well known. Those include the examples of ${\mathbb Z}_{2}$ and 
${\mathbb Z}_{4}$ four-dimensional orbifolds.

\noindent {\bf Example}. Consider  ${\mathbb Z}_{2}$ orbifolds of noncommutative tori $T_{\theta}^{d}$. 
The algebra of functions on this orbifold $T_{\theta}\rtimes {\mathbb Z}_{2}$ is generated by the torus generators $U_{\bf n}$ 
and a ${\mathbb Z}_{2}$ generator $W$ obeying the relations
\begin{equation} \label{z2}
W^{2} = 1 \, , \qquad WU_{\bf n} W = U_{-\bf n} \, . 
\end{equation}
A map $U_{\bf n} \mapsto U_{-\bf n}$ can be extended to an automorphism of the whole algebra $T_{\theta}^{d}$. 
We will denote this automorphism by $w$ (do not confuse it with the involution $*$ that is an antiautomorphism).  
A general element of algebra $T_{\theta}\rtimes {\mathbb Z}_{2}$ can be written as a formal linear combination 
$a_{0} + a_{1}W$ where $a_{0}, a_{1}\in T_{\theta}^{d}$. A multiplication of two such expressions is specified by the formula 
$$
(a_{0} + a_{1}W)(a_{0}' + a_{1}'W) = (a_{0}a_{0}' + a_{1}w(a_{1}')) + (a_{0}a_{1}' + a_{1}w(a_{1}'))W \, . 
$$

One can consider  a module over $T_{\theta}^{d}\rtimes {\mathbb Z}_{2}$ as  a module over 
$T_{\theta}$ equipped with an operator $W$ satisfying (\ref{z2}). 
As an example of  a module over  $T_{\theta}^{2}\rtimes {\mathbb Z}_{2}$ consider a module $E_{n,m}$ over $T_{\theta}^{2}$ 
defined in (\ref{U's}). An operator representing  $W$ can be defined as 
$$
W \phi_{j} (x) = \phi_{-j}(-x) \, . 
$$
One can check that the connection (\ref{2dconn}) is equivariant, i.e. satisfies 
$$
W\nabla_{j} W= -\nabla_{j} \, .
$$ 

One can generalize the previous example and consider   
\begin{equation} \label{ex}
W \phi_{j} (x) = exp\left( \pi i k_{1}\frac{a\theta - b}{n-m\theta}k_{2}\right) Z_{1}^{k_{1}}Z_{2}^{k_{2}} \phi_{-j}(-x) 
\end{equation}
where the operators $Z_{i}$ defined in (\ref{Z's}) are generators of endomophisms.
It is easy to check that thus defined $W$ also satisfies (\ref{z2}). 

The above constructions can be generalized to any Heisenberg module over $T_{\theta}^{d}$ as follows. 
In section \ref{Projsec} we defined Heisenberg modules over noncommutative tori by constructing 
an action of torus generators on  functions $f(x)\in {\cal S}(G)$. It is straightforward to check that 
the operator $W_{0} : f(x) \mapsto f(-x)$ satisfies the necessary condition (\ref{z2}). 
Moreover one can modify this operator and consider $W = Z_{(\nu, \tilde \nu)}W_{0}$ where  
$Z_{(\nu, \tilde \nu)}$, $(\nu, \tilde \nu)\in \Gamma^{*}$ is an endomorphism that acts on $f(x)$ according to (\ref{U})
 (see section \ref{Projsec} for details). For $d=2$ the example (\ref{ex}) above is precisely of this form.

%%%%%%%%%%%%%%%%%%%%%%%%%%%%%%%%%%%%%%%%%%%%%%%%%%%%%%%%%%%

\subsection{ K-theory of orbifolds } \label{Orb2sec}
%%%%%%%%%%%%%% Baum-Connes 
In this section we  discuss K-theory of commutative orbifolds. 
 Consider a compact manifold $\cal M$ and a right action of a finite group $G$ on it.
The $K$-theory of the corresponding orbifold is   called a $G$-equivariant K-theory on $\cal M$. 
(Everywhere we  consider the complex K-theory.) 
The equivariant K-theory on $\cal M$ is equivalent to K-theory of noncommutative algebra 
$C({\cal M})\rtimes G$ (a noncommutative $C^{*}$ algebra defined as a crossed product of a 
commutative 
algebra $C({\cal M })$ of  functions on $\cal M$ and the group $G$ acting on this algebra).
The formal definition in terms of the Grothendieck construction discussed in section  \ref{Ksec} 
can  be applied directly to the algebra $C({\cal M})\rtimes G$ to define an orbifold $K$-group. 
However there is more to be said about the relation of $K^{0}({\cal M})$ with the orbifold 
$K$-group.

The group $K(C({\cal M})\rtimes G)\otimes {\mathbb C}$ can be expressed in homological terms. 
(Multiplication by $\mathbb C$ means that we disregard finite order elements in the K-group.)
For the standard K-theory the Chern character defines a map  
\begin{equation} 
{\rm ch}: K^{0}({\cal M}) \to  H^{even}({\cal M}, {\mathbb Z}) \, . 
\end{equation}  
To define an analogue of this map in the equivariant case one should consider a kind of 
equivariant cohomology (``delocalized'' equivariant cohomology) $H^{i}_{G}({\cal M})$. 
Following \cite{BaumConnes} we will give a direct geometric definition of this 
cohomology. (It can also be defined  algebraically as cyclic cohomology of 
$C({\cal M})\rtimes G$.) The definition can be given in the following way. For $g \in G$
we define ${\cal M}^{g}$ as the set of points $x\in {\cal M}$ obeying $x g = x$. Let us 
define $\hat {\cal M}$ as a disjoint union $\hat {\cal M} = \cup_{g \in G} {\cal M}^{g}$. 
One can construct an action of the group $G$ on $\hat {\cal M}$. (If $x g = x$ then 
$(xh)(h^{-1}g h)=xh$. Therefore, one can say that an element $h\in G$ specifies a map from ${\cal M}^{g}$ 
to ${\cal M}^{h^{-1}g h}$.) The group $H^{even}_{G}({\cal M} )$ can be defined as the $G$-invariant 
part of even-dimensional cohomology of $\hat {\cal M}$: 
$$
H^{even}_{G}({\cal M}) = H^{even}(\hat {\cal M}; {\mathbb C}) = \left(  \bigoplus_{g \in G} 
H^{even}({\cal M}^{g}, {\mathbb C})\right)^{G} \, .
$$ 
One can define the equivariant Chern character 
\begin{equation} \label{eqch}
ch_{G} : K^{0}_{G}({\cal M}) \to H^{even}_{G}({\cal M})
\end{equation}
and prove that this character induces an isomorphism between $K^{0}_{G}({\cal M})\otimes {\mathbb C}$ and
$H^{even}_{G}({\cal M})$. The definition of $ch_{G}$ can be given in the following way. 
We represent an element of $K^{0}_{G}({\cal M})$ as a $G$-equivariant vector bundle $E$ over $\cal M$. 
This means that an element $g\in G$ acts on $E$ defining a linear map of a fiber over 
$x\in {\cal M}$ into a fiber over $xg\in {\cal M}$. In particular if $x\in {\cal M}^{g}$ then the fiber $E_{x}$ 
over $x$ is mapped into $E_{x}$ itself  by $g$:
$$
g : E_{x} \to E_{x} \, , \quad x\in {\cal M}^{g} \, .
$$
Let $\lambda_{1}, \lambda_{2}, \dots , \lambda_{s}$ be the set of distinct eigenvalues of this linear transformation.
Then, we have a direct sum  decomposition into eigenspaces: $E_{x} = E_{x}^{1}\oplus E_{x}^{2} \oplus \dots \oplus E^{s}_{x}$. 
At the level of vector bundles one has $E|_{{\cal M}^{g}} = E^{1}\oplus E^{2}\oplus \dots \oplus E^{s}$. 
The element $g$ acts on $E^{i}$ by multiplication by $\lambda_{i}$. Following \cite{BaumConnes} we define 
$ch_{G}^{g}(E)\in  H^{even}({\cal M}^{g}; {\mathbb C})$ by 
\begin{equation} \label{char}
ch_{G}^{g}(E)=\sum_{i=1}^{s}\lambda_{i} ch(E^{i})
\end{equation}
where $ch(E_{i})$ is the ordinary (non-equivariant) Chern character of $E^{i}$. 
Then the equivariant Chern character  (\ref{eqch}) is defined as the direct sum 
$$
ch_{G}(E)=\bigoplus_{g \in G} ch_{G}^{g}(E) \, . 
$$

Let us illustrate this construction in the case when 
${\cal M}=T^{d}$ and $G={\bf Z}_{2} = \{ \b{0},\b{1}\} $ where $\b{1}$ acts as $x\mapsto -x$. 
In this case ${\cal M}^{g}$ is equal to $T^{d}$ for $g = \b{0}$ and ${\cal M}^{g}$ contains
$2^{d}$  points for $g = \b{1}$. We see that $H^{even}(\hat {\cal M})$ consists of even-dimensional 
cohomology of $T^{d}$ and of $2^{d}$ summands $H^{0}(pt ; {\mathbb C}) = {\mathbb C}$. All elements 
of $H^{even}(\hat {\cal M})$ are ${\bf Z}_{2}$-invariant. Therefore, $H^{even}_{\bf Z_{2}}(T^{d} )$ has the dimension 
$2^{d-1} + 2^{d} = 3\cdot 2^{d-1}$.  Therefore (as the K-group at hand has no torsion)  
$$
K^{0}_{{\mathbb Z}_{2}}(T^{d}) = K_{0}(C(T^{d})\rtimes {\mathbb Z}_{2}) = {\mathbb Z}^{3\cdot 2^{d-1}} \, .
$$
 
In the above considerations we talked about the group $K^{0}$. However one can prove  corresponding 
statements for the group $K^{1}$ and cohomology groups $H^{odd}$. In particular 
$H^{odd}_{{\mathbb Z}_{2}}(T^{d})=0$. This agrees with the fact that $K^{1}_{{\mathbb Z}_{2}}(T^{d})=0$. 

%%%%%%%%%%%%%%%%%%%%%%%%%%%%%%%%%%%%%%%%%%%%%%%%%%%%%%%%%%%%%%%%%%%%%%%%%%%%%%%%%%%%%%%%%%%%%%%%%%%%%%%%%%
\subsection{K-theory of noncommutative ${\mathbb Z}_{2}$ orbifolds} \label{Orb3sec}
In this section we will explain how $K$-theoretic invariants can be constructed for noncommutative 
toroidal ${\mathbb Z}_{2}$ orbifolds. For shortness we will adopt a new notation for the algebra 
of functions on these orbifolds $B_{\theta}^{d} \equiv T_{\theta}\rtimes {\mathbb Z}_{2}$. 
The action of ${\mathbb Z}_{2}$ on $T_{\theta}^{d}$ is   specified by $w : U_{\bf n} \mapsto U_{-\bf n}$. 

First of all one can calculate the K-groups of  $B_{\theta}^{d}$ \cite{FW}. As usual it does not change under 
continuous variations of the matrix $\theta$ and 
\begin{eqnarray*} 
&& K_{0}(B_{\theta}^{d}) = K_{0}(B_{\theta=0}^{d}) = K_{0}(C(T^{d})\rtimes {\mathbb Z}_{2}) = {\mathbb Z}^{3\cdot 2^{d-1}} \, , \\
&& K_{1}(B_{\theta}^{d}) = K_{1}(B_{\theta=0}^{d}) =  K_{1}(C(T^{d})\rtimes {\mathbb Z}_{2}) = 0  \, . 
\end{eqnarray*}
We explained already that a module over $B_{\theta}^{d}$ is a module over $T_{\theta}^{d}$ equipped with an operator $W$ 
satisfying (\ref{z2}). In this construction projective modules over $T_{\theta}^{d}$ correspond to projective modules 
over $B_{\theta}^{d}$. At the end of section \ref{Orb1sec} we introduced $B_{\theta}^{d}$-modules constructed via 
Heisenberg modules over $T_{\theta}^{d}$. It seems that all other  $B_{\theta}^{d}$-modules can be obtained 
as direct sums of these modules (assuming $\theta$ is irrational). For $d=2$ this can be derived from the results of \cite{Walters}.

Of course topological numbers of the underlying  $T_{\theta}^{d}$-module can be considered as topological numbers 
of the $B_{\theta}^{d}$-module. However this remark gives only $2^{d}$ topological numbers out of 
$3\cdot 2^{d-1}$ that we need. The remaining $2^{d-1}$ topological numbers can be constructed in the following way.
First one notices that in addition to the trace $\tau$ inherited from the canonical trace $\rm Tr$ on $T_{\theta}^{d}$: 
$$
\tau ( a_{0} + a_{1}W ) = {\rm Tr}(a_{0})   	
$$
we have $2^{d-1}$ new unbounded traces on $B_{\theta}^{d}$ denoted $\tau_{\epsilon}$. 
They are labeled by a sequence $\epsilon = (\epsilon_{1}, \dots , \epsilon_{d})$ where each 
$\epsilon_{i} = 0,1$. The traces are defined by the following formula 
\begin{equation} \label{taueps1}
\tau_{\epsilon} (a_{0} + a_{1}W) = 2\phi_{\epsilon}(a_{1})
\end{equation}
where $\phi_{\epsilon}$ is a linear functional on $T_{\theta}^{d}$ specified by the rule 
\begin{equation}\label{taueps2}
\phi_{\epsilon}(U_{\bf n}) = \prod_{i=1}^{d} (-1)^{\epsilon_{i}n_{i}} \, . 
\end{equation}
One can check that indeed the functionals $\tau_{\epsilon}$ satisfy the defining property of trace. 
One essential difference of this traces from the trace $\tau$ is that they are not positive definite. 

In the commutative situation when $\theta = 0$ and $U_{\bf n} = e^{i{\bf n}\cdot {\bf x}}$ the functional 
$\phi_{\epsilon}$ is nothing but the functional evaluating the value of the function at the fixed point 
$(\pi \epsilon_{1}, \dots , \pi \epsilon_{d})$.

Using these traces  we can construct new invariants of projective modules over $B_{\theta}^{d}$: 
\begin{equation} \label{newinv}
\tau_{\epsilon}(E) \equiv \tau_{\epsilon}(P)
\end{equation}
 where $P$ is the projector specifying $E$. One can prove that these invariants are always integer. 
They are related to the numbers of D-branes ``sitting at orbifold singularities''.

(Recall that any projective module $E$ over $B_{\theta}^{d}$ can be represented as a direct summand in a free module 
$(B_{\theta}^{d})^{N}$. In other words $E = P (B_{\theta}^{d})^{N}$ where $P$ is an orthogonal  projection: 
$P^{2} = P$, $P^{*} = P$. This projector is an element of a matrix algebra $Mat_{N}(B_{-\theta}^{d})$. 
Any trace on the algebra $B_{\theta}^{d}$ gives rise when combine with a matrix trace to a trace  
on the matrix algebra at hand. 
The value of this trace on the projector  $P$ is a $K$-theoretic invariant.)

In the commutative case one can relate the Chern character $ch_{G}$, $G={\mathbb Z}_{2}$ considered in the previous section  with invariants 
$ \tau_{\epsilon}(E)$. Namely    the invariants $\tau_{\epsilon}(E)$ correspond to characters  of the orbifold group 
representations on the fibers over fixed points in accordance with (\ref{char}).

\section{Literature}
In the present review we did not try to provide the main text with historically correct references. Instead we 
concentrated on references to papers complementing the material we discuss. However in this section we will attempt 
to review to a certain extent the existing literature on the topics considered in the text. We apologize to those authors 
whose work we unintentionally forgot to mention. 

The BFSS matrix model of M-theory was introduced in paper \cite{BFSS} and is reviewed in 
\cite{Banks_rev}, \cite{Banks_TASI}, \cite{Susskind_rev}, \cite{Taylor_rev1}, \cite{Taylor_rev2}, \cite{Bigatti}.
The IKKT matrix model of type IIB string theory was introduced in \cite{IKKT} and further developed in \cite{IIBstrfield}, 
\cite{IIBspace-time}, \cite{Mak1}. See \cite{IIBrev} for a review.

The book \cite{Connesbook} is up to date the most comprehensive textbook on noncommutative geometry. See also 
papers \cite{Connesrev1}, \cite{Connesrev2} for recent reviews. 
The smooth structure on noncommutative tori, connections and Chern character were first introduced in the seminal paper 
\cite{Connes1}. 
Most of the standard results 
about noncommutative tori we have been quoting throughout the review we learned from   papers \cite{Connes1}, \cite{RieffelC*}, 
\cite{RieffelProj},  \cite{ConnesRieffel}, \cite{PV}, \cite{Elliott}.
 See also \cite{RieffelRev} for a review of 
results about noncommutative tori. The results about the duality group $SO(d,d|{\mathbb Z})$ 
that governs Morita equivalence of noncommutative 
tori discussed in sections \ref{Moritasec}, \ref{GMoritasec}  
were obtained in papers \cite{ASMorita}, \cite{RS}. A  duality theorem generalizing $SO(d,d|{\mathbb Z})$ duality 
 was proposed in \cite{ASncsusy}. 
The classification of modules admitting constant curvature connections 
(sections \ref{Ksec}, \ref{cccsec}) was worked out  in \cite{KS} (Appendix D) and \cite{AstSchw}. 
The picture of Heisenberg modules over noncommutative tori as deformed vector bundles (section \ref{Deformsec}) 
and its relation with Morita equivalence induced duality was developed in \cite{Ho}, \cite{MorZum}, \cite{BrMorZum}, \cite{BrMorZum2}.

Compactification of M(atrix) Theory on noncommutative tori was originally considered in \cite{CDS} and further studied in the forthcoming 
papers \cite{DougHull}, \cite{ChKr}, \cite{KOk}, \cite{Li}, \cite{Berkooz}, \cite{Casalb}, \cite{HoWu}, \cite{HoWuWu}. 
The emergence of noncommutative  geometry from the string theory point of view was understood in \cite{Schomerus},
\cite{ChuHo}, \cite{SeibWitt}. The last paper  contains a number of important insights into the question and  was  
especially influential.

A  background of the kind that is used  in section \ref{Backgrsec} was first considered in  \cite{BFSS}, \cite{branes}. 
The M(atrix) theory in this background was shown to lead to a noncommutative Yang-Mills theory in papers  \cite{Li1}, \cite{ncIIB}, \cite{biloc} 
(section \ref{Backgrsec}). 
The approach to noncommutative Yang-Mills theory in terms of  a limit of finite-dimensional matrices was developed in \cite{Mak2}, 
\cite{Mak3}, \cite{Mak4}.

The BPS spectrum of Matrix theory compactified on noncommutative tori was studied in papers \cite{Ho}, \cite{HofVerI}, 
\cite{HofVerII}, \cite{KS}, \cite{BrMor}, \cite{susyalg}, \cite{1/4}, \cite{PiolineAS}. Our discussion  is based on 
papers \cite{KS} (section \ref{BPSsec}, \cite{susyalg} (sections \ref{Salgsec} - \ref{234sec}, \ref{BPSMoritasec}), \cite{HofVerII} 
(sections \ref{Translsec}, \ref{BPSMoritasec}). Geometric quantization is reviewed in \cite{Kirillov}. Connection between  
geometric quantization and topological terms in field theory was studied in \cite{Gawedzki}.

Instantons in noncommutative Yang-Mills theory were first considered in \cite{NS}; see \cite{Nikita_rev} for a recent review. 
Our  elementary discussion of instantons on a noncommutative four-torus  
 (sections \ref{Inst1sec}, \ref{Inst2sec}) closely follows first sections of paper \cite{AstNekSchw} in which a noncommutative 
geometry generalization of Nahm transform was studied.

Compactification of M(atrix) theory on noncommutative toroidal orbifolds (sections \ref{Orb1sec} - \ref{Orb3sec}) 
was studied in papers \cite{z2}, \cite{moduli}. Our exposition of equivariant K-theory in section \ref{Orb2sec} 
follows \cite{BaumConnes}. For $d=2$ the invariants (\ref{newinv}) were found and there properties were studied in 
\cite{Walters}. For a general case see \cite{z2}.

%%%%%%%%%%%%%%%%%%%%%%%%%%%%%%%%%%%%%%%%%%%%%%%%%%%%%%%%%%%%%%%%%%%%%%%%%%%%%%%%%%%%%%%%%%%%%%%%%%%%%%%%%%%%%%%%%%%%%%%%%

\end{document}